\documentclass[10pt,journal,compsoc]{IEEEtran}
\usepackage{setspace}
\usepackage{amsmath}
\usepackage{amsfonts}
\usepackage{CJK}
\usepackage{indentfirst}
\usepackage{bm}
\usepackage{graphicx}
\usepackage{subfigure}
\usepackage{booktabs}
\usepackage{multirow}
\usepackage{amssymb}
\usepackage{color}
\usepackage{float}
\usepackage{multicol}
\usepackage{fancyhdr}
\ifCLASSINFOpdf
\else
\fi

\begin{document}
\title{Optimal Transport for Unsupervised Denoising Learning}

\author{Wei Wang*, 
        Fei Wen*,
        Zeyu Yan*,
        and Peilin Liu
\IEEEcompsocitemizethanks{
\IEEEcompsocthanksitem *Contributed equally to this work. Corresponding author: Fei Wen.\\
W. Wang, F. Wen, Z. Yan and P. Liu are with the School of Electronic Information and Electrical Engineering, Shanghai Jiao Tong University, Shanghai, China, 200240.
E-mail: wangwei0803@sjtu.edu.cn; wenfei@sjtu.edu.cn; zeyuyan@sjtu.edu.cn; liupeilin@sjtu.edu.cn.
}
}


\IEEEtitleabstractindextext{%
\begin{abstract}
Recently, much progress has been made in unsupervised denoising learning. However, existing methods more or less rely on some assumptions on the signal and/or degradation model, which limits their practical performance. How to construct an optimal criterion for unsupervised denoising learning without any prior knowledge on the degradation model is still an open question. Toward answering this question, this work proposes a criterion for unsupervised denoising learning based on the optimal transport theory. This criterion has favorable properties, e.g., approximately maximal preservation of the information of the signal, whilst achieving perceptual reconstruction. Furthermore, though a relaxed unconstrained formulation is used in practical implementation, we prove that the relaxed formulation in theory has the same solution as the original constrained formulation. Experiments on synthetic and real-world data, including realistic photographic, microscopy, depth, and raw depth images, demonstrate that the proposed method even compares favorably with supervised methods, e.g., approaching the PSNR of supervised methods while having better perceptual quality. Particularly, for spatially correlated noise and realistic microscopy images, the proposed method not only achieves better perceptual quality but also has higher PSNR than supervised methods. Besides, it shows remarkable superiority in harsh practical conditions with complex noise, e.g., raw depth images. Code is available at \textit{https://github.com/wangweiSJTU/OTUR}.
\end{abstract}

}

\maketitle\thispagestyle{fancy}

\IEEEdisplaynontitleabstractindextext

\section{Introduction}
\label{sec:introduction}

\IEEEPARstart{I}{mage} restoration is a fundamental problem in low-level computer vision, which is crucial for many subsequent high-level tasks. The past few years have witnessed great progress in image restoration, which largely benefits from sophisticated deep neural networks and large amounts of paired training data \cite{dncnn,mao2016image,zhang2020residual}. However, in many real-world applications, it is difficult or expensive to collect noisy-clean image pairs for supervised learning. In this case, synthesized data can be used, but the gap between synthesized and real-world data would fundamentally limit the performance of the learned restoration model on realistic data. 

Traditional restoration methods, which do not rely on data driven learning, are typically designed by exploiting certain structures of the signal or properties of noise. Widely used signal structures include sparsity on certain basis, low-rankness, spatiotemporal smoothness, or self-similarity, while popular noise properties include Gaussianity, or spatially independent i.i.d. noise. Particularly, the representative non-local means (NLM) methods, e.g., BM3D \cite{nlm},\cite{bm3d}, have demonstrated excellent effectiveness in denoising i.i.d. Gaussian noise. However, the performance of these methods is limited when the prior assumption on the signal and noise deviates from real-world data. Moreover, the performance of these methods is closely related to hyper-parameter setting, e.g., the degree of sparsity, low-rankness or smoothness. More recently, deep image prior (DIP) \cite{dip} has been proposed to use a network as handcrafted prior, which has achieved impressive results but requires optimizing the network for each noisy image.

To address the difficulty in collecting paired noisy-clean training data in some practical applications, data driven leaning methods without requiring paired training data have recently attracted much attention. For example, the noise-to-noise (N2N) method \cite{n2n} learns to restore images by only looking at corrupted noisy samples. To further remove the requirement of paired noisy training data in N2N, the noise-to-void (N2V) and noise-to-self (N2S) methods \cite{n2v,n2s} learn restoration models in a self-supervised manner. Moreover, a number of variants of these methods have been developed, e.g., \cite{mbvd,s2s,esur,laine2019high,wu2020unpaired}.

Though much progress has been made in restoration learning in the condition without paired training data, existing methods more or less rely on some assumptions of the signal and/or degradation model. For example, the N2N method \cite{n2n}, as well as \cite{mbvd,esur}, requires paired noisy training data, whilst the self-supervised methods \cite{n2v,n2s,laine2019high,wu2020unpaired} assume that the noise is spatially independent and/or the noise type is a priori known. Hence, the performance of these methods is limited in some practical applications, especially when the noise or degradation is complex.

Unsupervised restoration learning in the absence of any prior knowledge on the degradation model is still a challenging problem. This work is motivated by the following open question:

\begin{itemize}
	\item[$\bullet$] \emph{How to construct an optimal criterion for unsupervised restoration learning in the absence of any prior knowledge on the degradation model?}
\end{itemize}

Toward answering this question, this work proposes an unsupervised learning criterion for restoration from the optimal transport theory. The proposed criterion not only has good theoretical properties, but also has favorable practical performance even compared with supervised methods in denoising various synthetic and real-world images. 

The main contributions of this work are as follows.

First, we propose an unsupervised criterion for restoration learning based on the optimal transport theory, which minimizes the transport cost between the input and reconstruction output under the constraint that the reconstruction output has the same distribution as natural clean samples.

Second, we compare the proposed criterion with the ideal supervised criterion to analyze its properties, which shows that it can well preserve the information of the signal whilst achieving perceptual reconstruction.

Third, the proposed criterion is relaxed in implementation with the use of GAN. In theory we show that, with Wasserstein-1 distance, the relaxed formulation has the same solution as the original criterion.

Finally, we apply the new method to various denoising applications, including synthesized images with different noise conditions, and real-world photographic, microscopy, depth, and raw depth images. The results demonstrate that the new method even compares favorably with supervised methods, with PSNR approaching that of supervised methods while having better perceptual quality. Noteworthily, for spatially correlated noise and realistic microscopy images, it not only achieves better perceptual quality to preserve more details but also has higher PSNR than supervised methods. Particularly, in denoising raw depth images with severe complex noise, the new method shows remarkable performance with significant superiority over existing methods.

Our method not only does not use any hand-crafted model for the signal or any prior knowledge on the noise, but also has sound theoretic foundation from the optimal transport theory. The proposed criterion can serve as an optimal criterion for unsupervised restoration learning in the absence of any prior knowledge on the degradation model.

\section{Related Work}

We review major categories of restoration methods which do not require paired training data, along with GAN based methods that related to this work. 

\textbf{Hand-crafted model based methods}. 
Most traditional model based methods do not rely on data driven learning but use prior model on the signal and/or noise, e.g., sparsity, low-rankness, spatio-temporal smoothness, or self-similarity of the signal \cite{buades2005review, mip_deblur}. For example, traditional image filters, such as Gaussian, median, bilateral and other filters, use local averaging to reduce the noise level. The NLM method \cite{nlm} estimates the central pixel of each patch by weighted average of the central pixels from similar patches. The BM3D method \cite{bm3d} filters out the high-frequency parts of stacked similar patches to achieve more robust denoising. The performance of these methods relies heavily on hyper-parameter selection \cite{lebrun2012analysis}. More recently, the DIP \cite{dip} method uses a network as handcrafted prior.

\textbf{Supervised learning with noisy pairs}. 
Instead of using noisy-clean image pairs, the N2N method \cite{n2n} uses noisy-noisy pairs for training. In principle, for zero-mean noise, N2N can achieve the same performance as supervised learning using clean targets. Extension of N2N to video denoising has been considered in \cite{mbvd}. When paired noisy images are not available, Noisier2Noise \cite{nr2n} and Noisy-as-clean \cite{nac} adopt a strategy that adds synthetic noise to noisy images to construct noisier-noisy pairs for training. Moreover, the GAN2GAN method \cite{gan2gan} uses GAN to simulate the noise and then generates training pairs for supervised restoration learning.

\textbf{Self-supervised methods}. The self-supervised methods N2V \cite{n2v} and N2S \cite{n2s} only use noisy images for training without requiring noisy pairs. Specifically, such methods use a blind-spot network to predict the masked pixel by neighboring pixels. Besides, additional information such as noise model can been incorporated to further improve the performance \cite{krull2020probabilistic,laine2019high,wu2020unpaired}. Moreover, the Self2Self method \cite{s2s} uses Bernoulli sampling to construct paired training data from noisy images, whilst the Neighbor2Neighbor method \cite{neighbor2neighbor} proposes a random neighbor down-sampler to generate the training pairs. These methods commonly use the assumption of independent noise between pixels and the smoothness of natural images. The independent noise assumption limits the application in complex real-world noise conditions. 

\textbf{Unsupervised learning with Stein’s unbiased risk estimator (SURE)}. In \cite{soltanayev2018training}, SURE has been applied to unsupervised training of Gaussian denoiser, which has been shown to perform comparably with methods trained with ground-truth. An extension to the case that paired noisy images are available has been considered in \cite{esur}. These SURE-based methods are developed under the assumption that the noise is i.i.d. Gaussian.

\textbf{Conditional GAN (CGAN) based methods}. Naturally, the denoising problem can be treated as a conditional generation problem, i.e., generating a natural clean image conditioned on a noisy input by CGAN \cite{cgan}. However, in the absence of paired training data, CGAN without supervision can disregard the input and randomly output natural samples far from the input. Accordingly, the RoCGAN \cite{RoCGAN} uses ground-truth data for supervised training of generation from noisy images. Though AmbientGAN \cite{ambientgan} does not rely on paired training data for conditional generation, it assumes that the degradation model is known (pre-defined) and easy to sample. NR-GAN \cite{nrgan} can ease this requirement, but it still requires either the noise distribution type is a priori known, e.g., zero-mean Gaussian, or the noise is transformation invariant.

As introduced above, there exist a large number of methods which do not rely on paired noisy-clean training data. However, existing methods more or less rely on some specific models or assumptions on the signal and/or the degradation model. In comparison, our method does not use any hand-crafted model for the signal or any prior knowledge on the degradation model.

\section{Problem Formulation and Preliminaries}

As visualized in Fig. 1, the restoration problem aims to restore a source signal  $X \sim {p_X}$ from its degraded observation $Y$. The restoration mapping can be denoted by $\hat X: = f(Y)$ with $f$ being a network for restoration. Typically, for denoising problems, the degradation model can be expressed as
\begin{equation}
	Y = X + N,
	\label{eq1}
\end{equation}
where $N$ denotes the noise. Note that, the model \eqref{eq1} will be used in later information-theoretic related analysis, but the proposed method does not assume an additive noise model as \eqref{eq1}. 

\begin{figure}[!t]
	\centering
	\includegraphics[width = 8cm]{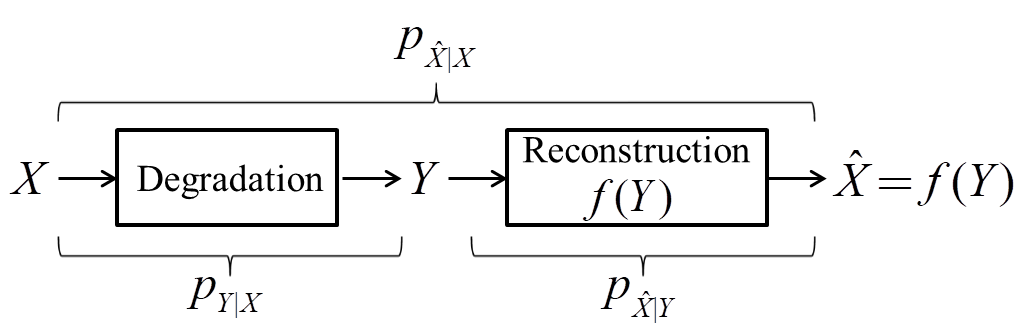}
	\caption{Problem setting. For a source $X \sim {p_X}$, a degraded version $Y$ is observed according to some conditional distribution ${p_{Y|X}}$. Given $Y$, a restoration $\hat X$ is obtained according to some conditional distribution ${p_{\hat X|Y}}$.  $f$ denotes a network for restoration.}
	\label{figure1}
\end{figure}

Generally speaking, the ideal goal of image restoration in the presence of noise or degradation is threefold:

\hangindent 2.5em
\emph{i) Noise suppression}: suppress the noise (or rectify the degradation) in the observation $Y$ as much as possible.

\hangindent 2.5em
\emph{ii) Maximally information preserving}: preserve the information of the source image $X$ contained in $Y$ as much as possible.

\hangindent 2.5em
\emph{iii) Perceptual reconstruction}: achieve high perceptual quality in reconstruction, i.e., the reconstruction $\hat X$ has good perceptual quality. The perceptual quality of a restored image is defined in terms of the extent to which it looks like a valid natural image from human’s perception.

An ideal restoration process to fulfill the above goal is to suppress the noise to maximally preserve the information of $X$ contained in $Y$, e.g., measured by the mutual information $I(Y;X)$, and generate an image with high perceptual quality from this information. Since from the data processing inequality for the Markov chain $X \to Y \to \hat X$ it follows that $\max I(\hat X;X) \le I(Y;X)$, an optimal restoration criterion to achieve the above goal can be expressed as
\begin{equation}
	\begin{array}{c}
		\mathop {\max }\limits_f I\left( {f(Y);X} \right)\\
		{\rm{subject \ to \ }}~~d({p_X},{p_{\hat X}}) \le 0,
	\end{array}
	\label{eq2}
\end{equation}
where $d( \cdot , \cdot )$ is some divergence between distributions, e.g., the Kullback-Leibler divergence or Wasserstein distance, which satisfies $d(p,q)\ge 0,\ d(p,q) = 0 \Leftrightarrow p = q$ for any distributions $p$ and $q$.
This criterion maximizes the mutual information between the reconstruction $f(Y)$ and the ground-truth $X$ while constraining the reconstruction to have the same distribution as natural clean images.

It has been shown that perceptual quality is associated with the deviation from natural sample statistics \cite{wang2006quality,wang2005reduced,mittal2012no,mittal2012making,moorthy2011blind,SaadBC12 }.  It can be conveniently defined in terms of the deviation between the distributions of restored and natural samples, e.g., $d({p_X},{p_{\hat X}})$ \cite{blau2018perception}. In light of this understanding, to achieve perceptual reconstruction, a natural way is to minimize the deviation $d({p_X},{p_{\hat X}})$ using GAN \cite{srgan,wang2018esrgan,blau2018perception,gan}.  The constraint $d({p_X},{p_{\hat X}}) \le 0$ enforces that the reconstruction has perfect perception quality, i.e., looks like natural samples from human’s perception.

For ease of implementation, the constrained formulation \eqref{eq2} is typically relaxed to an unconstrained formulation as
\begin{equation}
	\mathop {\max }\limits_f I(f(Y),X) + \lambda d({p_X},{p_{\hat X}}).
	\label{eq3}
\end{equation}
In practical applications, distortion plus adversarial loss based formulation is typically used for the implementation of \eqref{eq3} as 
\begin{equation}
	\mathop {\min }\limits_f {L_d}\left( {f(Y),X} \right) + \lambda {L_{adv}}({p_X},{p_{\hat X}}),
	\label{eq4}
\end{equation}
where ${L_d}$ is a distortion loss such as MSE, ${\ell _1}$-norm, or distance between deep features. ${L_{adv}}$ is an adversarial loss measuring the deviation between the two distributions ${p_{\hat X}}$ and ${p_X}$ (e.g., measured by a discriminator). $\lambda$ is a positive parameter that balances between the distortion and adversarial losses. When $\lambda  = 0$, it reduces to the widely used supervised method without considering perception quality as
\begin{equation}
	\mathop {\min }\limits_f {L_d}\left( {f(Y),X} \right),
	\label{eq5}
\end{equation}
which is the most popular formulation for supervised restoration learning. Recent studies have shown that minimizing distortion alone does not necessarily lead to high perceptual quality, while incorporating perception loss would lead to increase of the lowest achievable distortion, e.g. MSE \cite{blau2018perception}. 

In the formulations \eqref{eq2}--\eqref{eq5}, paired training data $\{ (Y,X)\}$ is required for supervision. In many practical applications, it is difficult to collect paired training data for supervised training. For such applications, synthesized paired data is widely used. However, the gap between synthesized and real-world data would fundamentally limit the performance on realistic data, especially when the real-world degradation is too complex to accurately simulate. 

To address this problem, numerous methods that do not require noisy-clean paired training data have been developed, as detailed in Section 2. Though much progress has been made recently, the success of most of these methods relies heavily on certain specific assumptions on the degradation, e.g., i.i.d. noise, spatially independent noise, or the noise type is \emph{a priori} known. Such assumptions limit the practical performance on real-world data with complex degradation.

\section{Proposed Unsupervised Criterion from Optimal Transport}

This section first presents the proposed criterion for unsupervised restoration learning. Then, it is relaxed into an unconstrained formulation for ease of implementation. Particularly, we show that in theory the relaxed formulation has the same solution as the original one. Furthermore, some favorable properties of the proposed formulation are analyzed in an information-theoretic view.

\subsection{An Unsupervised Criterion from Optimal Transport}

The optimal transport problem aims to find the most efficient transport map of transforming one distribution of mass to another whilst minimizing the cost, which has wide applications in signal processing, image processing, and machine learning \cite{kolouri2017optimal,Villani2003}. It can be traced back to the seminal work of Monge \cite{monge1781memoire} in 1781, with significant advancements by Kantorovich \cite{kantorovich1942translation} in 1942. 

Let ${\cal P}(X)$ and ${\cal P}(Y)$  be two sets of probability measures on $X$ and $Y$, respectively. Let $c:Y \times X \to [0, + \infty ]$ be a cost function that $c(y,x)$ measures the cost of transporting $y \in Y$ to $x \in X$. The optimal transport problem seeks the most efficient transport from $\nu  \sim {\cal P}(Y)$ to $\mu  \sim {\cal P}(X)$ that with the minimal cost.  

\textbf{Definition 1. (Transport map)}: $f:Y \to X$ is a transport map that transports $\nu  \sim {\cal P}(Y)$ to $\mu  \sim {\cal P}(X)$, if 
\begin{equation}
	\mu (A) = \nu ({f^{ - 1}}(A)) \nonumber,
\end{equation}
for all $\mu$-measurable sets $A$.

\textbf{Definition 2. (Monge’s optimal transport problem)}: For two probability measures $\nu  \sim {\cal P}(Y)$ and $\mu  \sim {\cal P}(X)$, find a $\nu$-measurable map $f:Y \to X$ by 
\begin{equation}
	\begin{array}{c}
		\mathop {\inf }\limits_f \int_Y {c\left( {f(y),y} \right)} d\nu (y)\\
		{\rm{subject \ to\ }}~~\mu  = {f_\# }\nu ,
	\end{array}
	\label{eq6}
\end{equation}
where ${f_\# }\nu$ denotes the transport of $\nu$ by $f$. A minimum ${f^*}$ to this problem, if there exists one, is called an optimal transport map.

Intuitively, the optimal transport problem finds a transport to turn the mass of $\nu $ into $\mu $ at the minimal cost measured by the cost function $c$, e.g. $c\left( {y,f(y)} \right): = {\left\| {f(y) - y} \right\|^\beta }$ with $\beta  \ge 1$.

For the reconstruction problem defined in Section 3, and in the absence of paired training data, we propose an unsupervised learning formulation as
\begin{equation}
	\begin{gathered}
		\mathop {\min }\limits_f {\mathbb{E}_{Y \sim {p_Y}}}\left(\left\|Y-f(Y)\right\|^\beta\right) \\ 
		{\text{subject \ to \ }}~~{p_{\hat X}} = {p_X},
	\end{gathered}
	\label{eq7}
\end{equation}
where $\beta  \ge 1$. It is easy to see that this problem is an implementation of the Monge’s optimal transport problem \eqref{eq6}.

At first glance, formulation \eqref{eq7} appears counter-intuitive since the objective pushes the reconstruction to the noisy input. However, an in-depth analysis can reveal its favorable properties. Specifically, given the degraded observation $Y$, it seeks a reconstruction map $f$ satisfying that:

\hangindent 2.5em
\emph{i) Perceptual reconstruction:} the constraint ${p_{\hat X}} = {p_X}$ ensures the reconstruction $\{ \hat X\}$ has the same distribution as the clean natural samples $\{ X\} $, hence has good perception quality, i.e., looks like natural samples.

\hangindent 2.5em
\emph{ii) Minimum distance transport:} the objective in \eqref{eq7} imposes the fidelity of the reconstruction to the observation $Y$, by which the formulation has a minimum distance transport property. As will be shown in the next subsection, this property makes the reconstruction map to achieve approximately maximal preservation of the information of $X$ contained in $Y$.

These properties make optimal transport a good criterion for unsupervised reconstruction learning in the circumstance that clean natural samples and degraded samples are available but paired noisy-clean data is unavailable. These properties accord well with the goal of perceptual reconstruction as mentioned in Section 3, while without the need of paired training data for supervision.

For ease of implementation, we relax the constrained formulation \eqref{eq7} into an unconstrained one as
\begin{equation}
	\mathop {\min }\limits_f {{\Bbb E}_{Y \sim {p_Y}}}\left(\left\|Y-f(Y)\right\|^\beta\right) + \lambda d({p_X},{p_{\hat X}}),
	\label{eq8}
\end{equation}
where $d({p_X},{p_{\hat X}})$  measures the deviation between ${p_{\hat X}}$ and ${p_X}$, and $\lambda  > 0$ is a balance parameter. Under mild conditions, this formulation with a penalty parameter $\lambda $ is equivalent to the formulation with a constraint $d({p_X},{p_{\hat X}}) \le {\mu _\lambda }$ for some ${\mu _\lambda } > 0$. As $\lambda $ increases, the value of ${\mu _\lambda }$ decreases. As $\lambda  \to \infty $, the solution of \eqref{eq8} satisfies that $d({p_X},{p_{\hat X}}) \to 0$, i.e. ${p_{\hat X}} \to {p_X}$, and the first fidelity term in \eqref{eq8} can be ignored since its weight, which is proportional to $1/\lambda$, tends to be
0. In this case formulation \eqref{eq8} can be implemented as GAN, i.e., to generate natural samples by
\begin{equation}
	\mathop {\min }\limits_f d({p_X},{p_{\hat X}}).
	\label{eq9}
\end{equation}
However, it does not satisfy the minimum distance transport property. For example, the generator can disregard the input $Y$ and randomly output samples from the distribution ${p_X}$ to satisfy ${p_X} = {p_{\hat X}}$. As a consequence, the proposed method can significantly outperform GAN, as will be shown in experiments. 

It is worth noting that, the standard CGAN \cite{cgan} can avoid this degeneration problem of generating samples far away from $Y$ by discriminating between data pairs $(Y,X)$ and $(Y,\hat X)$, but it requires paired training data for supervision. Similarly, RoCGAN \cite{RoCGAN} and the distortion plus adversarial loss based method \eqref{eq4} also require paired training data. AmbientGAN does not require paired training data and can be used for conditional generation, but it assumes that the degradation function is known (pre-defined) and easy to sample \cite{ambientgan}. This limitation is alleviated in NR-GAN \cite{nrgan}, but it requires either the noise distribution type is a priori known (e.g., zero-mean Gaussian) or the noise is rotation, channel shuffle, or color-inverse-invariant. In contrast, though also based on GAN for training, our method avoids the above limitations.

Though relaxed, next we show that the formulation \eqref{eq8} has good theoretical property, e.g., it has the same solution as the constrained formulation \eqref{eq7} under certain conditions.

\textbf{Theorem 1.} Suppose that $d(\cdot, \cdot)$ is the Wasserstein-1 distance, denoted by $W_1(\cdot, \cdot)$, and $\beta=1$. If $\lambda>1$, then the optimal solution set of (8) is the same as that of (7).

\textit{Proof:} 
To justify theorem 1, we first consider the case of $\lambda=1$, with which and $\beta=1$, formulation (8) can be rewritten as
\begin{align}
	\mathop {\min }\limits_{f}\mathbb{E} {\left\|Y-f(Y)\right\|}+W_1({p_X},{p_{f(Y)}}). \label{lambda1loss}
\end{align}
Let $f^*$ be an optimal solution to \eqref{lambda1loss} and $L^*$ be the corresponding objective value. It is easy to see that $f^*$ is also an optimal solution of
\begin{equation}
	\begin{array}{c}
		\mathop {\min }\limits_{f}\ \int {\left\|Y-f(Y)\right\|}dP_Y(y) \\ 
		{\text{subject \ to \ }}~~d(p_{f(Y)}, p_{f^*(Y)})=0,
	\end{array}
\end{equation}
which means $\mathbb{E} {\left\|Y-f^*(Y)\right\|}=W_1(p_Y, p_{f^*(Y)})$. Thus $L^*=W_1(p_X, p_{f^*(Y)})+W_1({p_Y},{p_{f^*(Y)}})$.

Next, we show that $L^*\ge W_1(p_X, p_Y)$. Suppose $p_{X,Y}^*$ minimizing $\int_{x,y} p_{X,Y}(x,y)\left\|X-Y\right\|dxdy$, then
\begin{equation}
	\begin{aligned}
		\begin{split}
			W_1(p_X, p_Y)&=\int_{x,y} p_{X,Y}^*(x,y)\left\|x-y\right\|dxdy\\
			&=\int_{x,y,z} p_{X,Y,Z}^*(x,y,z)\left\|x-z+z-y\right\|dzdxdy\\
			&\le\int_{x,y,z} p_{X,Y,Z}^*(x,y,z) \Big(\left\|x-z\right\|\\
			&\qquad\qquad\qquad\qquad\qquad\quad+\left\|z-y\right\| \Big) dzdxdy\\
			&=\int_{x,z} p_{X,Z}^*(x,z)\left\|x-z\right\|dzdx\\
			&\quad+ \int_{y,z} p_{Y,Z}^*(y,z)\left\|z-y\right\|dzdy\\
			&=W_1(p_X, p_Z)+W_1(p_Y,p_Z). \label{secstep}
		\end{split}
	\end{aligned}
\end{equation}
Let $z=f^*(y)$, we have $L^*\ge W_1(p_X, p_Y)$, which is obviously achievable when $f(y)=y$. Thus, $L^*=W_1(p_X, p_Y)$.

When $\lambda>1$ and with $\beta=1$, \eqref{eq8} can be written as
\begin{equation}
	\begin{aligned}
		\begin{split}
			&\quad \mathop {\min }\limits_{f}\mathbb{E} {\left\|Y-f(Y)\right\|}+\lambda W_1({p_X},{p_{f(Y)}})\\
			&\ge\mathop {\min }\limits_{f}\left(\mathbb{E} {\left\|Y-f(Y)\right\|}+ W_1({p_X},{p_{f(Y)}})\right)\\
			&\quad\quad\quad+(\lambda-1)\mathop {\min }\limits_{f}W_1({p_X},{p_{f(Y)}})\\
			&=L^*+(\lambda-1)\mathop {\min }\limits_{f}W_1({p_X},{p_{f(Y)}})\\
			&\ge W_1(p_Y,p_X). \label{laststep}
		\end{split}
	\end{aligned}
\end{equation}
Inequality \eqref{laststep} gives the infimum of \eqref{eq8} when $\lambda>1$. Now, we prove that any optimal solution to \eqref{eq7} is also an optimal solution to \eqref{eq8}. Let $g^*$ be an optimal solution of \eqref{eq7}, we have $\mathbb{E} {\left\|Y-g^*(Y)\right\|}+\lambda W_1({p_X},{p_{g^*(Y)}})=W_1(p_Y,p_X)$, since $p_{g^*(Y)}={p_X}$ and $W_1({p_X},{p_{g^*(Y)}})=0$. Then, it is easy to see from \eqref{laststep} that $g^*$ is optimal to \eqref{eq8}.

Last, we show that any optimal solution to \eqref{eq8} is also an optimal solution to \eqref{eq7}. The equality in \eqref{laststep} holds only if $W_1(p_X,p_{f(Y)})=0$, which means $p_X=p_{f(Y)}$. Thus, optimizing \eqref{eq8} is equivalent to optimizing
\begin{equation}
	\begin{array}{c}
		\mathop {\min }\limits_{f}\mathbb{E} {\left\|Y-f(Y)\right\|}+\lambda W_1({p_X},{p_{f(Y)}}) \\ 
		{\text{subject \ to \ }}~~p_X=p_{f(Y)},\label{trans8}
	\end{array}
\end{equation}
which is equivalent to \eqref{eq7}. Hence, the optimal solution set of \eqref{eq8} is the same as that of \eqref{eq7}, which results in theorem 1.

\emph{\textbf{Remark 1}:} 
Interestingly, this theorem shows that, for $d( \cdot , \cdot )$ being the Wasserstein-1 distance and $\beta=1$, the relaxed formulation \eqref{eq8} with any $\lambda > 1$ has the same solution set as the constrained formulation \eqref{eq7}. This result provides a solid theoretical foundation for the relaxation, which is used in practical implementation instead of \eqref{eq7}. However, it is worth noting that, in practice the WGAN used to compute the Wasserstein-1 distance can only approximate the Wasserstein-1 distance to some extent \cite{wgan}. As a consequence, to achieve satisfactory performance in practice, we still need to tune the parameter $\lambda $ rather than selecting an arbitrary value of it. 

\subsection{An Information-Theoretic View of the Proposed Formulation}

Here, from an information-theoretic view, we show the proposed formulation \eqref{eq7} has a favorable property that, the reconstruction map $f(Y)$ can approximately achieve maximal preservation of the information of $X$ contained in $Y$.

First, the ideal supervised learning criterion is given by \eqref{eq2}, which achieves perceptual reconstruction whilst maximizing the mutual information between $X$ and $\hat X$. In practical applications, mutual information is difficult to compute except for certain simple specific data distributions, hence MSE is widely used as the reconstruction loss, with which an implementation of the supervised criterion \eqref{eq2} can be written as
\begin{equation}
	\begin{gathered}
		\mathop {\min }\limits_f {\mathbb{E}_{(X,Y) \sim p_{X,Y}}}\left( {{{\left\| {f(Y) - X} \right\|}^2}} \right) \\ 
		{\text{subject \ to \ }}~~d({p_X},{p_{\hat X}}) \leqslant 0.
	\end{gathered}
	\label{eq10}
\end{equation}
While general ${\ell _\beta }$ loss with any $\beta  \geqslant 1$ can be alternative to MSE, here we consider MSE for convenience.

Under Gaussian distribution of ${p_{\hat X}}$ and ${p_Y}$, $\mathop {\min }\limits_f {\mathbb{E}_{Y \sim {p_Y}}}\left( {{{\left\| {f(Y) - Y} \right\|}^2}} \right)$ is equivalent to maximizing the mutual information $\mathop {\max }\limits_f I(f(Y);Y)$ \cite{cover1999elements}. Hence, with $\beta  = 2$, the proposed unsupervised learning formulation \eqref{eq7} can be viewed as a special case of the following information-theoretic criterion for unsupervised learning under Gaussian distribution
\begin{equation}
	\begin{gathered}
		\mathop {\max }\limits_f I(f(Y);Y) \\ 
		{\text{subject \ to \ }}~~d({p_X},{p_{\hat X}}) \leqslant 0.
	\end{gathered}
	\label{eq11}
\end{equation}
Next we show that the proposed unsupervised formulation is an approximation of the supervised formulation under certain conditions.

\textbf{Proposition 2.} Let $X$, $Y$, $\hat X$ and $N$ denote the source signal, degraded signal, restoration and noise, respectively, as defined in \eqref{eq1}. 
Suppose that both $X$ and $\hat X$ are independent with $N$, ${p_{\hat X}}$ and ${p_Y}$ are Gaussian distributed, then formulation \eqref{eq11} is equivalent to formulation \eqref{eq2}, and formulation \eqref{eq7} with $\beta  = 2$ is equivalent to formulation \eqref{eq10}.

\emph{Proof}: First, under the constraint $d({p_X},{p_{\hat X}}) \leqslant 0$, i.e. ${p_X} = {p_{\hat X}}$, we have
\begin{equation}
	\begin{aligned}
		\nonumber
		\mathbb{E}\left( {{{\| {\hat X - X} \|}^2}} \right) &= \mathbb{E}{\| {\hat X} \|^2} + \mathbb{E}{\left\| X \right\|^2} - 2\mathbb{E}({{\hat X}^T}X) \hfill \\ 
		&= const - 2\mathbb{E}({{\hat X}^T}X) \hfill .
	\end{aligned} 
\end{equation}
Similarly, it follows that
\begin{equation}
	\begin{aligned}
		\nonumber
		\mathbb{E}\left( {{{\| {\hat X - Y} \|}^2}} \right)  &= \mathbb{E}\left( {{{\| {\hat X - X - N} \|}^2}} \right) \\ 
		& \mathop   = \limits^{(a)} 2\mathbb{E}{\left\| X \right\|^2} + \mathbb{E}{\left\| N \right\|^2} - 2\mathbb{E}({{\hat X}^T}X) \\ 
		&  = const - 2E({{\hat X}^T}X), \\ 
	\end{aligned} 
\end{equation}
where (a) is due to $d({p_X},{p_{\hat X}}) \leqslant 0$ and that both $X$ and $\hat X$ are independent with $N$. Thus, under the condition that both $X$ and $\hat X$ are independent with $N$, the unsupervised formulation \eqref{eq7} with $\beta  = 2$ is equivalent to the supervised formulation \eqref{eq10}.

Furthermore, under the assumption of Gaussian distribution of ${p_{\hat X}}$ and ${p_Y}$, $\mathop {\min }\limits_f {\mathbb{E}_{Y \sim {p_Y}}}\left( {{{\left\| {f(Y) - Y} \right\|}^2}} \right)$ is equivalent to $\mathop {\max }\limits_f I(f(Y);Y)$. Hence, it is easy to see the equivalence between the two information-theoretic formulations \eqref{eq11} and \eqref{eq2}.

\emph{\textbf{Remark 2}:} From the data processing inequality for the Markov chain $X \to Y \to \hat X$, $I(\hat X;X)$ is upper bounded by $I(Y;X)$ as $I(\hat X;X) \leqslant I(Y;X)$. Naturally, if the restoration mapping $f$ can perfectly suppress the noise $N$ (hence $\hat X$ is independent with $N$), maximally preserving the information of $X$ contained in $Y$ can be achieved by maximizing the mutual information $I(\hat X;Y)$. The assumption that the clean data $X$ is independent with the noise $N$ is reasonable in most applications. However, the assumption of independence between the reconstruction $\hat X$ and the noise $N$ is impractical, since the noise component in the observation $Y$ cannot be guaranteed to be completely suppressed. In practice, when the denoising process $f(Y)$ can suppress the noise component in $Y$ to a large extent, the correlation between $\hat X$ and $N$ would be weak. In this scenario, the unsupervised criterion \eqref{eq11} can be viewed as an approximation of the supervised criterion \eqref{eq2}.

\section{Experimental Results}

\begin{figure}[!t]
	\centering
	\subfigure[Generator]{
		{\includegraphics[width = 8.2cm]{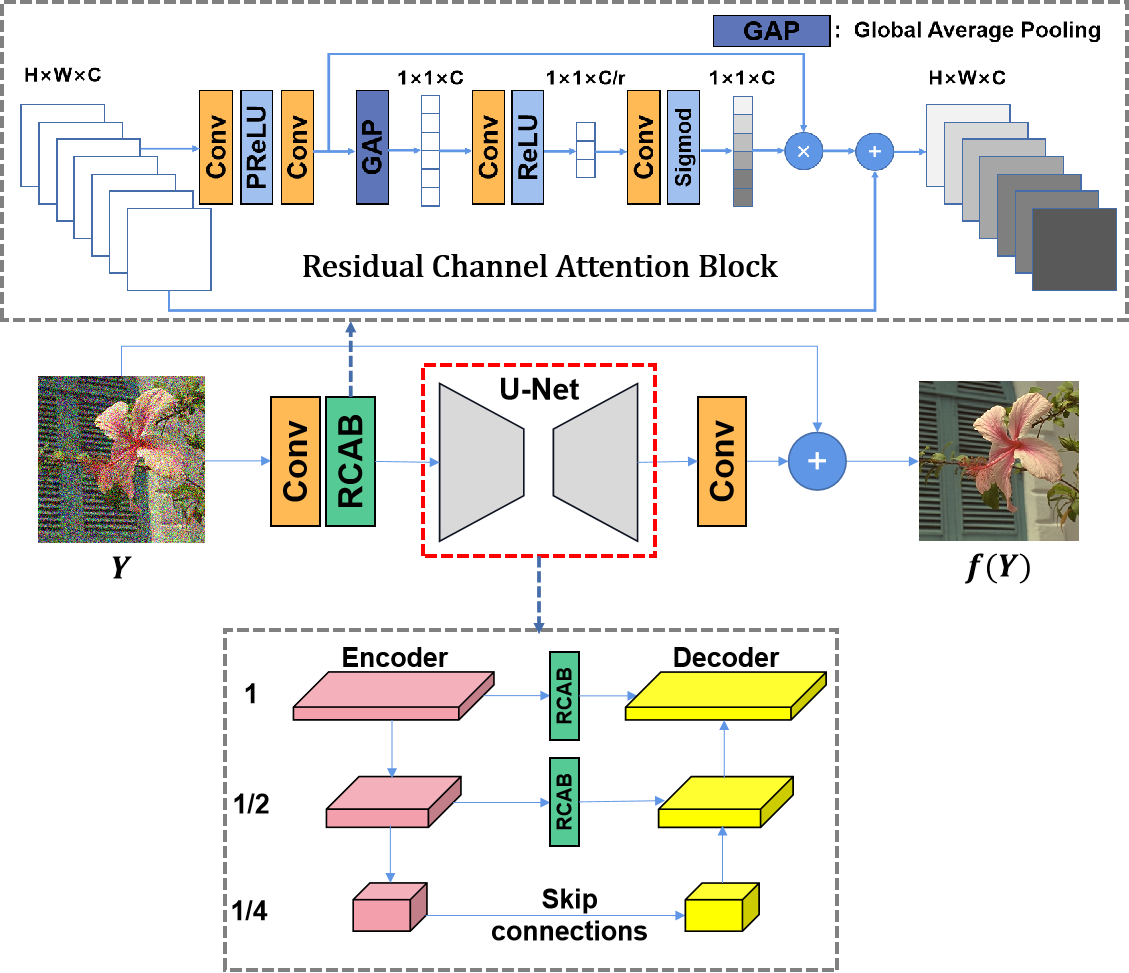}}}\\
	\subfigure[Discriminator]{
		{\includegraphics[width = 8.2cm]{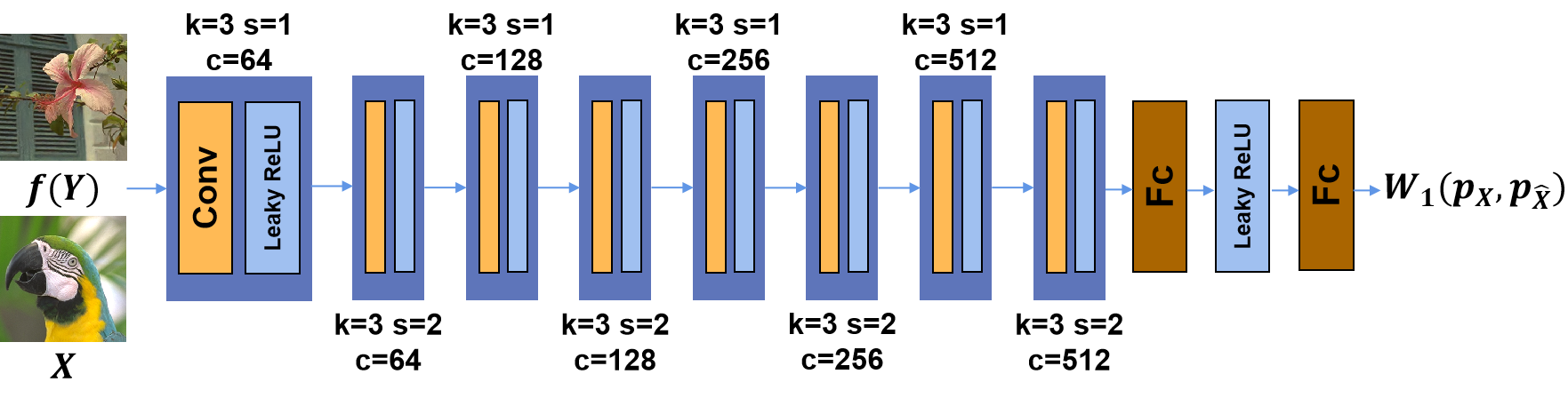}}}
	\caption{Network architecture used in the proposed method, (a) A residual U-Net as the generator, (b) A discriminator as the one used in \cite{srgan}.}
	\label{figure2}
\end{figure}

We evaluate the proposed method on synthetic and real-world noisy images in comparison with state-of-the-art supervised, self-supervised, and unsupervised methods. The considered real-world images include photographic, microscopy, and raw depth images, for which the ground-truth is difficult or impractical to collect. The compared methods include DnCNN \cite{dncnn}, RNAN \cite{rnan}, noise-to-clean (N2C), N2N \cite{n2n}, N2V \cite{n2v}, Nr2N \cite{nr2n}, N2S \cite{n2s}, S2S \cite{s2s}, N2Same \cite{n2same}, NAC \cite{nac}, BM3D \cite{bm3d}, and CGAN (9). DnCNN, RNAN and N2C are supervised methods but use different network models. For fair comparison, N2C, N2N, CGAN and the proposed method use the same network model as detailed below in Section 5.1.
The performance is evaluated in terms of both restoration distortion and perceptual quality measures, including peak signal  to  noise  ratio  (PSNR), structural similarity (SSIM), perception index (PI) \cite{pi} and learned perceptual image patch similarity (LPIPS) \cite{lpips}.

\begin{table*}[!t]
	\renewcommand\arraystretch{2}
	\footnotesize
	\caption{Quantitative distortion comparison (PSNR/SSIM) on synthetic noisy RGB images with Gaussian, Poisson and Brown Gaussian noise.}
	\centering
	\begin{tabular}{|c|c|c|c|c|c|c|}
		\hline
		\multicolumn{2}{|c|}{\multirow{2}{*}{Method}}  & \multicolumn{3}{c|}{Gaussian}                                                                                                                                                        & Poisson               & Brown Gaussian                                                   \\ \cline{3-7}  
		\multicolumn{2}{|c|}{}   
		& $\sigma\ $= 15                                                           & $\sigma\ $= 25                                                  & $\sigma\ $= 50                                                  & ${\lambda _p}$= 30                & $5\times5$ filter, $\sigma=50 $                                             \\ \hline
		\multicolumn{2}{|c|}{Noisy}                  & 24.69/0.5078         & 20.17/0.3288                                          & 14.15/0.1549                                           & 18.65/0.3041         & 25.53/0.6186                                                    \\ \hline
		\multirow{4}{*}{Supervised}   & DnCNN \cite{dncnn}           & 34.60/0.9209                                                     & 32.14/0.8775                                            & 28.95/0.7917                                            & 30.53/0.8645          & 27.67/0.7384                                                     \\ \cline{2-7} 
		& RNAN \cite{rnan}           & 34.90/\textbf{0.9256} &   32.33/0.8840
		& 29.40/\textbf{0.8116}                                   & 31.15/0.8706 & 31.78/0.9149                                            \\ \cline{2-7} 
		& N2C            & \textbf{34.92}/0.9253 & \textbf{32.50}/\textbf{0.8852}                                   & \textbf{29.43}/0.8072                                  & \textbf{31.63}/\textbf{0.8745} & 31.86/\textbf{0.9216}                                            \\ \cline{2-7} 
		& N2N \cite{n2n}             & 34.89/0.9251                                                     & 32.48/0.8848                                            & 29.42/0.8069                                            & 31.59/0.8712          & 28.96/0.8658                                                     \\ \hline
		\multirow{6}{*}{Self-supervised}      & N2V \cite{n2v}            & 27.71/0.7507          & 28.19/0.7558                                            & 26.36/0.7293                                            & 28.66/0.7996          & 26.73/0.6984                                                     \\ \cline{2-7}
		& Nr2N \cite{nr2n}            & 32.19/0.8905          & 30.29/0.8107                                            & 26.98/0.7042                                            & -          & -                                                     \\ \cline{2-7}
		& N2S \cite{n2s}            & 30.29/0.8635          & 29.53/0.8329                                            & 26.54/0.7299                                            & 28.65/0.8230          & 26.55/0.6986                                                     \\ \cline{2-7}
		& S2S \cite{s2s}            & 31.64/0.8799          & 30.08/0.8381                                            & 27.43/0.7664                                            & 28.74/0.8282          & 26.67/0.7057                                                    \\ \cline{2-7}
		& N2Same \cite{n2same}            & 29.15/0.8520          & 27.85/0.7427                                            & 25.22/0.7018                                            & 27.62/0.7788          & 27.80/0.7372                
		\\ \cline{2-7}
		& NAC \cite{nac}            & 30.23/0.9053          & 28.82/0.8354                                            & 24.46/0.6224                                            &  27.82/0.8281         &     26.85/0.6975              
		\\ \hline
		\multirow{3}{*}{Unsupervised} & BM3D \cite{bm3d}           & 34.44/0.9172                                                     & 31.87/0.8687                                            & 28.66/0.7789                                            & 30.44/0.8278          & 29.20/0.7864                                                     \\ \cline{2-7} 
		& CGAN             & 29.97/0.8602         & 28.67/0.8058  & 25.93/0.6659 & 29.47/0.8304          & 30.52/0.8832                                                     \\ \cline{2-7} 
		& Ours                    & 33.99/0.9115                                                     & 31.27/0.8581                                            & 28.11/0.7668  & 30.59/0.8501          & \textbf{32.08}/0.9105                                                     \\  \hline
	\end{tabular}
\end{table*}

\begin{table*}[!t]
	\renewcommand\arraystretch{2}
	\footnotesize
	\caption{Quantitative perceptual quality comparison (PI/LPIPS) on synthetic noisy RGB images with Gaussian, Poisson and Brown Gaussian noise.}
	\centering
	\begin{tabular}{|c|c|c|c|c|c|c|}
		\hline
		\multicolumn{2}{|c|}{\multirow{2}{*}{Method}}  & \multicolumn{3}{c|}{Gaussian}                                                                                                                                                        & Poisson               & Brown Gaussian                                                   \\ \cline{3-7}  
		\multicolumn{2}{|c|}{}   
		& $\sigma\ $= 15                                                           & $\sigma\ $= 25                                                  & $\sigma\ $= 50                                                  & ${\lambda _p}$= 30                & $5\times5$ filter, $\sigma=50 $                                             \\ \hline
		\multirow{4}{*}{Supervised}   & DnCNN \cite{dncnn}            & 2.34/0.024                                                     & 2.45/0.040                                            & 2.70/0.073                                            & 2.46/0.041          & 2.43/0.103                                                     \\ \cline{2-7} 
		& RNAN \cite{rnan}            & 2.35/0.022 &   2.60/0.038
		& 2.95/0.068                                   & 2.45/0.037 & 2.33/0.036                                            \\ \cline{2-7} 
		& N2C             & 2.41/0.022 & 2.57/0.037                                   & 2.96/0.068                                 & 2.59/0.037 & 2.33/0.034                                            \\ \cline{2-7} 
		& N2N \cite{n2n}              & 2.44/0.022                                                     & 2.58/0.037                                            & 2.97/0.069                                            & 2.66/0.042          & 2.39/0.056                                                     \\ \hline
		\multirow{6}{*}{Self-supervised}      & N2V \cite{n2v}             & 2.66/0.088          & 2.33/0.080                                            & 2.88/0.113                                            & 2.23/0.071          & 2.77/0.131                                                     \\ \cline{2-7}
		& Nr2N \cite{nr2n}             & 3.05/0.053          & 4.12/0.085                                            & 5.68/0.134                                            & -          & -                                                     \\ \cline{2-7}
		& N2S \cite{n2s}             & 2.93/0.050          & 2.94/0.059                                            & 3.46/0.097                                            & 2.99/0.061          & 3.38/0.111                                                     \\ \cline{2-7}
		& S2S \cite{s2s}            & 3.54/0.054          & 3.93/0.070                                            & 4.56/0.109                                            & 4.09/0.075          & 3.46/0.118                                                    \\ \cline{2-7}
		& N2Same \cite{n2same}            & 4.60/0.113          & 4.67/0.116                                            & 4.77/0.125                                            & 4.93/0.116          & 3.30/0.103
		\\ \cline{2-7}
		& NAC \cite{nac}             & 4.21/0.065          & 3.43/0.056                                            & 4.27/0.094                                            &    4.63/0.128       &  4.34/0.121
		\\ \hline
		\multirow{3}{*}{Unsupervised} & BM3D \cite{bm3d}            & 2.38/0.030                                                     & 2.65/0.050                                            & 3.19/0.087                                            & 3.07/0.072          & 3.82/0.108                                                     \\ \cline{2-7} 
		& CGAN             & 2.07/0.019        & 2.45/0.068  &  3.28/0.104 & 2.03/0.048          & 2.32/0.043                                                     \\ \cline{2-7} 
		& Ours                    & \textbf{2.06}/\textbf{0.017}                                                     & \textbf{2.05}/\textbf{0.031}                                            & \textbf{2.00}/\textbf{0.067} & \textbf{2.02}/\textbf{0.034}          & \textbf{2.29}/\textbf{0.030}                                                     \\  \hline
	\end{tabular}
\end{table*}

For real-world depth image denoising, the widely used colorization \cite{color}, cross-bilateral filter, nearest-neighbor hole-filling methods, and the Gaussian filter, median filter and bilateral filter denoising methods are also compared. Since the widely used Kinect camera only provide pre-processed images, we additionally conduct an experiment using another commercial time-of-flight (ToF) camera which can provide raw depth data without any pre-processing.

\subsection{Network Settings}

The proposed formulation \eqref{eq8} is implemented in an adversarial training framework using WGAN-gp \cite{wgangp}, with $\beta = 1$ and $\lambda $ being tuned for each task. Note that although our formulation requires the access to clean images $X$, it is unsupervised since the clean images and the noisy ones are not required to be paired, which is similar to the unpaired image restoration task \cite{unpair}.
We use a U-Net architecture with skip connection for the generator, as shown in Fig. 2(a). It consists of two down-sampling CNN layers in the encoder and two up-sampling CNN layers in the decoder. Residual channel attention block (RCAB) \cite{cab} is used for each down-sampling and up-sampling layer. The discriminator is the same as that in \cite{srgan}. We remove the batch norm layers and sigmoid activation of the discriminator as \cite{wgangp} did. The architecture is shown in Fig. 2(b). We train the generator and the discriminator alternatingly for a total of 200 epochs and use the RMSProp optimizer with a learning rate of $1 \times {10^{ - 4}} $ for the generator and $0.5 \times {10^{ - 4}} $ for the discriminator. The learning rate is decayed by a factor of 10 after 100 epochs. N2C, N2N, CGAN and the proposed method use the same restoration model in Fig. 2(a).

For adversarial training in each task, we collect 20,000--60,000 patches of size $64 \times 64 $ from clean images as the clean target domain, and the same number of patches of noisy images as the noisy source domain. For real-world photographic and microscopy images, the clean images are collected by averaging over tens of realizations of the same scenes, while clean depth images are generated by 3D modeling \cite{suncg}. Moreover, ground-truth clean targets of noisy patches are collected to train N2C.

\begin{figure*}[!t]
	\renewcommand{\thesubfigure}
	\centering
	\subfigure[Test image]{
		\includegraphics[width=0.19\linewidth]{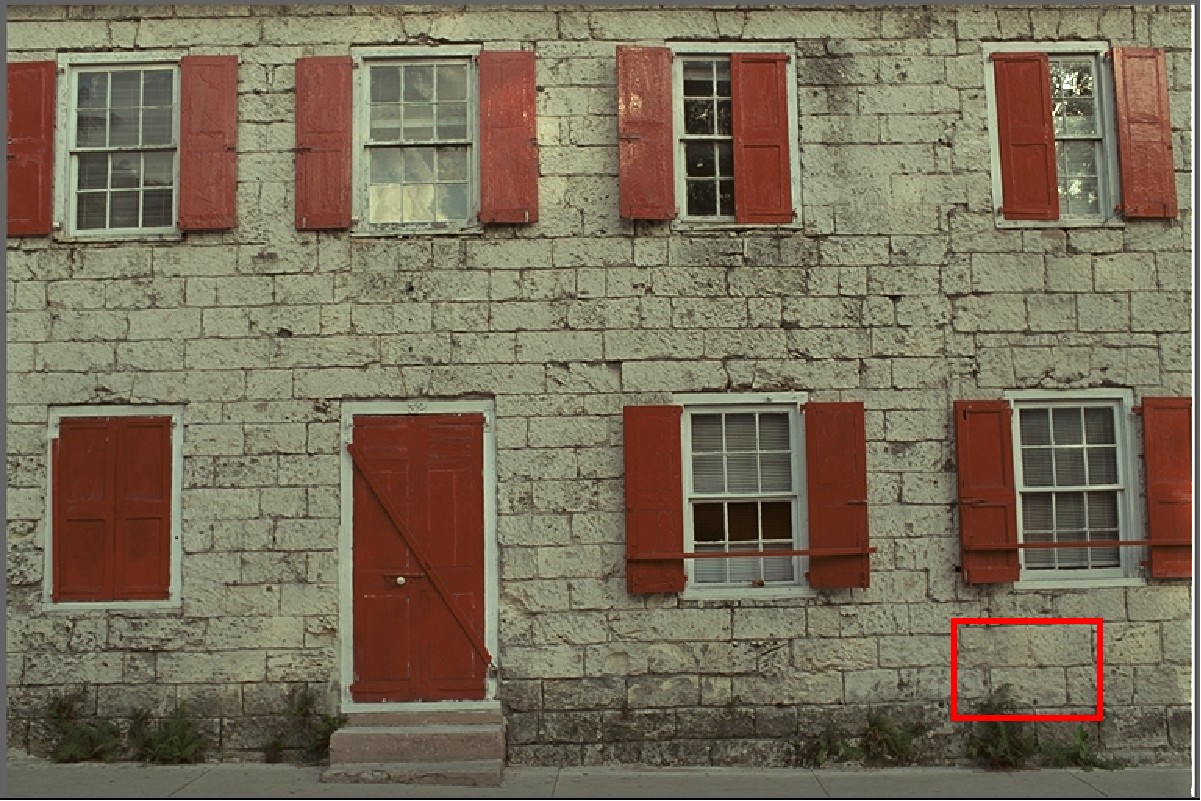}
	}\subfigure[Ground truth]{
		\includegraphics[width=0.19\linewidth]{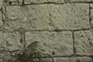}
	}\subfigure[Noisy (20.28 dB)]{
		\includegraphics[width=0.19\linewidth]{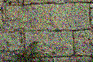}
	}\subfigure[DnCNN (29.66/2.30/0.045)]{
		\includegraphics[width=0.19\linewidth]{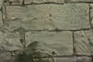}
	}\subfigure[N2C (\textbf{29.86}/2.36/0.040)]{
		\includegraphics[width=0.19\linewidth]{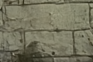}
	}
	\subfigure[N2N (29.84/2.40/0.041)]{
		\includegraphics[width=0.19\linewidth]{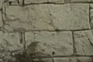}
	}\subfigure[BM3D (29.36/2.53/0.059)]{
		\includegraphics[width=0.19\linewidth]{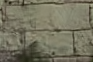}
	}\subfigure[CGAN (26.69/2.71/0.079)]{
		\includegraphics[width=0.19\linewidth]{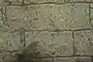}
	}\subfigure[N2V (27.73/2.38/0.068)]{
		\includegraphics[width=0.19\linewidth]{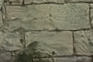}
	}\subfigure[Ours (28.96/\textbf{2.09}/\textbf{0.029})]{
		\includegraphics[width=0.19\linewidth]{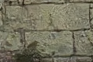}
	}
	\centering
	(a) Gaussian noise with $\sigma = 25$
	
	\subfigure[Test image]{
		\includegraphics[width=0.19\linewidth]{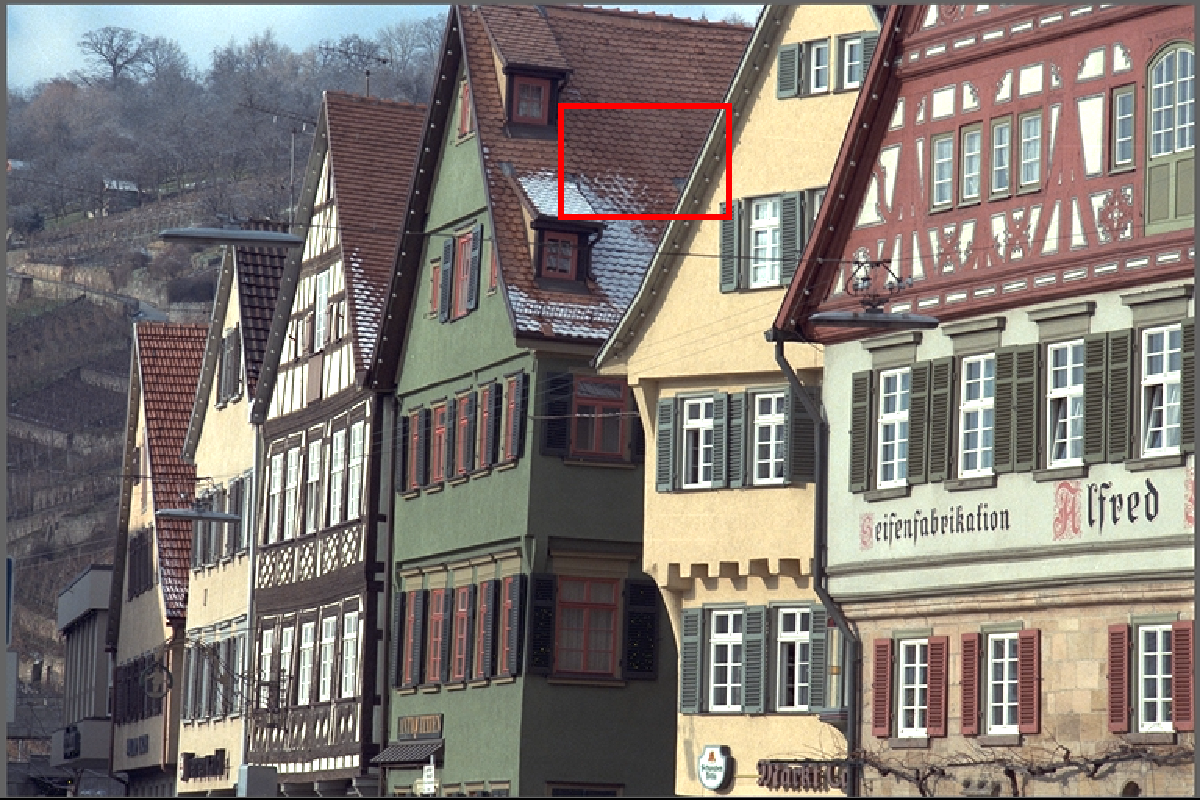}
	}\subfigure[Ground truth]{
		\includegraphics[width=0.19\linewidth]{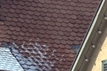}
	}\subfigure[Noisy (18.64 dB)]{
		\includegraphics[width=0.19\linewidth]{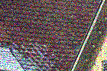}
	}\subfigure[DnCNN (28.86/3.30/0.035)]{
		\includegraphics[width=0.19\linewidth]{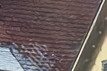}
	}\subfigure[N2C (\textbf{29.03}/3.47/0.034)]{
		\includegraphics[width=0.19\linewidth]{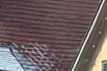}
	}
	\subfigure[N2N (28.96/3.55/0.037)]{
		\includegraphics[width=0.19\linewidth]{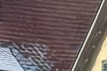}
	}\subfigure[BM3D (27.30/3.31/0.046)]{
		\includegraphics[width=0.19\linewidth]{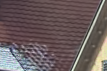}
	}\subfigure[CGAN (26.53/2.92/0.043)]{
		\includegraphics[width=0.19\linewidth]{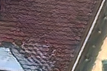}
	}\subfigure[N2V (26.79/3.08/0.059)]{
		\includegraphics[width=0.19\linewidth]{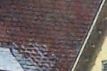}
	}\subfigure[Ours (28.05/\textbf{3.02}/\textbf{0.030})]{
		\includegraphics[width=0.19\linewidth]{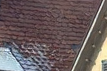}
	}
	
	(b) Poisson noise with ${\lambda _p}$= 30
	
	\subfigure[Test image]{
		\includegraphics[width=0.19\linewidth]{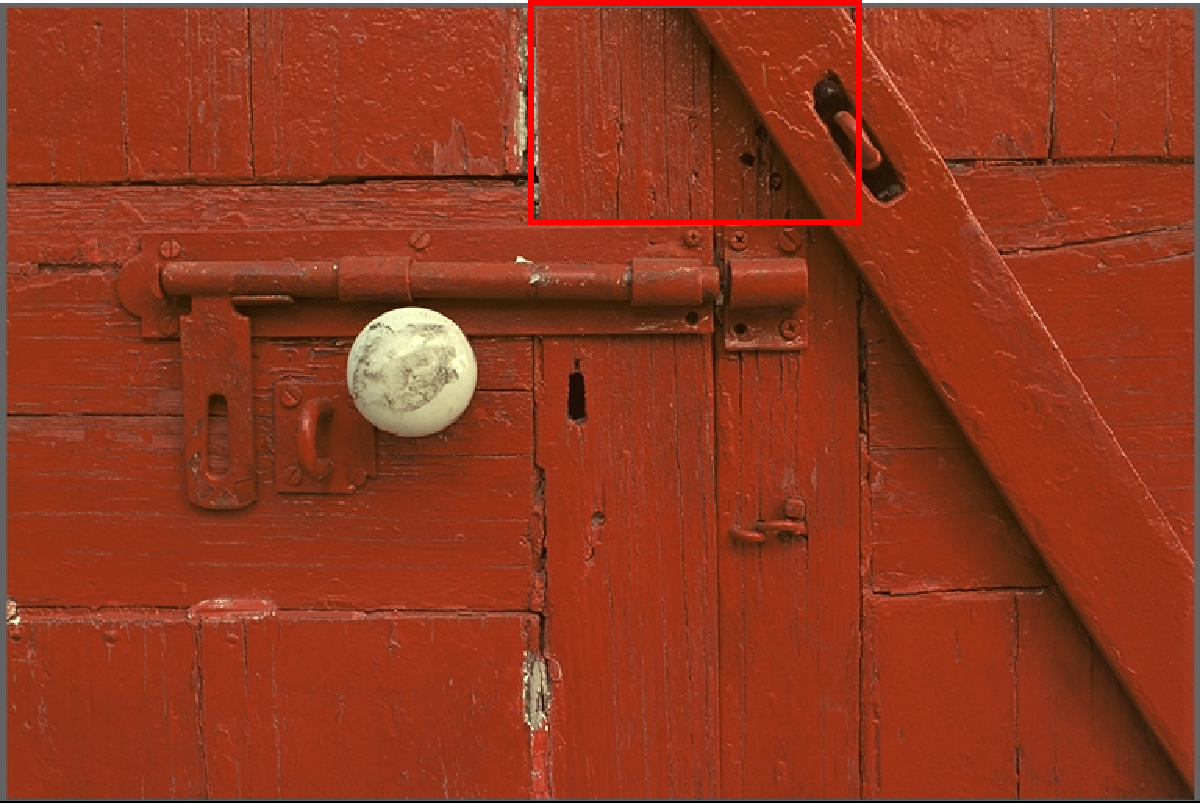}
	}\subfigure[Ground truth]{
		\includegraphics[width=0.19\linewidth]{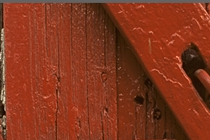}
	}\subfigure[Noisy (25.50 dB)]{
		\includegraphics[width=0.19\linewidth]{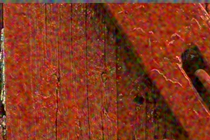}
	}\subfigure[DnCNN (27.31/2.51/0.119)]{
		\includegraphics[width=0.19\linewidth]{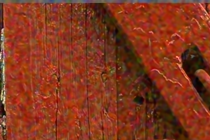}
	}\subfigure[N2C (32.50/2.42/0.060)]{
		\includegraphics[width=0.19\linewidth]{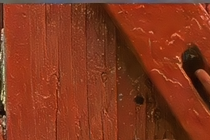}
	}
	\subfigure[N2N (29.64/2.49/0.097)]{
		\includegraphics[width=0.19\linewidth]{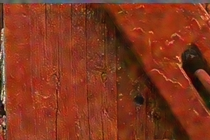}
	}\subfigure[BM3D (29.84/4.02/0.162)]{
		\includegraphics[width=0.19\linewidth]{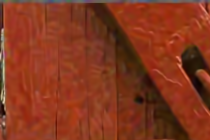}
	}\subfigure[CGAN (29.50/2.53/0.086)]{
		\includegraphics[width=0.19\linewidth]{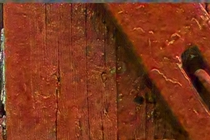}
	}\subfigure[N2V (26.69/3.24/0.122)]{
		\includegraphics[width=0.19\linewidth]{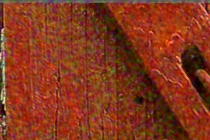}
	}\subfigure[Ours (\textbf{32.75}/\textbf{1.85}/\textbf{0.055})]{
		\includegraphics[width=0.19\linewidth]{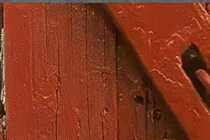}
	}
	
	(c) Brown Gaussian noise with $5\times5$ Gaussian filter applied to Gaussian noise ($\sigma  = 50$)
	
	\caption{Visual comparison on synthetic noisy images with different noise conditions. The PSNR/PI/LPIPS results are provided in the brackets. The images are enlarged for clarity.}
	\label{figure3}
\end{figure*}

\begin{figure*}[!t]
	\centering
	\subfigure[Test image]{
		\includegraphics[width=0.19\linewidth]{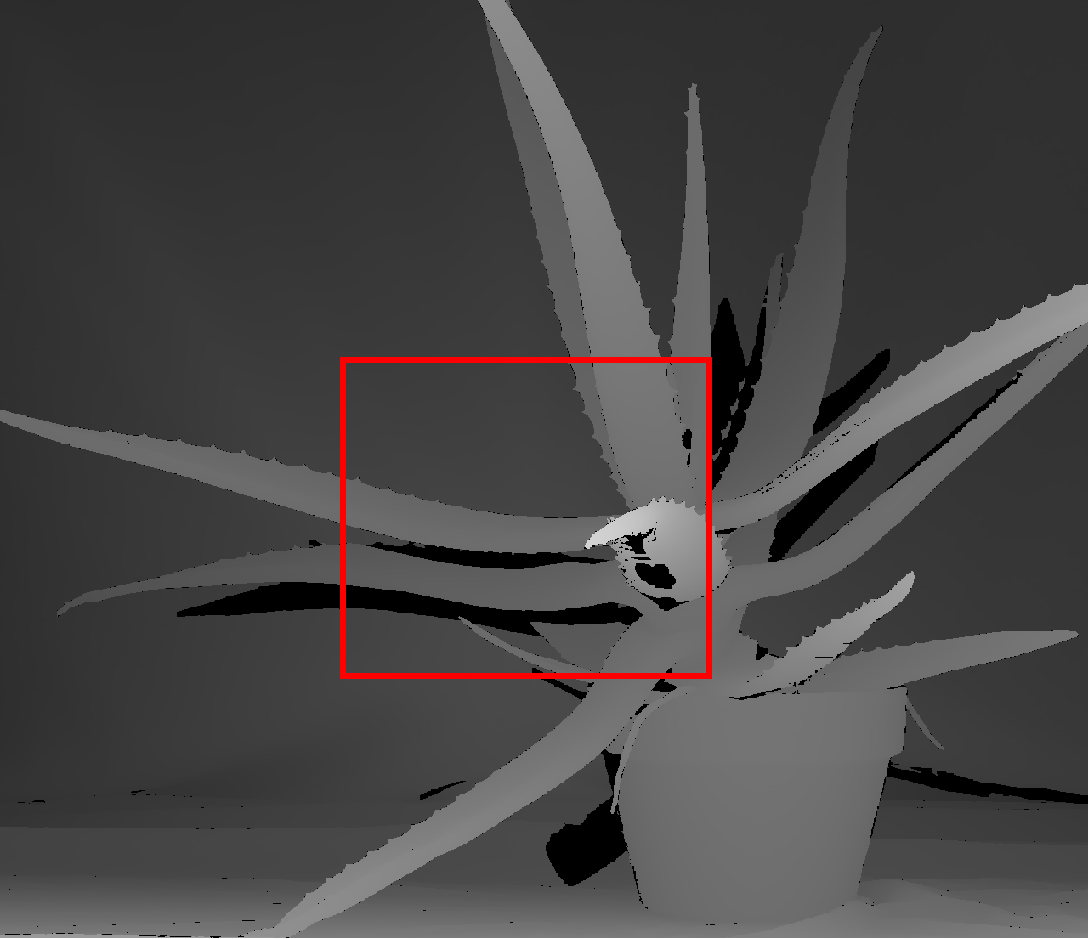}
	}\subfigure[Ground truth]{
		\includegraphics[width=0.19\linewidth]{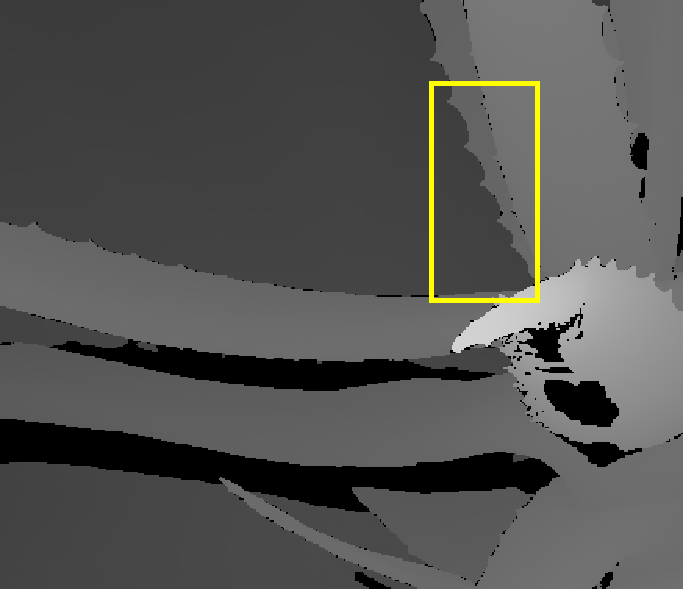}
	}\subfigure[Noisy (20.34 dB)]{
		\includegraphics[width=0.19\linewidth]{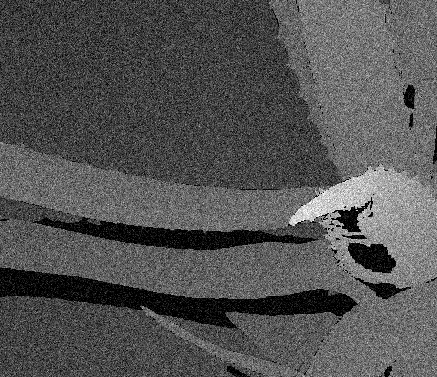}
	}\subfigure[DnCNN (\textbf{36.63 dB})]{
		\includegraphics[width=0.19\linewidth]{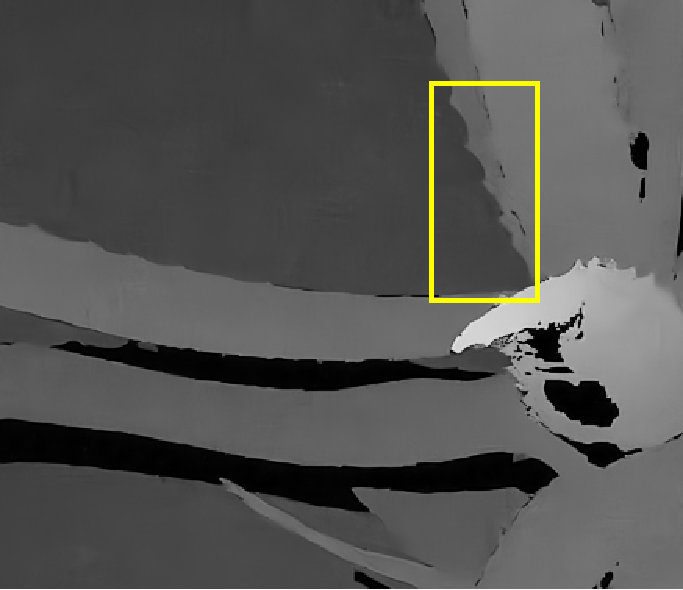}
	}\subfigure[N2C (36.54 dB)]{
		\includegraphics[width=0.19\linewidth]{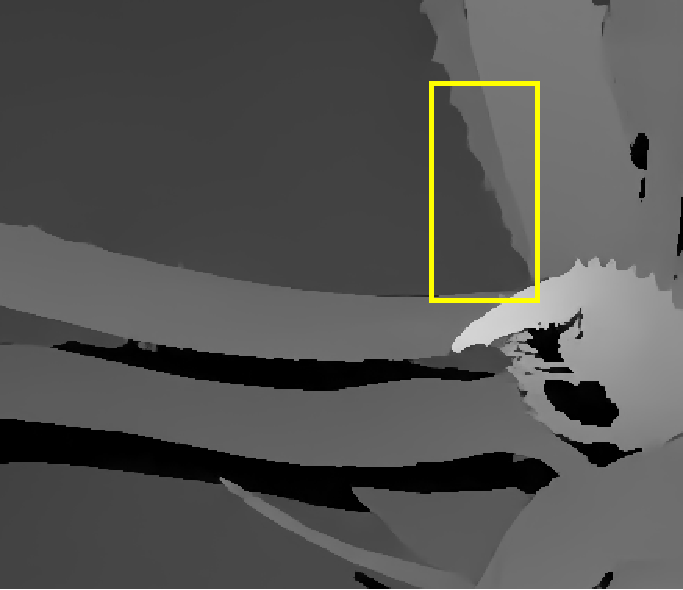}
	}
	\subfigure[N2N (36.43 dB)]{
		\includegraphics[width=0.19\linewidth]{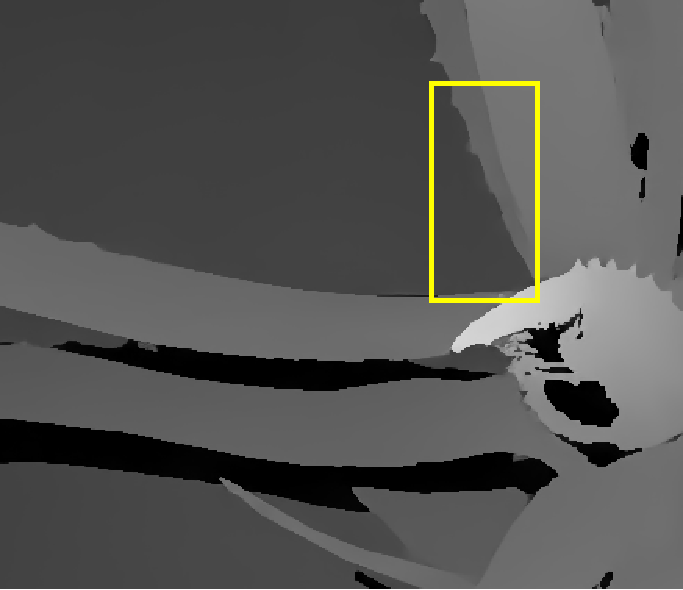}
	}\subfigure[BM3D (36.23 dB)]{
		\includegraphics[width=0.19\linewidth]{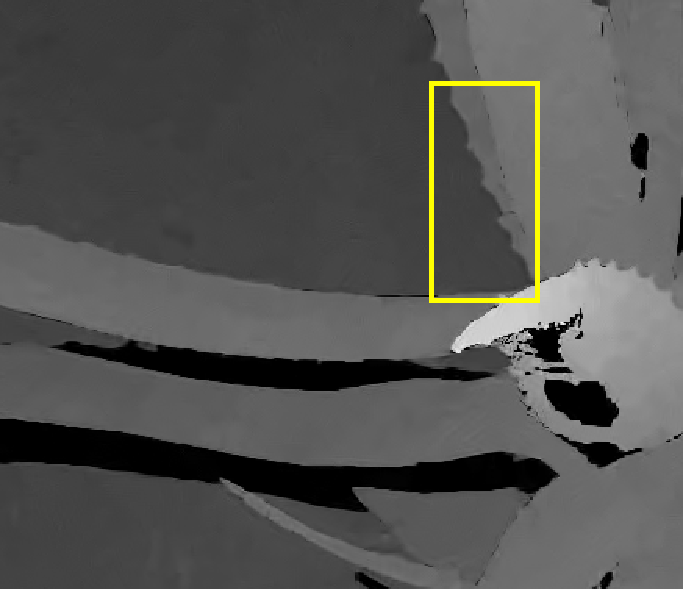}
	}\subfigure[CGAN (31.71 dB)]{
		\includegraphics[width=0.19\linewidth]{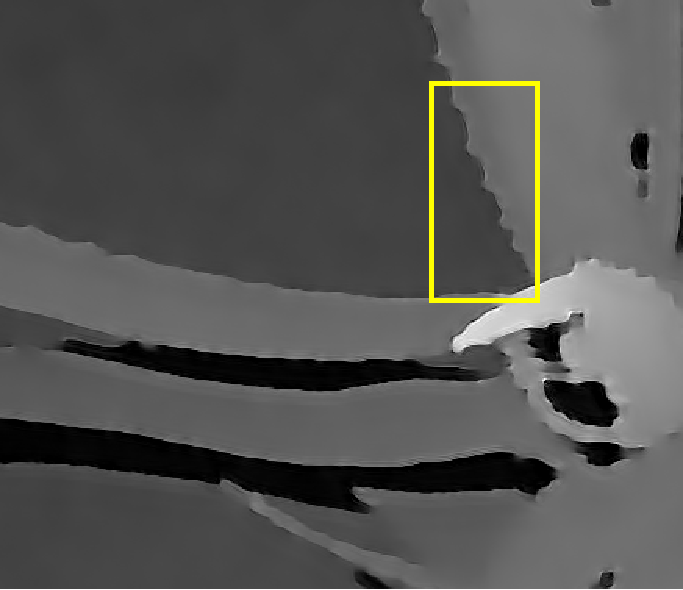}
	}\subfigure[N2V (32.58 dB)]{
		\includegraphics[width=0.19\linewidth]{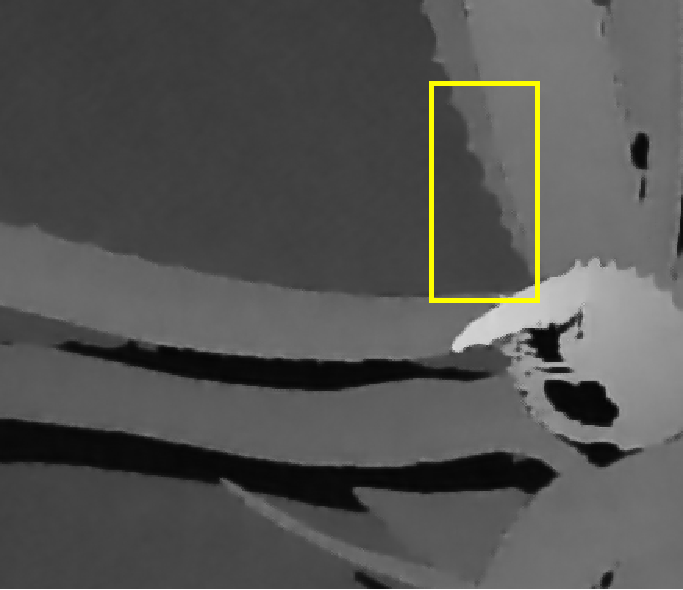}
	}\subfigure[Ours (36.27 dB)]{
		\includegraphics[width=0.19\linewidth]{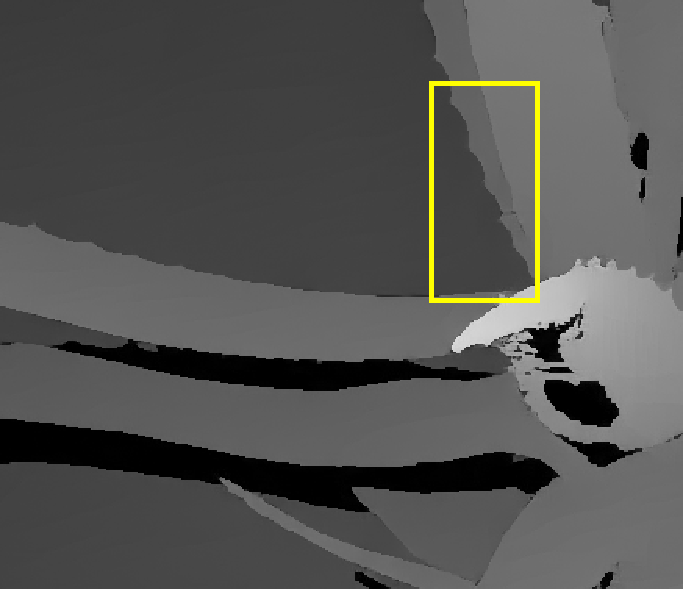}
	}
	\caption{Visual comparison on synthetic noisy depth images in the case of Gaussian noise with $\sigma=25$. The images are enlarged for clarity.}
	\label{figure4}
\end{figure*}

\subsection{Synthetic Noisy RGB Images}

The considered synthetic noise types include additive Gaussian noise, Poisson noise and Brown Gaussian noise \cite{n2n}. For Gaussian noise, different noise levels with standard deviation $\sigma  \in \{ 15,25,50\} $ are considered. The Brown Gaussian noise is generated by applying a Gaussian filter (with kernel size $ 5 \times 5 $) to Gaussian noise with standard deviation of $\sigma  = 50$. For Poisson noise, we set the maximum event count ${\lambda _p} = 30$ and the noise is implemented as
${Y_i} = {\rm{Poisson(}}{\lambda _p}{X_i}{\rm{)/}}{\lambda _p} $ where $i$ is the color channel.

The BSDS500 dataset \cite{arbelaez2010contour} is used for training. Evaluation results are provided on the KODAK24 dataset \cite{kodak}, which contain 24 outdoor images. We train the N2C, N2N, and CGAN models using the same network as the proposed method. N2N is trained using noisy pairs with 2 realizations of each scene. 
N2V, N2S, S2S, N2Same, NAC are retrained using their original network models.
The original DnCNN model trained in \cite{dncnn} is directly used as it is blind to noise.
Nr2N \cite{nr2n} is only tested for Gaussian noise, as only its model trained in Gaussian noise is available.
For RNAN, the results are obtained by the original models trained in \cite{rnan}, except for the case of Poisson and Brown Gaussian noise where it is retrained.

Table 1 shows the PSNR and SSIM results of the compared methods, 
whilst Table 2 shows the PI and LPIPS scores. 
PI and LPIPS are two perception quality measures that accord well with human perception. 
It can be seen that, N2C outperforms DnCNN, which is mainly due to more sophisticated network architecture using residual channel attention blocks. 
For spatially independent Gaussian and Poisson noise, the PSNR of our method is about 1 dB lower than the supervised method N2C. 
Our method outperforms BM3D in the cases of Poisson and Brown Gaussian noise, 
and the advantage is more conspicuous in Brown Gaussian noise. 
Particularly, for Brown Gaussian noise, which is spatially correlated, 
our method even has better performance than the supervised method N2C in terms of PSNR. 
Besides, it can be seen from Table 2 that our method achieves the best PI and LPIPS scores, 
which demonstrate that it can yield the best perceptual quality.

\begin{table*}[!t]
	\renewcommand\arraystretch{2}
	\footnotesize
	\centering
	\caption{Results of our method for different values of $\lambda $ in the condition of Gaussian noise with $\sigma  = 15$.}
	\begin{tabular}{|c|c|c|c|c|c|c|c|}
		\hline
		$1/\lambda $ & 0 (CGAN) & 1 & 10 & 20 & 30 & 50 & 100 \\ \hline
		PSNR/SSIM & 29.97/0.8602 & 31.44/0.8867 & 32.85/0.8998 & 33.53/0.9049 & \textbf{33.99}/\textbf{0.9115} & 33.80/0.9108 & 30.97/0.8679 \\ \hline
	\end{tabular}
\end{table*}

Fig. 3 compares the visual quality of typical restoration by the methods. It is clear that our method can achieve significant better perceptual quality than DnCNN, N2C, N2N, N2V and BM3D, as it preserves more detail information while DnCNN, N2C, N2N and BM3D yield over-smoothed images. It is especially obvious when the image has abundant details. This advantage is due to the fact that the optimal transport criterion seeks a restoration that has the same distribution as natural clean images.

On the whole, in terms of both distortion (PSNR/SSIM) and perceptual quality (PI/LPIPS), our method even compares favorably with state-of-the-art supervised methods, especially in the case of complex noise, e.g., spatially correlated Brown Gaussian noise. The results also accord well with recent findings on the distortion-perception trade-off that, better perception quality can only be achieved at some increase of the distortion, e.g., MSE, PSNR and SSIM \cite{blau2019rethinking,yan2021perceptual}.

Moreover, Table 3 shows the results of the proposed method for different values of $\lambda $ in the condition of Gaussian noise with $\sigma  = 15$. While Theorem 1 shows that any $\lambda  > 1$ is an optimal selection, the empirical results show that a value of $\lambda  < 1$ achieves the highest PSNR. This is because in practice the used WGAN can only approximate the Wasserstein-1 distance.

\subsection{Synthetic Noisy Depth Images}

Depth images only capture the geometry of scenes, which is much different from RGB images. We consider the denoising of depth images with synthetic Gaussian noise with different standard deviation $\sigma  \in \{ 15,25,50\} $. The modes are trained on a simulated depth image dataset SUNCG \cite{suncg}, which contains 45,622 images of scenes with realistic room and furniture layouts.

\begin{table}[!t]
	\renewcommand\arraystretch{2}
	\footnotesize
	\centering
	\caption{Quantitative comparison on synthetic noisy depth images with Gaussian noise.}	
	\begin{tabular}{|c|c|c|c|}
		\hline
		Method        & $\sigma = 15 $                    & $\sigma = 25 $                  & $\sigma = 50 $                    \\ \hline
		Noisy         & 24.61/0.2510          & 20.17/0.1218          & 14.15/0.0451          \\ \hline
		RNAN \cite{rnan} & \textbf{41.67}/\textbf{0.9962} & \textbf{39.73}/\textbf{0.9890} & \textbf{35.85}/\textbf{0.9770}          \\ \hline
		DnCNN \cite{dncnn} & 40.97/0.9828 & 38.67/0.9722 & 34.57/0.9477          \\ \hline
		N2C           & 40.62/0.9904         & 38.23/0.9853          & 34.88/0.9746 \\ \hline
		N2N \cite{n2n}   & 40.40/0.9899          & 38.17/0.9850          & 34.72/0.9735          \\ \hline
		N2V \cite{n2v}   & 33.66/0.9517          & 31.95/0.9400          & 29.37/0.9258          \\ \hline
		N2S \cite{n2s}         & 35.17/0.8919          & 33.87/0.8803          & 29.46/0.7700 \\ \hline
		S2S \cite{s2s}           & 34.86/0.8933          & 33.05/0.8763          & 28.95/0.7625 \\ \hline
		N2Same \cite{n2same}           & 30.36/0.8649          & 27.48/0.7599          & 24.17/0.6803 \\ \hline
		NAC \cite{nac}           & 32.23/0.8737          & 29.91/0.8352          & 27.28/0.6344 \\ \hline
		BM3D \cite{bm3d}  & 40.38/0.9895          & 38.17/0.9778          & 33.66/0.9409          \\ \hline
		CGAN          & 31.04/0.9577          & 30.73/0.9549          & 29.77/0.9462          \\ \hline
		Ours          & 38.93/0.9843          & 37.11/0.9775          & 34.43/0.9595          \\ \hline
	\end{tabular}
\end{table}

Table 4 shows the PSNR and SSIM results of the compared methods on 27 images from the Middlebury dataset \cite{middle1,middle2}. Since the distribution of depth images is much simpler than that of color images, for the same level of noise each method can yield better PSNR and SSIM results than in the RGB experiments. It can be seen that, as the noise strength increases, the PSNR and SSIM of our method get closer to that of supervised methods. For example, in the case of noise strength $\sigma  = 50$, the PSNR of N2C is 34.88 dB, while that of our method is 34.43 dB.

Fig. 4 compares the visual quality of the methods on a typical restoration example in the case of  $\sigma  = 25$. Again, our method yields better perceptual quality than the supervised methods and BM3D, as it preserves more details. The better perception quality is achieved at some necessary increase of PSNR \cite{blau2019rethinking,yan2021perceptual}.

\subsection{Real-world Microscopy Images}

Next, we consider several real-world denoising tasks, where the ground-truth is difficult to collect. The first is microscopy image denoising. Microscopy images are an important source of data for biological and medical research. However, due to factors such as illumination and equipment during acquisition, microscopy images are inevitably corrupted by noise, which would affect subsequent high-precision analysis. Besides, since clean reference images are unavailable, unsupervised or self-supervised methods are desired.

\begin{table}[!t]
	\renewcommand\arraystretch{2}
	\footnotesize
	\centering
	\caption{Quantitative comparison on real-world microscopy images.}
	\begin{tabular}{|c|c|c|c|}
		\hline
		\multicolumn{2}{|c|}{Method}           & PSNR/SSIM   &PI/LPIPS \\ \hline
		\multicolumn{2}{|c|}{Noisy}             & 28.17/0.5685  &- \\ \hline
		\multirow{2}{*}{Supervised}   & N2C     & 35.28/\textbf{0.9291}  &7.50/0.122\\ \cline{2-4} 
		& N2N \cite{n2n}       & 35.07/0.9162 &6.93/0.130\\ \hline
		\multirow{3}{*}{Self-supervised}   & N2S \cite{n2s}              & 30.58/0.6474 &7.49/0.452\\ \cline{2-4} 
		& S2S \cite{s2s}              & 30.76/0.6950 &7.16/0.449\\ \cline{2-4}
		& N2Same \cite{n2same}       & 33.05/0.8332 &6.53/0.169\\ \hline
		\multirow{4}{*}{Unsupervised} & VST+BM3D \cite{vst} & 33.95/0.8061 &6.61/0.199\\ \cline{2-4} 
		& VST+NLM \cite{nlm} & 32.59/0.7664 &7.32/0.217\\ \cline{2-4}
		& CGAN       & 31.79/0.8285 &6.32/0.182\\ \cline{2-4} 
		& Ours (clean)              & 34.49/0.8969 &5.97/0.118\\ \cline{2-4}
		& Ours (pseudo)              & \textbf{35.73}/0.9051 &\textbf{5.93}/\textbf{0.069}\\ \hline
	\end{tabular}
\end{table}

We use a real fluorescence microscopy dataset, FMD \cite{micro}, for training and testing. The dataset consists of 12,000 real fluorescence microscopy images of different samples obtained with different microscopes. It uses an image averaging strategy to collect pseudo ground-truth. Specifically, for each scene, 50 images are captured and averaged to get the pseudo ground-truth. N2C is trained with the supervision of this pseudo ground-truth. We use the noisy images in the training set to train CGAN, N2V, and our model. For N2N, we randomly select two from the 50 noisy images of each scene to generate training pairs. In this experiment, we consider an improved version of BM3D for microscopy images, VST+BM3D \cite{vst}. 
Two variants of our method are considered.
The first uses the averaged images in FMD as clean source, denoted by “ours (pseudo)”.
The second uses clean images from another microscopy image dataset collected by the approach of FairSIM \cite{fair},
denoted by “ours (clean)”.


Table 5 shows the quantitative results on the test set of FMD.  
It is worth noting that all these results are computed based on the pseudo ground-truth. 
We also provide the PI and LPIPS scores to evaluate the perceptual quality of restored microscopy images by treating them as color images.
It can be seen that our method achieves the highest PSNR, best PI and LPIPS scores, 
while N2C yields the highest SSIM. 
Even though using clean images from another source degrades the performance of our method a bit, 
it still gets the highest PSNR/SSIM among the compared self-supervised and unsupervised methods,
and at the meantime achieves the best PI/LPIPS scores other than “ours (pseudo)”.
The results demonstrate the feasibility and effectiveness of using a similar but different source of clean images in training our method. 
This means our method can be easily applied to the applications in which clean source images are unavailable.

Fig. 5 presents visual comparison on a typical restoration example. 
For visual clarity, we process the grayscale images of the R, G, B three channels separately, 
and then merge the three restored channels into a color image. 
Clearly, the restoration of our method is much clearer than that of VST+BM3D, N2C and N2N, 
which demonstrates its much better perceptual quality. 
More examples are provided in the supplemental material.

\begin{figure*}[!t]
	\centering
	\subfigure[Test image]{
		\includegraphics[width=0.24\linewidth]{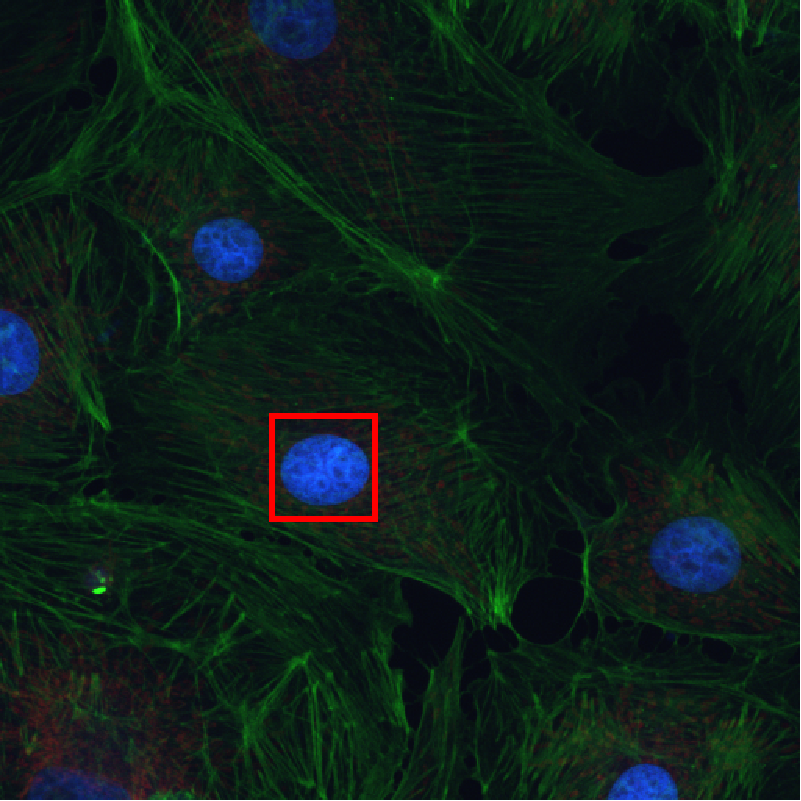}
	}\subfigure[Ground truth (average)]{
		\includegraphics[width=0.24\linewidth]{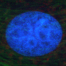}
	}\subfigure[Noisy (26.55 dB)]{
		\includegraphics[width=0.24\linewidth]{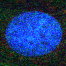}
	}\subfigure[VST+BM3D (33.87/7.87/0.099)]{
		\includegraphics[width=0.24\linewidth]{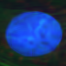}
	}
	\subfigure[VST+NLM (33.10/8.53/0.143)]{
		\includegraphics[width=0.24\linewidth]{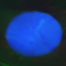}
	}\subfigure[N2C (34.50/5.40/0.072)]{
		\includegraphics[width=0.24\linewidth]{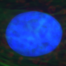}
	}\subfigure[N2N (34.36/5.27/0.070)]{
		\includegraphics[width=0.24\linewidth]{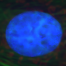}
	}\subfigure[Ours (\textbf{34.69}/\textbf{5.16}/\textbf{0.048})]{
		\includegraphics[width=0.24\linewidth]{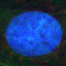}
	}
	\caption{Visual comparison on real-world microscopy images, where the pseudo ground-truth is obtained by averaging over 50 realizations of each scene. The PSNR/PI/LPIPS results are provided in the brackets. The images are enlarged for clarity.}
	\label{figure5}
\end{figure*}

\begin{figure*}[!t]
	\centering
	\subfigure[Test image]{
		\includegraphics[width=0.24\linewidth]{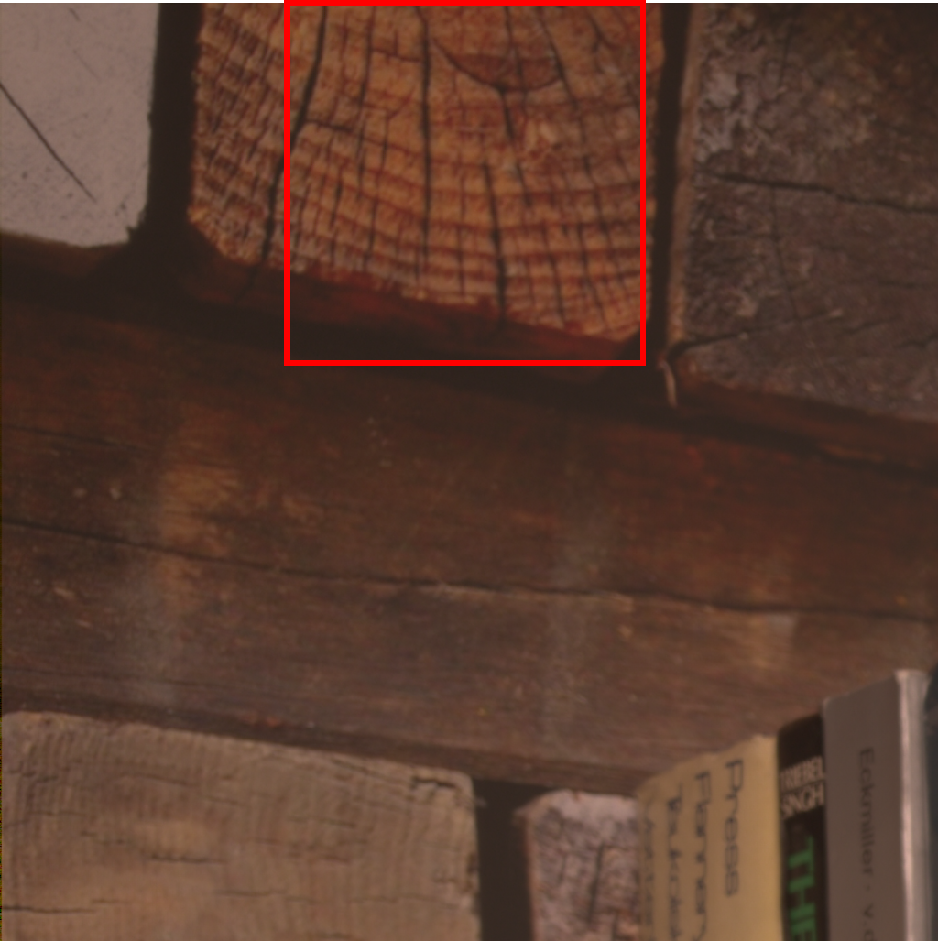}
	}\subfigure[Ground truth (average)]{
		\includegraphics[width=0.24\linewidth]{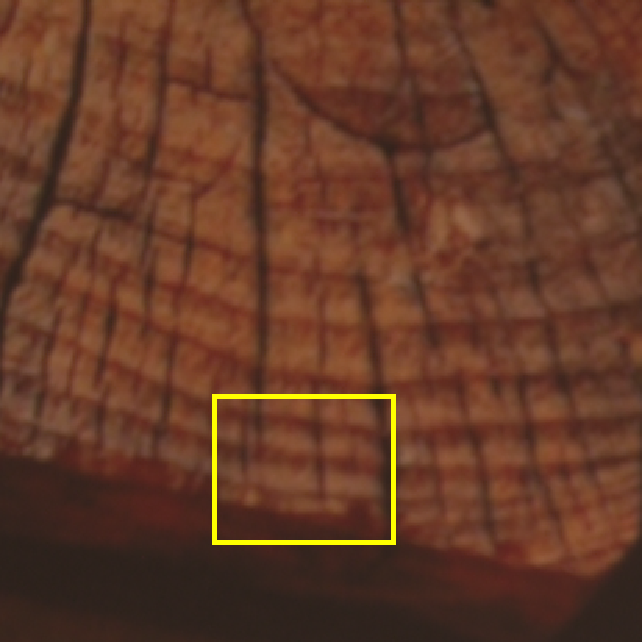}
	}\subfigure[Noisy (31.00 dB)]{
		\includegraphics[width=0.24\linewidth]{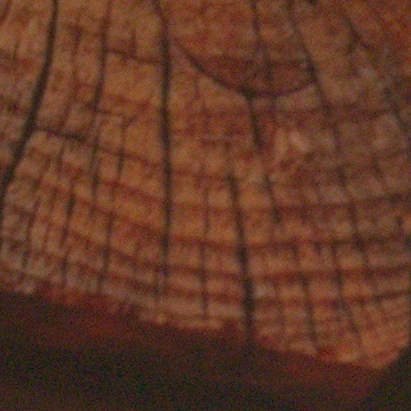}
	}\subfigure[BM3D (38.87/8.04/0.028)]{
		\includegraphics[width=0.24\linewidth]{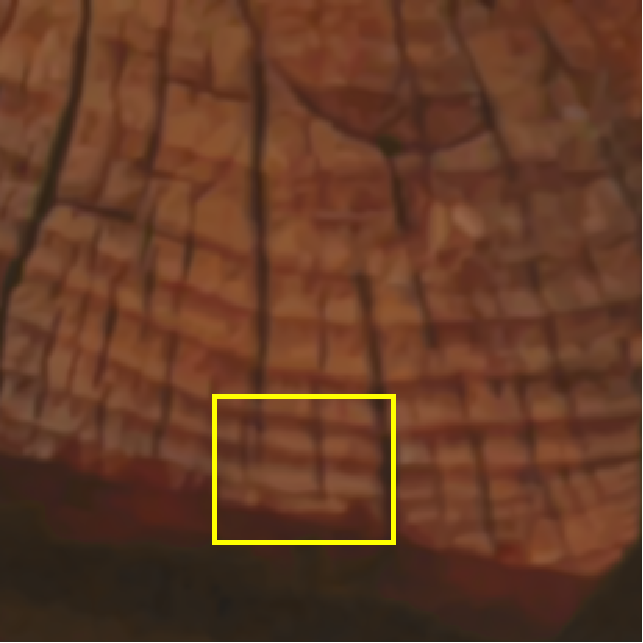}
	}
	\subfigure[N2C (\textbf{41.04}/7.90/0.013)]{
		\includegraphics[width=0.24\linewidth]{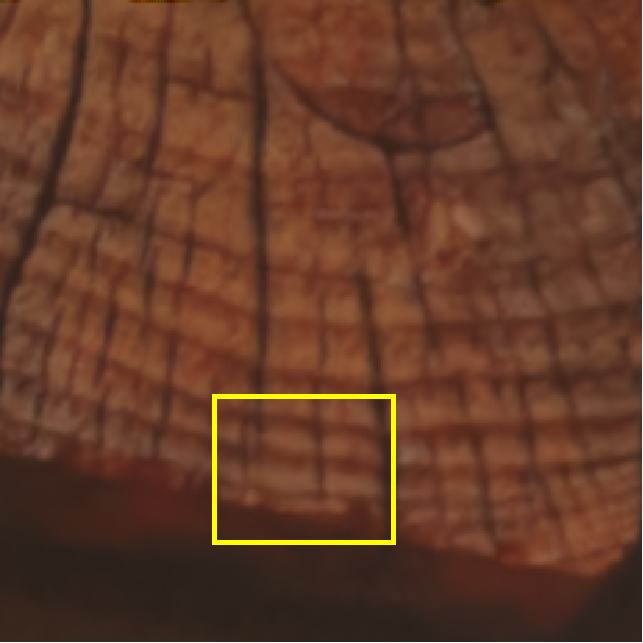}
	}\subfigure[N2N (40.91/8.03/0.014)]{
		\includegraphics[width=0.24\linewidth]{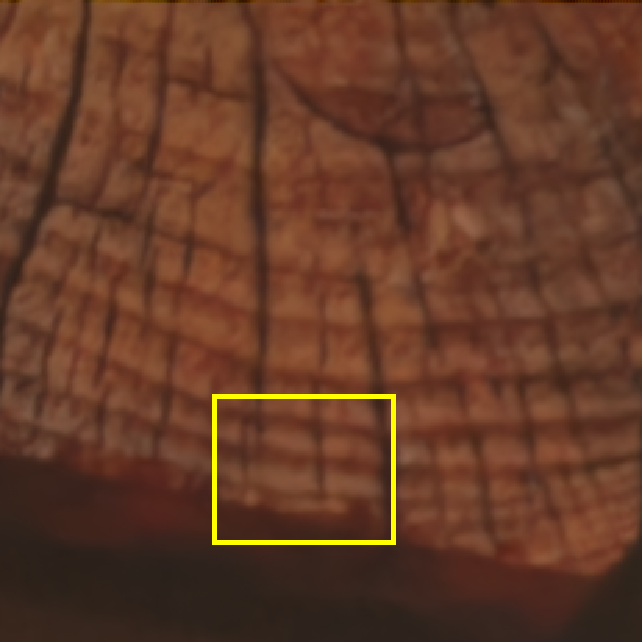}
	}\subfigure[CGAN (38.90/7.77/0.013)]{
		\includegraphics[width=0.24\linewidth]{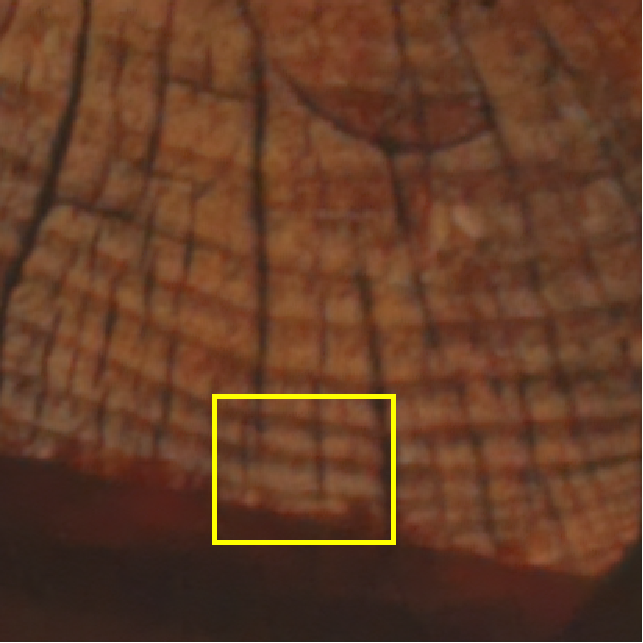}
	}\subfigure[Ours (40.48/\textbf{7.48}/\textbf{0.009})]{
		\includegraphics[width=0.24\linewidth]{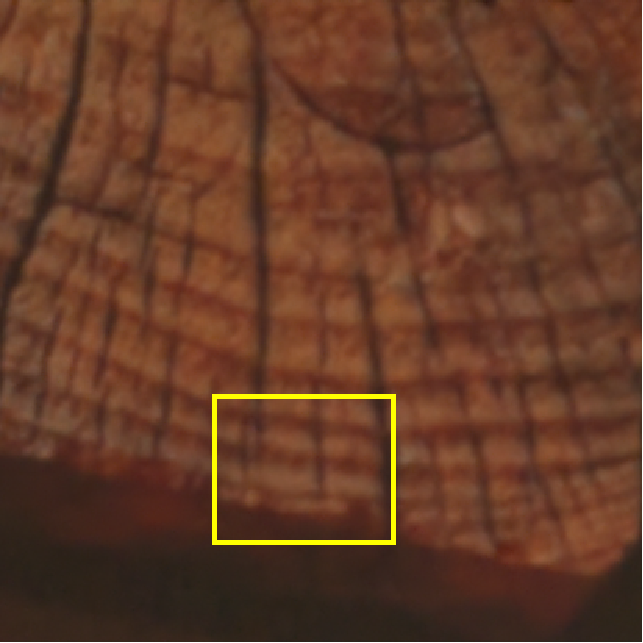}
	}
	\caption{Visual comparison on real-world photographic images, where the pseudo ground-truth is obtained by averaging over 10 realizations of each scene. The PSNR/PI/LPIPS results are provided in the brackets. The images are enlarged for clarity.}
	\label{figure6}
\end{figure*}

\subsection{Real-world Photographic Images}

\begin{figure*}[!t]
	\centering
	\subfigure[RGB of the scene ]{
		\includegraphics[width=0.324\linewidth]{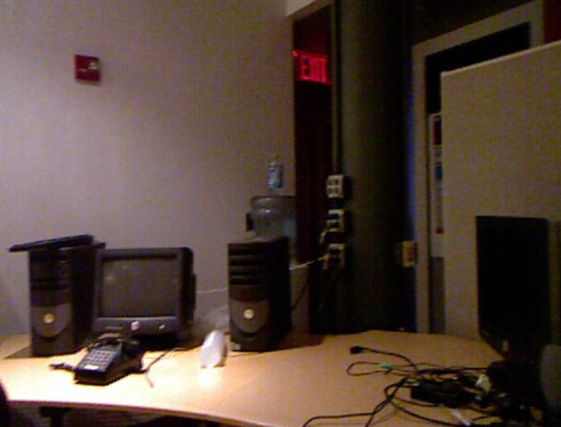}
	}\subfigure[Depth of Kinect]{
		\includegraphics[width=0.324\linewidth]{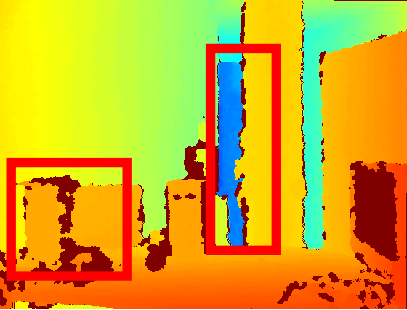}
	}\subfigure[Colorization ]{
		\includegraphics[width=0.324\linewidth]{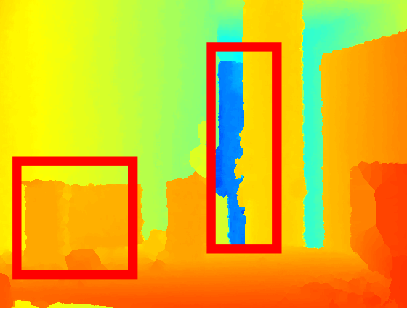}
	}
	\subfigure[Cross-bilateral filter]{
		\includegraphics[width=0.324\linewidth]{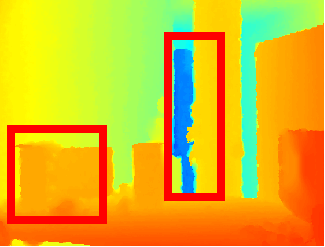}
	}\subfigure[Nearest-neighbor ]{
		\includegraphics[width=0.324\linewidth]{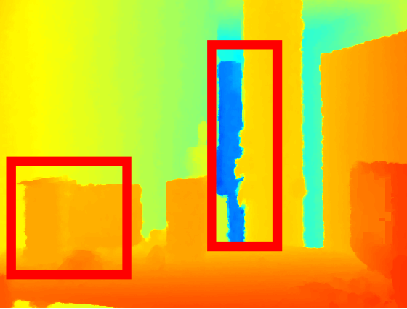}
	}\subfigure[Ours ]{
		\includegraphics[width=0.324\linewidth]{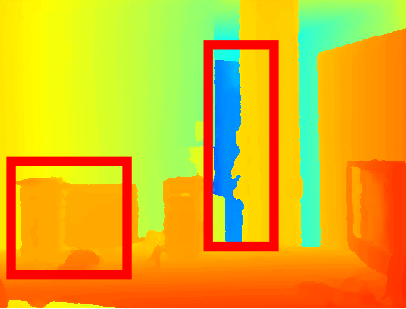}
	}
	\caption{Visual comparison on real-world depth images from the NYU dataset \cite{nyu} captured by Kinect. The colorization and the cross-bilateral filter methods are RGB-D fusion methods.}
	\label{figure7}
\end{figure*}

This experiment evaluates the methods on denoising real-world photographic noisy images. 
In practical photography, due to factors such as ambient light and exposure time, 
the captured photos often have noise, which leads to unsatisfactory visual quality. 
Besides, because of poor aperture and photoreceptor used in mobile phone cameras, 
noise is common in the captured images.

\begin{table}[!t]
	\renewcommand\arraystretch{2}
	\footnotesize
	\centering
	\caption{Quantitative comparison on real-world photographic images.}
	\begin{tabular}{|c|c|c|c|}
		\hline
		\multicolumn{2}{|c|}{Method}     & PSNR/SSIM        & PI/LPIPS     \\ \hline
		\multicolumn{2}{|c|}{Noisy}        & 30.16/0.6349    & -      \\ \hline
		\multirow{2}{*}{Supervised}   & N2C      & \textbf{40.90}/\textbf{0.9651}  & 8.53/0.033 \\ \cline{2-4} 
		& N2N \cite{n2n}  & 40.88/0.9648      & 8.49/0.033   \\ \hline
		\multirow{4}{*}{Self-supervised}               & N2V \cite{n2v}  & 33.42/0.8226      &  7.87/0.149  \\ \cline{2-4}
		& N2S \cite{n2s}  & 32.98/0.7403      &  8.13/0.172  \\ \cline{2-4} 
		& S2S \cite{s2s}  & 38.41/0.9459      &  7.55/0.035  \\ \cline{2-4}
		& N2Same \cite{n2same}  & 35.81/0.9429      &  8.23/0.034  \\ \hline
		\multirow{4}{*}{Unsupervised} & BM3D \cite{bm3d} & 37.97/0.9495      &  8.45/0.032  \\ \cline{2-4} 
		& CGAN         & 38.20/0.9487      & 7.78/0.030    \\ \cline{2-4} 
		& Ours (clean)         & 38.67/0.9502      & 7.53/0.027    \\ \cline{2-4} 
		& Ours (pseudo)         & 39.69/0.9527      & \textbf{7.42}/\textbf{0.015}    \\ \hline
	\end{tabular}
\end{table}

The N2C, N2N, CGAN, N2V, N2S, S2S, N2Same, and our models are trained on a real photograph image dataset SIDD \cite{sidd}, which is collected by smartphone cameras and contains about 30,000 noisy images from 10 scenes under different lighting conditions. For each scene, the pseudo ground-truth is collected as follows: first continuously shoot multiple images of the scene, then impose image processing (such as image registration, abnormal image removal, etc.), and finally adopt weighted average to obtain the ground-truth of this scene. This pseudo ground-truth is used for training N2C. For N2N, we randomly select two from the noisy images of each scene to generate training pairs. Since the size of the test images is too large (e.g. $4000\times3000$), we crop the test images into patches of size $1024\times1024$ for testing. 
Similar to the microscopy image deoising experiment, 
two variants of our method are considered.
The first uses the averaged images in SIDD as the clean source, denoted by “ours (pseudo)”.
The second uses clean images from a different source collected by a HUAWEI Mate 40 smartphone, denoted by “ours (clean)”.

Table 6 presents the quantitative results on the test set of SIDD. 
The ground-truth generated by post-processing and averaging is used to compute the PSNR and SSIM results. 
Accordingly, the results do not exactly reflect the difference from the real clean images. 
Our method achieves a PNSR of 39.69 dB, which is about 1.2 dB lower than that of N2C but about 1.7 dB higher than that of BM3D.
Again, our method gets the best PI and LPIPS scores, which demonstrates that it can achieve the best perceptual quality.
For our method, using clean images from another source leads to some degradation of performance, 
but it still outperforms all other self-supervised and unsupervised methods in PSNR/SSIM,
and at the meantime outperforms all the supervised, self-supervised and unsupervised methods other than “ours (pseudo)” in terms of PI/LPIPS scores.

Fig. 6 presents visual comparison on a restoration example. 
Since the noise level is small and the restoration PSNRs of the methods are quite high, e.g., near 40 dB, 
the difference in visual quality is not distinct. 

\subsection{Real-world Depth Images}

\begin{figure*}[!t]
	\centering
	\subfigure[RGB of the scene]{
		\includegraphics[width=0.24\linewidth]{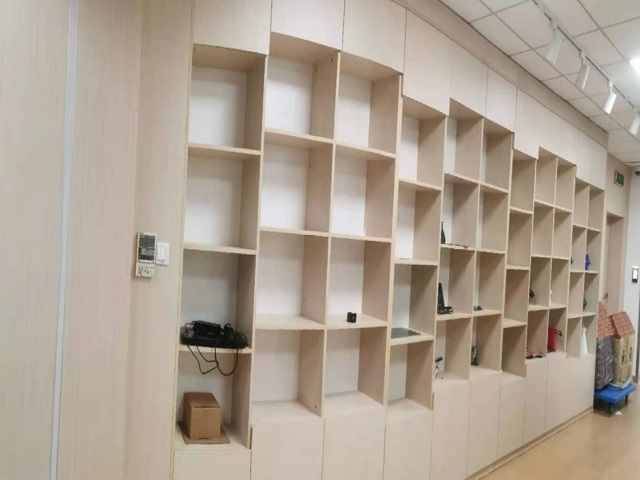}
	}\subfigure[Raw depth]{
		\includegraphics[width=0.24\linewidth]{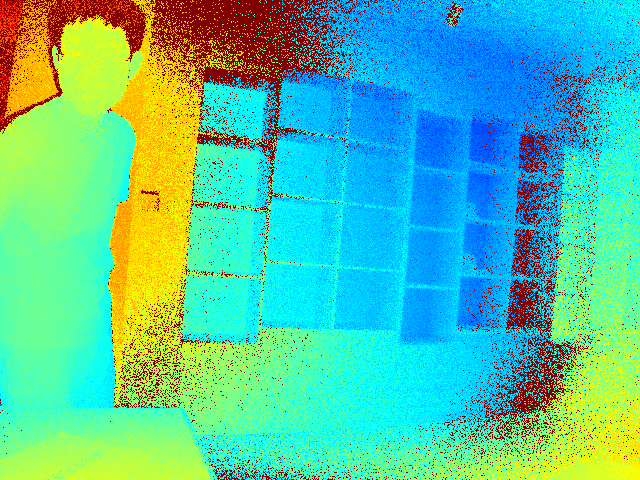}
	}\subfigure[Nearest-neighbor (NN)]{
		\includegraphics[width=0.24\linewidth]{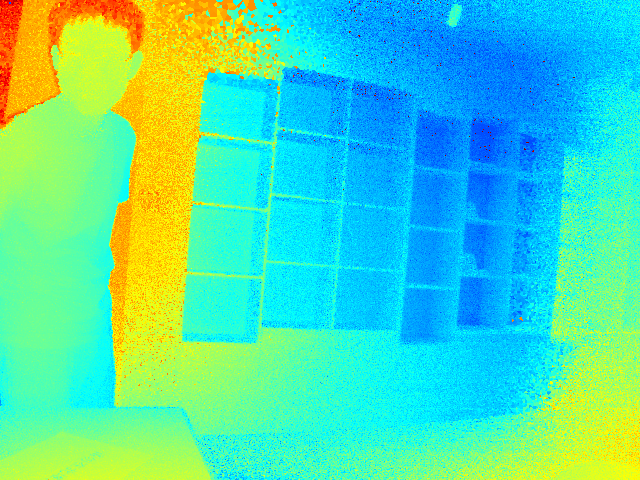}
	}\subfigure[CGAN]{
		\includegraphics[width=0.24\linewidth]{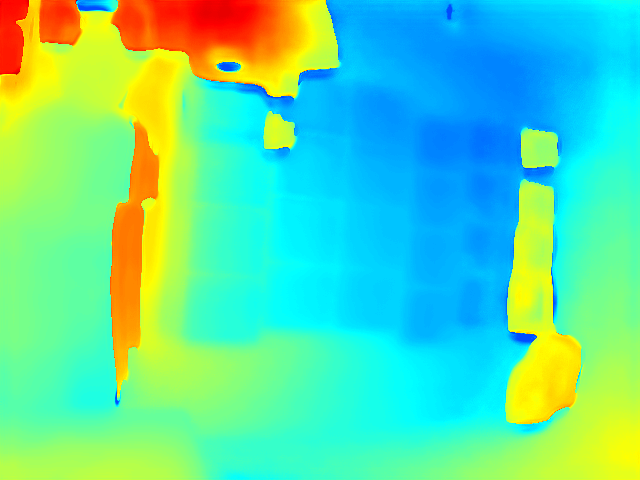}
	}
	\subfigure[NN+Gaussian filter]{
		\includegraphics[width=0.24\linewidth]{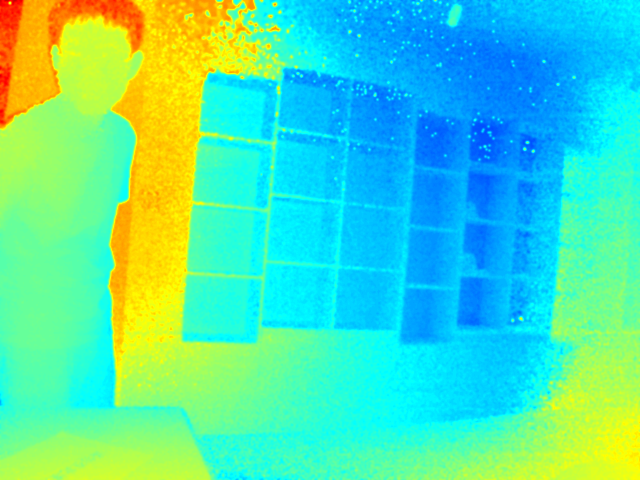}
	}\subfigure[NN+Bilateral filter]{
		\includegraphics[width=0.24\linewidth]{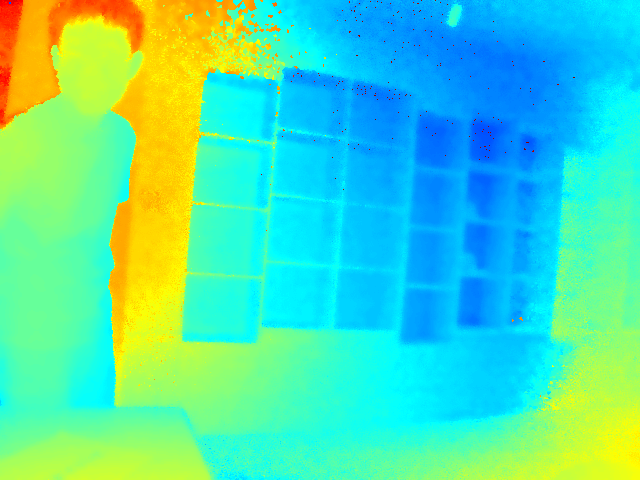}
	}\subfigure[NN+Median filter]{
		\includegraphics[width=0.24\linewidth]{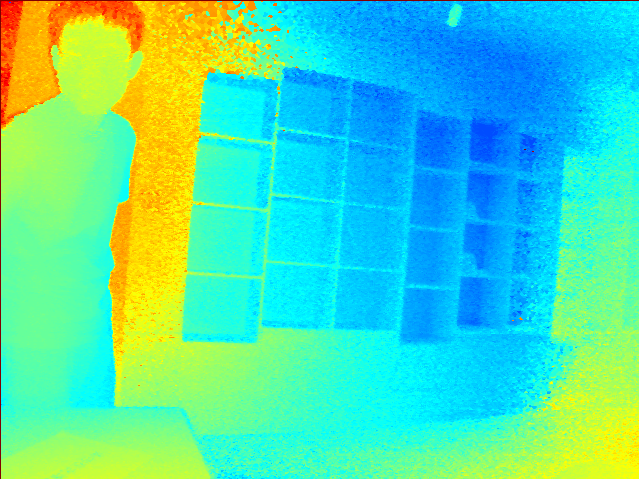}
	}\subfigure[NN+N2V]{
		\includegraphics[width=0.24\linewidth]{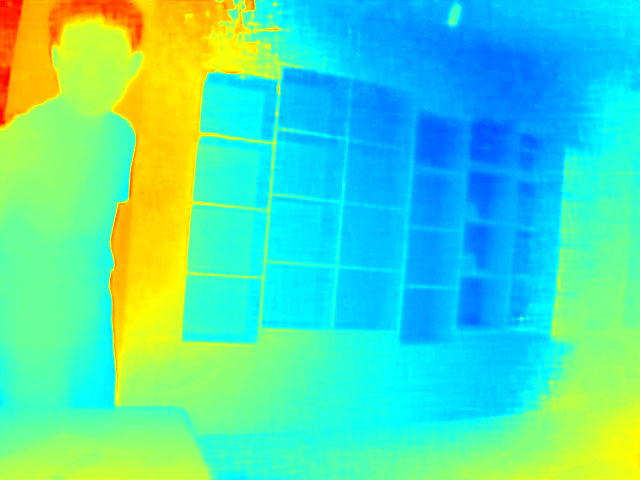}
	}
	\subfigure[NN+BM3D]{
		\includegraphics[width=0.24\linewidth]{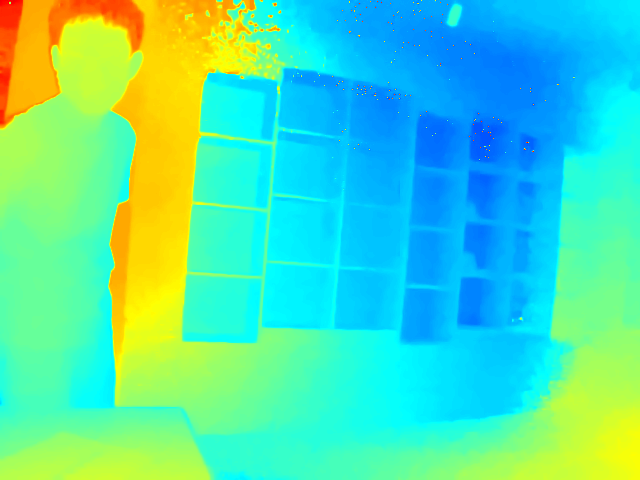}
	}\subfigure[Ours]{
		\includegraphics[width=0.24\linewidth]{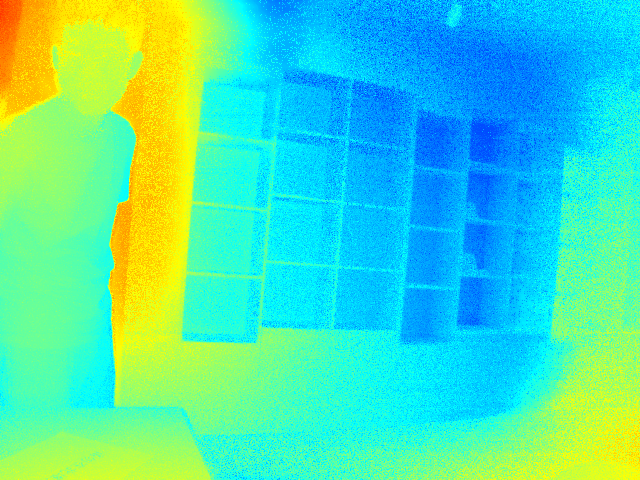}
	}\subfigure[Ours+Gaussian filter]{
		\includegraphics[width=0.24\linewidth]{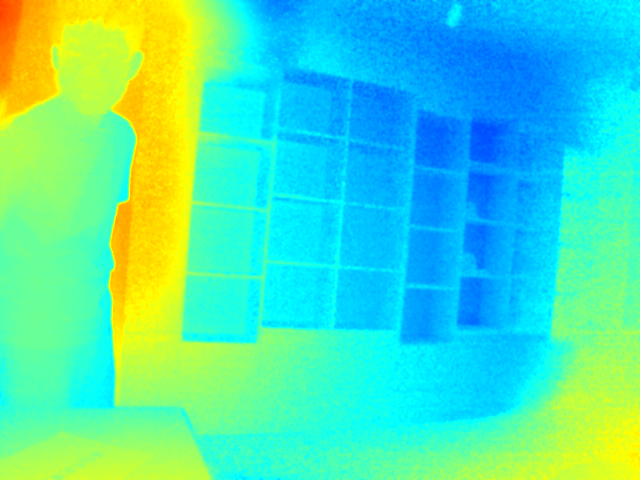}
	}\subfigure[Ours+Bilateral filter]{
		\includegraphics[width=0.24\linewidth]{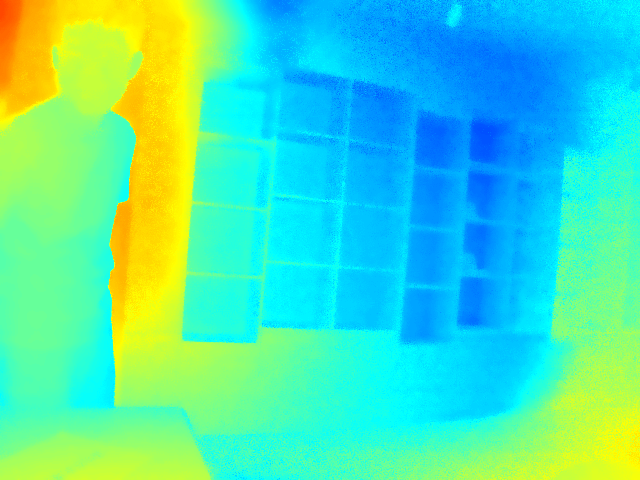}
	}
	\subfigure[Ours+Median filter]{
		\includegraphics[width=0.24\linewidth]{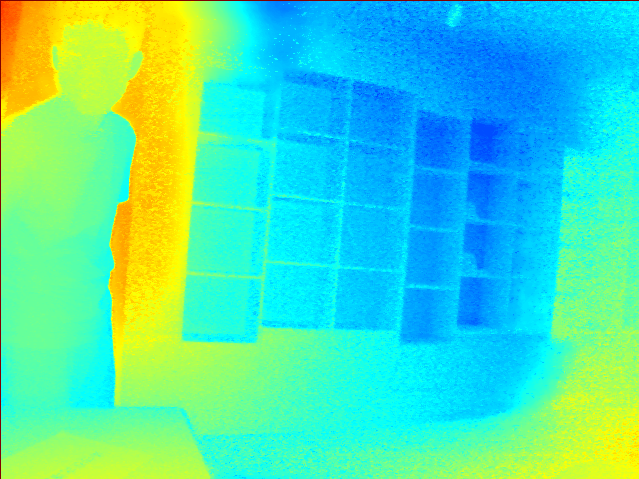}
	}\subfigure[Ours+N2V]{
		\includegraphics[width=0.24\linewidth]{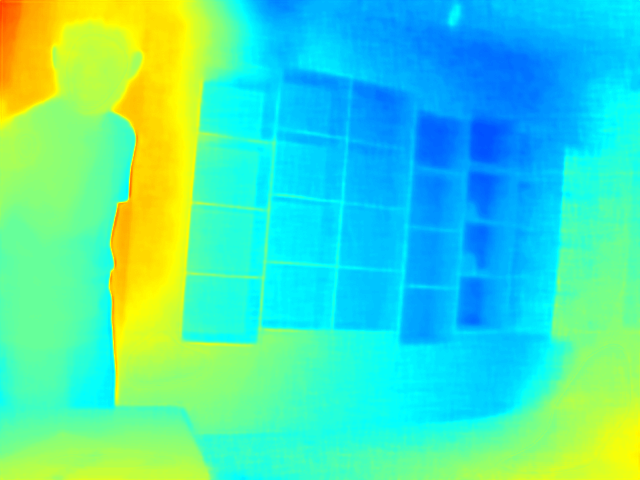}
	}\subfigure[Ours+BM3D]{
		\includegraphics[width=0.24\linewidth]{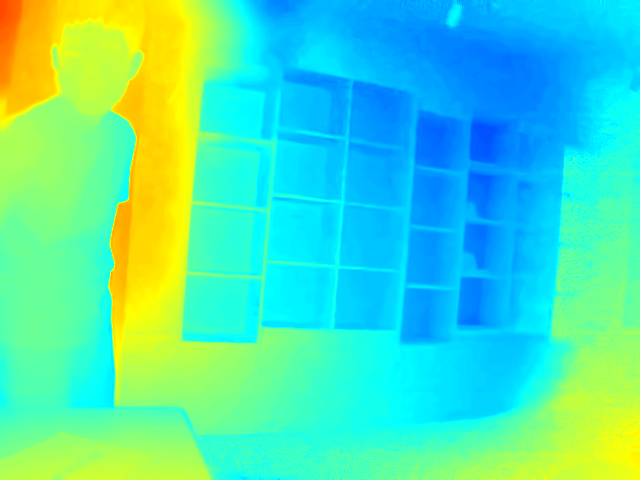}
	}\subfigure[Ours+OT denoising]{
		\includegraphics[width=0.24\linewidth]{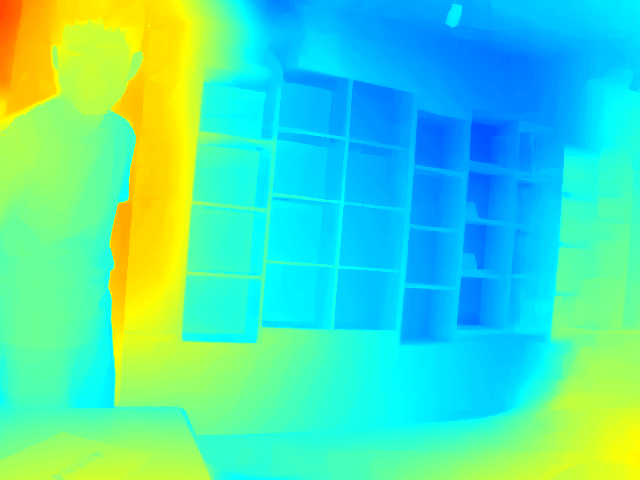}
	}
	\caption{Visual comparison on a real-world raw depth image captured by a commercial ToF camera. The RGB image is only used as a reference of the scene, which is not aligned with the depth image. “OT denoising” denotes our model trained on synthetic depth images with Gaussian noise as in Section 5.3.}
	\label{figure8}
\end{figure*}

Recently, depth cameras have become increasingly popular, e.g., it is widely equipped in latest mobile phones and tablets. Due to its imaging mechanism, depth images have much larger noise than RGB images. Besides, there commonly exist holes (invalid pixels) in depth images due to low reflectivity and transparency of objects in the scene. 

We first consider the denoising of Kinect depth images. As the Kinect camera is equipped with built-in filters, the output depth images have less noise but still have some holes. For training our model, we use the NYU dataset \cite{nyu} as the noisy sample domain, and randomly collect the same amount of patches from the SUNCG dataset \cite{suncg} as the clean target domain. The NYU dataset contains 1,449 depth images of indoor scenes and the corresponding RGB images captured by a Kinect camera. 

Since we have neither the corresponding ground-truth images nor noisy image pairs on the same scenes, the supervised methods N2C and N2N are not compared. We mainly compare the proposed method with representative and widely used inpainting methods for depth images, including the colorization, cross-bilateral filter, and nearest-neighbor methods. The classic nearest-neighbor method only uses a single depth image as input, whilst the colorization and cross-bilateral filter methods are RGB-D fusion methods, which take depth and corresponding RGB images as input.

Fig. 7 presents the restoration on a typical example. We colorize the depth image for visual clarity, in which the dark-red color represents invalid pixels with values of zero, i.e., holes. Obviously, our method achieves significantly better restoration quality than the nearest-neighbor method and the RGB-D fusion methods, colorization and cross-bilateral filter. The restoration of our method, which has straighter and sharper edges, is more consistent with the geometry of the scene (see the RGB of the scene). More examples are provided in the supplemental material.

\subsection{Real-world Raw Depth Images}

As the Kinect camera only provides pre-processed images, we additionally consider a raw depth denoising experiment using another commercial ToF camera which provides raw depth data without any pre-processing. We collect 1,430 raw depth images using this camera as the noisy sample domain. For the clean target domain, we also randomly collect the same amount of patches from the SUNCG dataset \cite{suncg}.

Since raw depth images have severe noise and invalid pixels, the denoising methods are used in combination with the nearest neighbor hole-filling methods. The compared denoising methods include N2V, BM3D, and three widely used methods for depth image denoising, Gaussian filter, median filter and bilateral filter. Our model can naturally learn to simultaneously fill holes and denoise at the same time. However, intensive extra experiments show that the following two-stage process can yield better performance than one-stage process. Specifically, since raw depth images contain many outliers and invalid pixels, we use the $\ell_1$ loss in \eqref{eq8} to train our model, which is robust to outliers. Then, as the restoration by this model still contain some noise, as shown in Fig. 8 (j), we cascade a post-denoiser to our model to obtain a refinement of the restoration. The evaluated post-denoisers include Gaussian filter, Bilateral filter, Median filter, N2V, BM3D and our model trained on synthetic depth images with Gaussian noise in Section 5.3.

Fig. 8 compares the restoration on a typical example. The depth images are colorized for visual clarity, in which the dark-red color represents zero-value invalid pixels. It can be seen from Fig. 8 that the raw depth image contains severe complex noise. Our method with a post-denoiser significantly outperforms the NN+BM3D, NN+N2V and the three filter based methods, as it can fill in the holes more naturally while preserving more edge details of the images. Particularly, the best performance is achieved when our model is paired with the denoiser trained on synthetic depth images with Gaussian noise. We provide more examples in the supplemental material. 

A significant feature of the depth denoising task is that, it is generally impossible to collect ground-truth reference images for supervised training. The methods \cite{DBLP:conf/cvpr/ZhangF18,DBLP:conf/eccv/JeonL18} construct paired training data by first reconstructing the 3D geometry of the scene. But 3D geometry reconstruction usually suffers from errors. Besides, there exists another challenge of pixel-to-pixel alignment in constructing paired training data from 3D reconstruction. Our method shows remarkable superiority in denoising raw depth data but requires neither paired training data nor any prior information on the degradation model.

\section{Conclusions}

An unsupervised restoration learning criterion has been proposed from the optimal transport theory, which does not assume any prior knowledge on the degradation model. Favorable properties of this criterion have been revealed, e.g., approximately maximal preservation of the information of the signal, whilst achieving perceptual reconstruction. Moreover, for the relaxed formulation of this criterion, which is used in practical implementation, we proved that in theory the relaxation has the same solution as the original criterion. 
Experiment results on synthetic and real-world data demonstrated that the new method even compares favorably with supervised methods. 
Particularly, the results demonstrated remarkable superiority of the new method in denoising raw depth images with complex noise, 
which may be preferred in other practical applications with harsh circumstances in the wild but without any prior knowledge on the degradation model.

\ifCLASSOPTIONcaptionsoff
\newpage
\fi

\bibliographystyle{ieeetr}
\bibliography{ref}

\begin{thebibliography}{10}

\bibitem{dncnn}
K.~Zhang, W.~Zuo, Y.~Chen, D.~Meng, and L.~Zhang, ``Beyond a gaussian denoiser:
  Residual learning of deep cnn for image denoising,'' {\em IEEE Transactions
  on Image Processing}, vol.~26, no.~7, pp.~3142--3155, 2017.

\bibitem{mao2016image}
X.~Mao, C.~Shen, and Y.-B. Yang, ``Image restoration using very deep
  convolutional encoder-decoder networks with symmetric skip connections,'' in
  {\em Advances in Neural Information Processing Systems}, pp.~2802--2810,
  2016.

\bibitem{zhang2020residual}
Y.~Zhang, Y.~Tian, Y.~Kong, B.~Zhong, and Y.~Fu, ``Residual dense network for
  image restoration,'' {\em IEEE Transactions on Pattern Analysis and Machine
  Intelligence}, vol.~43, no.~7, pp.~2480--2495, 2020.

\bibitem{nlm}
A.~Buades, B.~Coll, and J.-M. Morel, ``A non-local algorithm for image
  denoising,'' in {\em IEEE Conference Computer Vision and Pattern Recognition
  (CVPR)}, vol.~2, pp.~60--65, 2005.

\bibitem{bm3d}
K.~Dabov, A.~Foi, V.~Katkovnik, and K.~Egiazarian, ``Image denoising by sparse
  3-d transform-domain collaborative filtering,'' {\em IEEE Transactions on
  Image Processing}, vol.~16, no.~8, pp.~2080--2095, 2007.

\bibitem{dip}
D.~Ulyanov, A.~Vedaldi, and V.~Lempitsky, ``Deep image prior,'' in {\em IEEE
  Conference Computer Vision and Pattern Recognition (CVPR)}, pp.~9446--9454,
  2018.

\bibitem{n2n}
J.~Lehtinen, J.~Munkberg, J.~Hasselgren, S.~Laine, T.~Karras, M.~Aittala, and
  T.~Aila, ``Noise2noise: Learning image restoration without clean data,'' in
  {\em International Conference on Machine Learning (ICML)}, pp.~4620--4631,
  2018.

\bibitem{n2v}
A.~Krull, T.-O. Buchholz, and F.~Jug, ``Noise2void-learning denoising from
  single noisy images,'' in {\em IEEE Conference Computer Vision and Pattern
  Recognition (CVPR)}, pp.~2129--2137, 2019.

\bibitem{n2s}
J.~Batson and L.~Royer, ``Noise2self: Blind denoising by self-supervision,'' in
  {\em International Conference on Machine Learning (ICML)}, pp.~524--533,
  PMLR, 2019.

\bibitem{mbvd}
T.~Ehret, A.~Davy, J.-M. Morel, G.~Facciolo, and P.~Arias, ``Model-blind video
  denoising via frame-to-frame training,'' in {\em IEEE Conference Computer
  Vision and Pattern Recognition (CVPR)}, pp.~11369--11378, 2019.

\bibitem{s2s}
Y.~Quan, M.~Chen, T.~Pang, and H.~Ji, ``Self2self with dropout: Learning
  self-supervised denoising from single image,'' in {\em IEEE Conference
  Computer Vision and Pattern Recognition (CVPR)}, pp.~1890--1898, 2020.

\bibitem{esur}
M.~Zhussip, S.~Soltanayev, and S.~Y. Chun, ``Extending stein's unbiased risk
  estimator to train deep denoisers with correlated pairs of noisy images,'' in
  {\em Advances in Neural Information Processing Systems}, pp.~1465--1475,
  2019.

\bibitem{laine2019high}
S.~Laine, T.~Karras, J.~Lehtinen, and T.~Aila, ``High-quality self-supervised
  deep image denoising,'' {\em Advances in Neural Information Processing
  Systems}, vol.~32, pp.~6970--6980, 2019.

\bibitem{wu2020unpaired}
X.~Wu, M.~Liu, Y.~Cao, D.~Ren, and W.~Zuo, ``Unpaired learning of deep image
  denoising,'' in {\em European Conference on Computer Vision (ECCV)},
  pp.~352--368, Springer, 2020.

\bibitem{buades2005review}
A.~Buades, B.~Coll, and J.-M. Morel, ``A review of image denoising algorithms,
  with a new one,'' {\em Multiscale Modeling \& Simulation}, vol.~4, no.~2,
  pp.~490--530, 2005.

\bibitem{mip_deblur}
F.~Wen, R.~Ying, Y.~Liu, P.~Liu, and T.-K. Truong, ``A simple local minimal
  intensity prior and an improved algorithm for blind image deblurring,'' {\em
  IEEE Transactions on Circuits and Systems for Video Technology}, pp.~1--15,
  2020.

\bibitem{lebrun2012analysis}
M.~Lebrun, ``An analysis and implementation of the bm3d image denoising
  method,'' {\em Image Processing on Line}, vol.~2012, pp.~175--213, 2012.

\bibitem{nr2n}
N.~Moran, D.~Schmidt, Y.~Zhong, and P.~Coady, ``Noisier2noise: Learning to
  denoise from unpaired noisy data,'' in {\em IEEE Conference Computer Vision
  and Pattern Recognition (CVPR)}, pp.~12064--12072, 2020.

\bibitem{nac}
J.~Xu, Y.~Huang, M.-M. Cheng, L.~Liu, F.~Zhu, Z.~Xu, and L.~Shao,
  ``Noisy-as-clean: learning self-supervised denoising from corrupted image,''
  {\em IEEE Transactions on Image Processing}, vol.~29, pp.~9316--9329, 2020.

\bibitem{gan2gan}
S.~Cha, T.~Park, B.~Kim, J.~Baek, and T.~Moon, ``Gan2gan: Generative noise
  learning for blind denoising with single noisy images,'' in {\em
  International Conference on Learning Representations}, 2020.

\bibitem{krull2020probabilistic}
A.~Krull, T.~Vi{\v{c}}ar, M.~Prakash, M.~Lalit, and F.~Jug, ``Probabilistic
  noise2void: Unsupervised content-aware denoising,'' {\em Frontiers in
  Computer Science}, vol.~2, p.~5, 2020.

\bibitem{neighbor2neighbor}
S.~Soltanayev and S.~Y. Chun, ``Training deep learning based denoisers without
  ground truth data,'' in {\em Advances in Neural Information Processing
  Systems}, pp.~3261--3271, 2018.

\bibitem{soltanayev2018training}
S.~Soltanayev and S.~Y. Chun, ``Training deep learning based denoisers without
  ground truth data,'' in {\em Advances in Neural Information Processing
  Systems}, pp.~3261--3271, 2018.

\bibitem{cgan}
M.~Mirza and S.~Osindero, ``Conditional generative adversarial nets,'' {\em
  arXiv preprint arXiv:1411.1784}, 2014.

\bibitem{RoCGAN}
G.~G. Chrysos, J.~Kossaifi, and S.~Zafeiriou, ``Rocgan: Robust conditional
  gan,'' {\em International Journal of Computer Vision}, vol.~128, no.~10,
  pp.~2665--2683, 2020.

\bibitem{ambientgan}
A.~Bora, E.~Price, and A.~G. Dimakis, ``Ambientgan: Generative models from
  lossy measurements,'' in {\em International Conference on Learning
  Representations}, 2018.

\bibitem{nrgan}
T.~Kaneko and T.~Harada, ``Noise robust generative adversarial networks,'' in
  {\em IEEE Conference Computer Vision and Pattern Recognition (CVPR)},
  pp.~8404--8414, 2020.

\bibitem{wang2006quality}
Z.~Wang, G.~Wu, H.~R. Sheikh, E.~P. Simoncelli, E.-H. Yang, and A.~C. Bovik,
  ``Quality-aware images,'' {\em IEEE Transactions on Image Processing},
  vol.~15, no.~6, pp.~1680--1689, 2006.

\bibitem{wang2005reduced}
Z.~Wang and E.~P. Simoncelli, ``Reduced-reference image quality assessment
  using a wavelet-domain natural image statistic model,'' in {\em Human Vision
  and Electronic Imaging X}, vol.~5666, pp.~149--159, 2005.

\bibitem{mittal2012no}
A.~Mittal, A.~K. Moorthy, and A.~C. Bovik, ``No-reference image quality
  assessment in the spatial domain,'' {\em IEEE Transactions on Image
  Processing}, vol.~21, no.~12, pp.~4695--4708, 2012.

\bibitem{mittal2012making}
A.~Mittal, R.~Soundararajan, and A.~C. Bovik, ``Making a “completely blind”
  image quality analyzer,'' {\em IEEE Signal Processing Letters}, vol.~20,
  no.~3, pp.~209--212, 2012.

\bibitem{moorthy2011blind}
A.~K. Moorthy and A.~C. Bovik, ``Blind image quality assessment: From natural
  scene statistics to perceptual quality,'' {\em IEEE Transactions on Image
  Processing}, vol.~20, no.~12, pp.~3350--3364, 2011.

\bibitem{SaadBC12}
M.~A. Saad, A.~C. Bovik, and C.~Charrier, ``Blind image quality assessment: A
  natural scene statistics approach in the dct domain,'' {\em IEEE Transactions
  on Image Processing}, vol.~21, no.~8, pp.~3339--3352, 2012.

\bibitem{blau2018perception}
Y.~Blau and T.~Michaeli, ``The perception-distortion tradeoff,'' in {\em IEEE
  Conference Computer Vision and Pattern Recognition (CVPR)}, pp.~6228--6237,
  2018.

\bibitem{srgan}
C.~Ledig, L.~Theis, F.~Husz{\'a}r, J.~Caballero, A.~Cunningham, A.~Acosta,
  A.~Aitken, A.~Tejani, J.~Totz, Z.~Wang, {\em et~al.}, ``Photo-realistic
  single image super-resolution using a generative adversarial network,'' in
  {\em IEEE Conference Computer Vision and Pattern Recognition (CVPR)},
  pp.~4681--4690, 2017.

\bibitem{wang2018esrgan}
X.~Wang, K.~Yu, S.~Wu, J.~Gu, Y.~Liu, C.~Dong, Y.~Qiao, and C.~Change~Loy,
  ``Esrgan: Enhanced super-resolution generative adversarial networks,'' in
  {\em European Conference on Computer Vision (ECCV) Workshops}, 2018.

\bibitem{gan}
I.~Goodfellow, J.~Pouget-Abadie, M.~Mirza, B.~Xu, D.~Warde-Farley, S.~Ozair,
  A.~Courville, and Y.~Bengio, ``Generative adversarial nets,'' in {\em
  Advances in Neural Information Processing Systems}, 2014.

\bibitem{kolouri2017optimal}
S.~Kolouri, S.~R. Park, M.~Thorpe, D.~Slepcev, and G.~K. Rohde, ``Optimal mass
  transport: Signal processing and machine-learning applications,'' {\em IEEE
  Signal Processing Magazine}, vol.~34, no.~4, pp.~43--59, 2017.

\bibitem{Villani2003}
C.~Villani, {\em Topics in Optimal Transportation}.
\newblock No.~58, American Mathematical Soc., 2003.

\bibitem{monge1781memoire}
G.~Monge, ``M{\'e}moire sur la th{\'e}orie des d{\'e}blais et des remblais,''
  {\em Histoire de l'Acad{\'e}mie Royale des Sciences de Paris}, 1781.

\bibitem{kantorovich1942translation}
L.~Kantorovich, ``On translation of mass,'' in {\em Dokl. AN SSSR}, vol.~37,
  p.~20, 1942.

\bibitem{wgan}
M.~Arjovsky, S.~Chintala, and L.~Bottou, ``Wasserstein generative adversarial
  networks,'' in {\em International Conference on Machine Learning (ICML)},
  pp.~214--223, 2017.

\bibitem{cover1999elements}
T.~M. Cover, {\em Elements of Information Theory}.
\newblock John Wiley \& Sons, 1999.

\bibitem{rnan}
Y.~Zhang, K.~Li, K.~Li, B.~Zhong, and Y.~Fu, ``Residual non-local attention
  networks for image restoration,'' in {\em International Conference on
  Learning Representations (ICLR)}, 2018.

\bibitem{n2same}
Y.~Xie, Z.~Wang, and S.~Ji, ``Noise2same: Optimizing a self-supervised bound
  for image denoising,'' {\em Advances in Neural Information Processing
  Systems}, vol.~33, 2020.

\bibitem{pi}
Y.~Blau, R.~Mechrez, R.~Timofte, T.~Michaeli, and L.~Zelnik-Manor, ``The 2018
  pirm challenge on perceptual image super-resolution,'' in {\em Proceedings of
  the European Conference on Computer Vision (ECCV) Workshops}, pp.~0--0, 2018.

\bibitem{lpips}
R.~Zhang, P.~Isola, A.~A. Efros, E.~Shechtman, and O.~Wang, ``The unreasonable
  effectiveness of deep features as a perceptual metric,'' in {\em Proceedings
  of the IEEE Conference on Computer Vision and Pattern Recognition (CVPR)},
  pp.~586--595, 2018.

\bibitem{color}
A.~Levin, D.~Lischinski, and Y.~Weiss, ``Colorization using optimization,'' in
  {\em ACM SIGGRAPH 2004 Papers}, pp.~689--694, 2004.

\bibitem{wgangp}
I.~Gulrajani, F.~Ahmed, M.~Arjovsky, V.~Dumoulin, and A.~C. Courville,
  ``Improved training of wasserstein gans,'' in {\em Advances in Neural
  Information Processing Systems}, pp.~5767--5777, 2017.

\bibitem{unpair}
A.~Lugmayr, M.~Danelljan, and R.~Timofte, ``Ntire 2020 challenge on real-world
  image super-resolution: Methods and results,'' in {\em Proceedings of the
  IEEE Conference on Computer Vision and Pattern Recognition (CVPR) Workshops},
  pp.~494--495, 2020.

\bibitem{cab}
Y.~Zhang, K.~Li, K.~Li, L.~Wang, B.~Zhong, and Y.~Fu, ``Image super-resolution
  using very deep residual channel attention networks,'' in {\em European
  Conference on Computer Vision (ECCV)}, pp.~286--301, 2018.

\bibitem{suncg}
S.~Song, F.~Yu, A.~Zeng, A.~X. Chang, M.~Savva, and T.~Funkhouser, ``Semantic
  scene completion from a single depth image,'' in {\em IEEE Conference on
  Computer Vision and Pattern Recognition (CVPR)}, pp.~1746--1754, 2017.

\bibitem{arbelaez2010contour}
P.~Arbelaez, M.~Maire, C.~Fowlkes, and J.~Malik, ``Contour detection and
  hierarchical image segmentation,'' {\em IEEE Transactions on Pattern Analysis
  and Machine Intelligence}, vol.~33, no.~5, pp.~898--916, 2010.

\bibitem{kodak}
R.~Franzen, ``Kodak lossless true color image suite,'' {\em source: http://r0k.
  us/graphics/kodak}, vol.~4, no.~2, 1999.

\bibitem{blau2019rethinking}
Y.~Blau and T.~Michaeli, ``Rethinking lossy compression: The
  rate-distortion-perception tradeoff,'' in {\em International Conference on
  Machine Learning (ICML)}, pp.~675--685, 2019.

\bibitem{yan2021perceptual}
Z.~Yan, F.~Wen, R.~Ying, C.~Ma, and P.~Liu, ``On perceptual lossy compression:
  The cost of perceptual reconstruction and an optimal training framework,'' in
  {\em International Conference on Machine Learning (ICML)}, pp.~11682--11692,
  2021.

\bibitem{middle1}
D.~Scharstein and C.~Pal, ``Learning conditional random fields for stereo,'' in
  {\em IEEE Conference Computer Vision and Pattern Recognition (CVPR)},
  pp.~1--8, 2007.

\bibitem{middle2}
H.~Hirschmuller and D.~Scharstein, ``Evaluation of cost functions for stereo
  matching,'' in {\em IEEE Conference Computer Vision and Pattern Recognition
  (CVPR)}, pp.~1--8, 2007.

\bibitem{vst}
M.~Makitalo and A.~Foi, ``Optimal inversion of the generalized anscombe
  transformation for poisson-gaussian noise,'' {\em IEEE Transactions on Image
  Processing}, vol.~22, no.~1, pp.~91--103, 2012.

\bibitem{micro}
Y.~Zhang, Y.~Zhu, E.~Nichols, Q.~Wang, S.~Zhang, C.~Smith, and S.~Howard, ``A
  poisson-gaussian denoising dataset with real fluorescence microscopy
  images,'' in {\em IEEE Conference Computer Vision and Pattern Recognition
  (CVPR)}, pp.~11710--11718, 2019.

\bibitem{fair}
M.~M{\"u}ller, V.~M{\"o}nkem{\"o}ller, S.~Hennig, W.~H{\"u}bner, and T.~Huser,
  ``Open-source image reconstruction of super-resolution structured
  illumination microscopy data in imagej,'' {\em Nature Communications},
  vol.~7, no.~1, pp.~1--6, 2016.

\bibitem{nyu}
A.~Janoch, S.~Karayev, Y.~Jia, J.~T. Barron, M.~Fritz, K.~Saenko, and
  T.~Darrell, ``A category-level 3d object dataset: Putting the kinect to
  work,'' in {\em Consumer Depth Cameras for Computer Vision}, pp.~141--165,
  Springer, 2013.

\bibitem{sidd}
A.~Abdelhamed, S.~Lin, and M.~S. Brown, ``A high-quality denoising dataset for
  smartphone cameras,'' in {\em IEEE Conference Computer Vision and Pattern
  Recognition (CVPR)}, pp.~1692--1700, 2018.

\bibitem{DBLP:conf/cvpr/ZhangF18}
Y.~Zhang and T.~Funkhouser, ``Deep depth completion of a single rgb-d image,''
  in {\em IEEE Conference Computer Vision and Pattern Recognition (CVPR)},
  pp.~175--185, 2018.

\bibitem{DBLP:conf/eccv/JeonL18}
J.~Jeon and S.~Lee, ``Reconstruction-based pairwise depth dataset for depth
  image enhancement using cnn,'' in {\em European Conference on Computer Vision
  (ECCV)}, pp.~422--438, 2018.

\end{thebibliography}

\clearpage

\onecolumn
\begin{center}
	\huge
	\textbf{Optimal Transport for Unsupervised Denoising Learning\\ Supplemental Material}
\end{center}
 
\hspace*{\fill} \\
\\ \hspace*{\fill} \\

This material presents more restoration examples of our method  compared with state-of-the-art supervised and unsupervised methods on the six experiments, including:

\hangindent 2.8em {(a)}
Synthetic RGB noisy images with Gaussian, Poisson, and Brown Gassian noise.

\hangindent 2.8em {(b)}
Synthetic depth noisy images with Gaussian noise.

\hangindent 2.8em {(c)}
Real-word microscopy images.

\hangindent 2.8em {(d)}
Real-word photographic images.

\hangindent 2.8em {(e)}
Real-word Kinect depth images.

\hangindent 2.8em {(f)}
Real-word raw depth images.

Codes for reproducing all the results of our method is available at https://github.com/wangweiSJTU/OTUR, including the training and test codes for each of the six experiments.

Since the Kinect camera only provides pre-denoised images, in the sixth experiment we consider a raw depth denoising task using another commercial ToF camera, which provides
raw depth data without any pre-denoising. The used camera is shown in Fig. 1.

We also provide a short video to demonstrate the performance of our method on denoising raw depth video, please see the attached \emph{Comparison\_on\_raw\_depth\_images.mp4}.

\begin{figure}[H]
	\centering
	\includegraphics[width=0.6\linewidth]{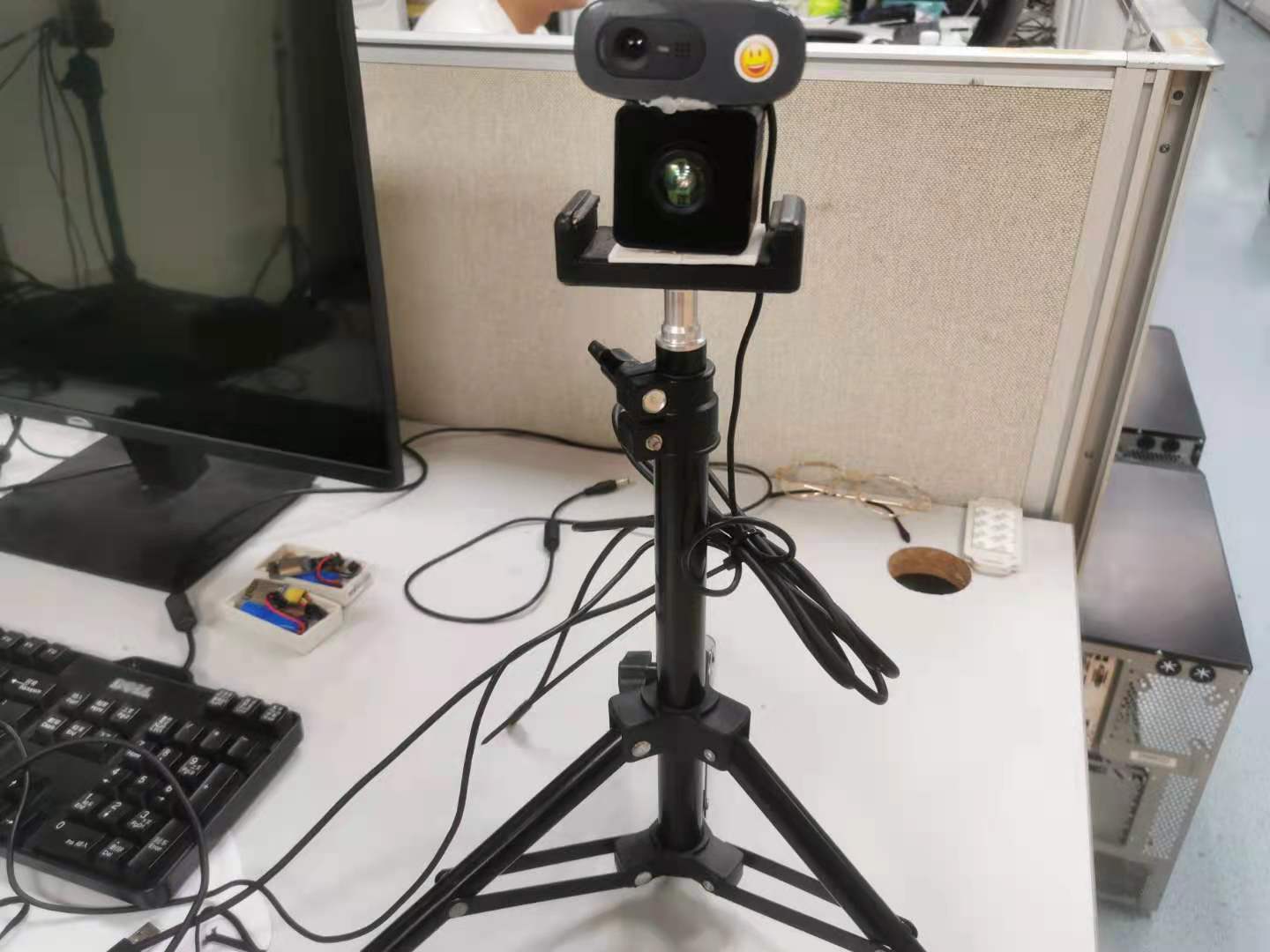}
	\caption{The ToF camera used in the raw depth denoising experiment for capturing raw depth data.}
\end{figure}

\appendices
\section{Synthetic Noisy RGB Images }
\subsection{Gaussian noise with $\sigma=25 $}

\begin{figure}[H]
	\centering
	\subfigure[Noisy (20.57 dB)]{
		\includegraphics[width=0.48\linewidth]{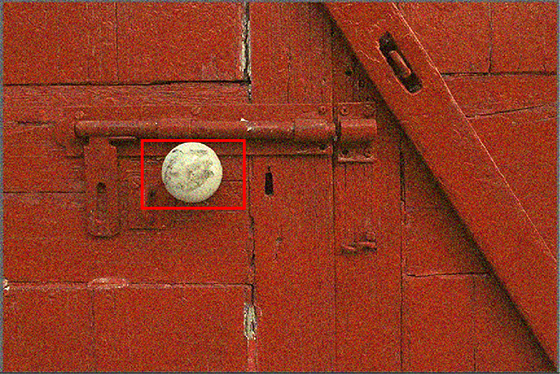}
	}
	\subfigure[Ground truth]{
		\includegraphics[width=0.48\linewidth]{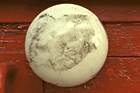}
	}
	\subfigure[N2C (\textbf{33.05}/2.98/0.085)]{
		\includegraphics[width=0.48\linewidth]{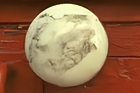}
	}
	\subfigure[N2N (33.04/2.86/0.088)]{
		\includegraphics[width=0.48\linewidth]{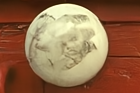}
	}
	\subfigure[BM3D (32.37/3.53/0.119)]{
		\includegraphics[width=0.48\linewidth]{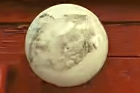}
	}
	\subfigure[Ours (32.12/\textbf{2.31}/\textbf{0.064})]{
		\includegraphics[width=0.48\linewidth]{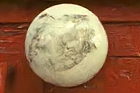}
	}
	\caption{Visual comparison on KODAK-2 with synthetic Gaussian noise. The PSNR/PI/LPIPS results are provided in the brackets. Our method achieves distinctly better perceptual quality than N2C, N2N and
		BM3D, as it preserves more details while N2C, N2N and
		BM3D yield over-smoothed restoration.}
\end{figure}
\clearpage

\begin{figure}[H]
	\centering
	\subfigure[Noisy (20.49 dB)]{
		\includegraphics[width=0.48\linewidth]{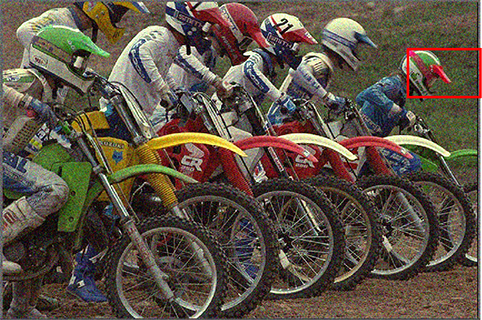}
	}
	\subfigure[Ground truth]{
		\includegraphics[width=0.48\linewidth]{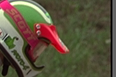}
	}
	\subfigure[N2C (\textbf{30.66}/3.43/0.029)]{
		\includegraphics[width=0.48\linewidth]{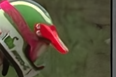}
	}
	\subfigure[N2N (30.63/3.48/0.030)]{
		\includegraphics[width=0.48\linewidth]{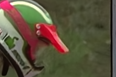}
	}
	\subfigure[BM3D (29.86/\textbf{3.03}/0.038)]{
		\includegraphics[width=0.48\linewidth]{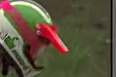}
	}
	\subfigure[Ours (29.27/3.13/\textbf{0.014})]{
		\includegraphics[width=0.48\linewidth]{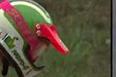}
	}
	\caption{Visual comparison on KODAK-5 with synthetic Gaussian noise. The PSNR/PI/LPIPS results are provided in the brackets. Our method achieves
		distinctly better perceptual quality than N2C, N2N and
		BM3D, as it preserves more details  while N2C, N2N and
		BM3D yield over-smoothed restoration.}
\end{figure}
\clearpage

\begin{figure}[H]
	\centering
	\subfigure[Noisy (20.37 dB)]{
		\includegraphics[width=0.48\linewidth]{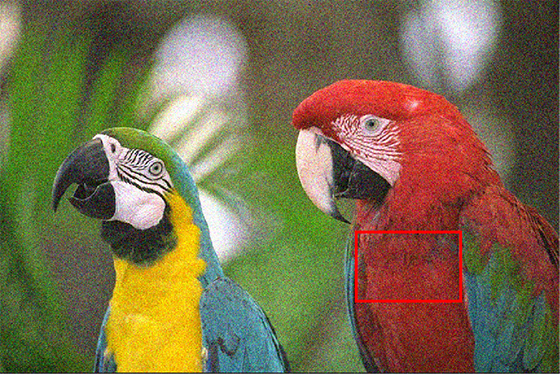}
	}
	\subfigure[Ground truth]{
		\includegraphics[width=0.48\linewidth]{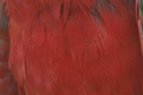}
	}
	\subfigure[N2C (\textbf{35.46}/3.60/0.045)]{
		\includegraphics[width=0.48\linewidth]{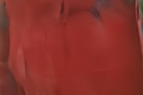}
	}
	\subfigure[N2N (35.43/3.69/0.046)]{
		\includegraphics[width=0.48\linewidth]{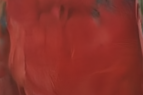}
	}
	\subfigure[BM3D (34.80/3.61/0.048)]{
		\includegraphics[width=0.48\linewidth]{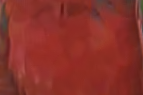}
	}
	\subfigure[Ours (34.05/\textbf{2.93}/\textbf{0.019})]{
		\includegraphics[width=0.48\linewidth]{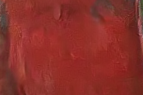}
	}
	\caption{Visual comparison on KODAK-23 with synthetic Gaussian noise. The PSNR/PI/LPIPS results are provided in the brackets. Our method achieves
		distinctly better perceptual quality than N2C, N2N and
		BM3D, as it preserves more details while N2C, N2N and
		BM3D yield over-smoothed restoration.}
\end{figure}
\clearpage

\subsection{Poisson noise with ${\lambda _p}= 30 $ }

\begin{figure}[H]
	\centering
	\subfigure[Noisy (18.65 dB)]{
		\includegraphics[width=0.48\linewidth]{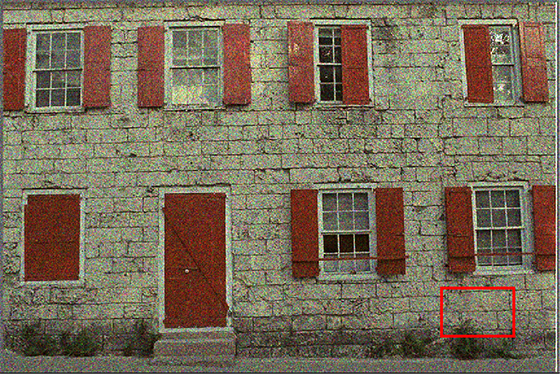}
	}
	\subfigure[Ground truth]{
		\includegraphics[width=0.48\linewidth]{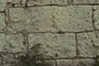}
	}
	\subfigure[N2C (\textbf{28.87}/2.39/0.044)]{
		\includegraphics[width=0.48\linewidth]{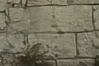}
	}
	\subfigure[N2N (28.86/2.39/0.051)]{
		\includegraphics[width=0.48\linewidth]{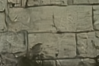}
	}
	\subfigure[BM3D (27.91/2.36/0.067)]{
		\includegraphics[width=0.48\linewidth]{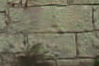}
	}
	\subfigure[Ours (28.05/\textbf{2.02}/\textbf{0.035})]{
		\includegraphics[width=0.48\linewidth]{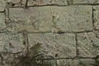}
	}
	\caption{Visual comparison on KODAK-1 with synthetic Poisson noise. The PSNR/PI/LPIPS results are provided in the brackets. Our method achieves
		distinctly better perceptual quality than N2C, N2N and
		BM3D, as it preserves more details while N2C, N2N and
		BM3D yield over-smoothed restoration. Meanwhile, the PSNR difference from the supervised methods N2C and N2N is less than 1 dB.}
\end{figure}
\clearpage

\begin{figure}[H]
	\centering
	\subfigure[Noisy (18.42 dB)]{
		\includegraphics[width=0.48\linewidth]{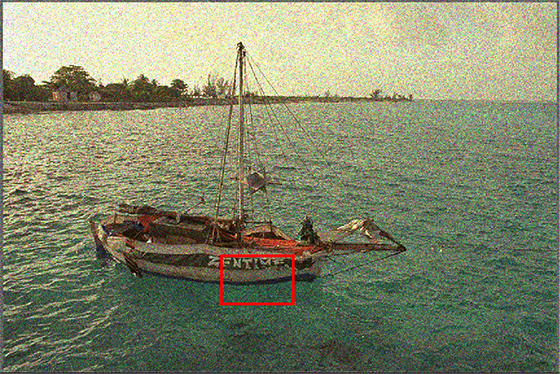}
	}
	\subfigure[Ground truth]{
		\includegraphics[width=0.48\linewidth]{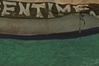}
	}
	\subfigure[N2C (\textbf{30.12}/2.32/0.032)]{
		\includegraphics[width=0.48\linewidth]{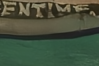}
	}
	\subfigure[N2N (30.11/2.34/0.037)]{
		\includegraphics[width=0.48\linewidth]{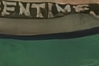}
	}
	\subfigure[BM3D (27.97/2.28/0.058)]{
		\includegraphics[width=0.48\linewidth]{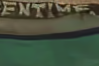}
	}
	\subfigure[Ours (29.29/\textbf{2.01}/\textbf{0.031})]{
		\includegraphics[width=0.48\linewidth]{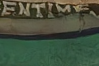}
	}
	\caption{Visual comparison on KODAK-6 with synthetic Poisson noise. The PSNR/PI/LPIPS results are provided in the brackets. Our method achieves
		distinctly better perceptual quality than N2C, N2N and
		BM3D, as it preserves more details while N2C, N2N and
		BM3D yield over-smoothed restoration. Meanwhile, the PSNR difference from the supervised methods N2C and N2N is less than 1 dB.}
\end{figure}
\clearpage

\begin{figure}[H]
	\centering
	\subfigure[Noisy (19.49 dB)]{
		\includegraphics[width=0.48\linewidth]{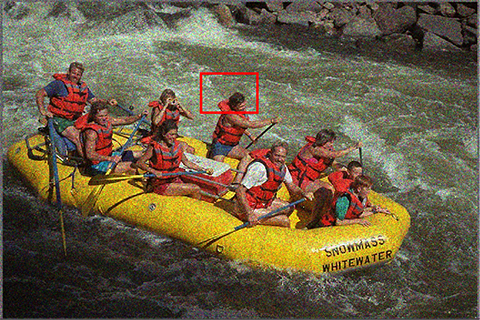}
	}
	\subfigure[Ground truth]{
		\includegraphics[width=0.48\linewidth]{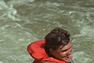}
	}
	\subfigure[N2C (\textbf{30.16}/1.78/0.033)]{
		\includegraphics[width=0.48\linewidth]{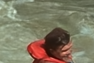}
	}
	\subfigure[N2N (30.15/1.81/0.037)]{
		\includegraphics[width=0.48\linewidth]{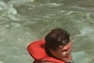}
	}
	\subfigure[BM3D (28.95/1.67/0.067)]{
		\includegraphics[width=0.48\linewidth]{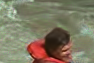}
	}
	\subfigure[Ours (29.43/\textbf{1.47}/\textbf{0.032})]{
		\includegraphics[width=0.48\linewidth]{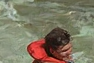}
	}
	\caption{Visual comparison on KODAK-14 with synthetic Poisson noise. The PSNR/PI/LPIPS results are provided in the brackets. Our method achieves
		distinctly better perceptual quality than N2C, N2N and
		BM3D, as it preserves more details while N2C, N2N and
		BM3D yield over-smoothed restoration. Meanwhile, the PSNR difference from the supervised methods N2C and N2N is less than 1 dB.}
\end{figure}
\clearpage

\subsection{Brown Gaussian noise with $5\times5$ Gaussian filter applied to Gaussian noise ($\sigma  = 50$) }

\begin{figure}[H]
	\centering
	\subfigure[Noisy (25.55 dB)]{
		\includegraphics[width=0.48\linewidth]{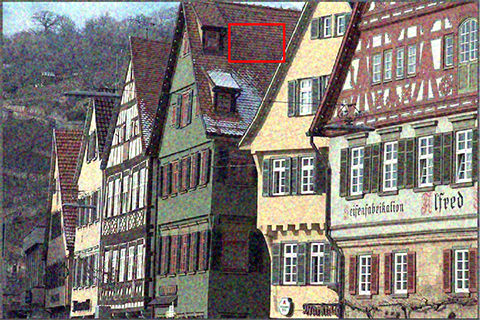}
	}
	\subfigure[Ground truth]{
		\includegraphics[width=0.48\linewidth]{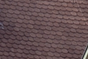}
	}
	\subfigure[N2C (30.55/3.42/0.026)]{
		\includegraphics[width=0.48\linewidth]{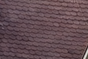}
	}
	\subfigure[N2N (27.52/\textbf{2.94}/0.041)]{
		\includegraphics[width=0.48\linewidth]{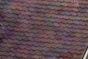}
	}
	\subfigure[BM3D (28.37/3.50/0.068)]{
		\includegraphics[width=0.48\linewidth]{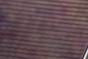}
	}
	\subfigure[Ours (\textbf{30.66}/3.19/\textbf{0.020})]{
		\includegraphics[width=0.48\linewidth]{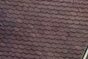}
	}
	\caption{Visual comparison on KODAK-8 with synthetic Brown Gaussian noise. The PSNR/PI/LPIPS results are provided in the brackets. Compared with N2C, N2N and
		BM3D, our method not only achieves better perceptual quality, but also achieves higher PSNR. The advantage over N2N and BM3D is especially conspicuous.}
\end{figure}
\clearpage

\begin{figure}[H]
	\centering
	\subfigure[Noisy (25.44 dB)]{
		\includegraphics[width=0.48\linewidth]{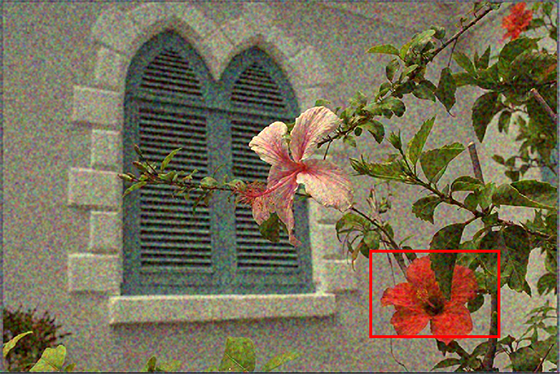}
	}
	\subfigure[Ground truth]{
		\includegraphics[width=0.48\linewidth]{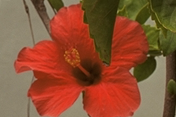}
	}1
	\subfigure[N2C (32.24/2.48/0.035)]{
		\includegraphics[width=0.48\linewidth]{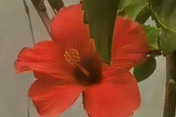}
	}
	\subfigure[N2N (29.11/2.52/0.052)]{
		\includegraphics[width=0.48\linewidth]{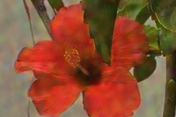}
	}
	\subfigure[BM3D (30.38/3.20/0.070)]{
		\includegraphics[width=0.48\linewidth]{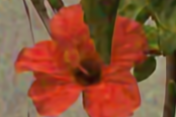}
	}
	\subfigure[Ours (\textbf{32.30}/\textbf{2.34}/\textbf{0.028})]{
		\includegraphics[width=0.48\linewidth]{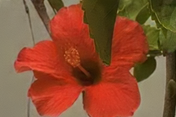}
	}
	\caption{Visual comparison on KODAK-7 with synthetic Brown Gaussian noise. The PSNR/PI/LPIPS results are provided in the brackets. Compared with N2C, N2N and
		BM3D, our method not only achieves better perceptual quality, but also achieves higher PSNR. The advantage over N2N and BM3D is especially conspicuous.}
\end{figure}
\clearpage

\begin{figure}[H]
	\centering
	\subfigure[Noisy (25.55 dB)]{
		\includegraphics[width=0.48\linewidth]{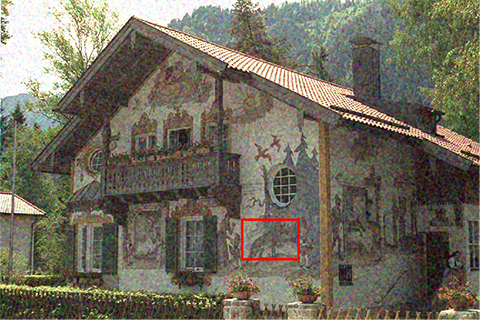}
	}
	\subfigure[Ground truth]{
		\includegraphics[width=0.48\linewidth]{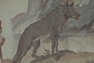}
	}
	\subfigure[N2C (30.25/1.82/0.036)]{
		\includegraphics[width=0.48\linewidth]{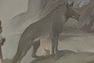}
	}
	\subfigure[N2N (27.83/1.84/0.044)]{
		\includegraphics[width=0.48\linewidth]{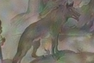}
	}
	\subfigure[BM3D (28.24/2.44/0.085)]{
		\includegraphics[width=0.48\linewidth]{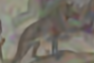}
	}
	\subfigure[Ours (\textbf{31.04}/\textbf{1.77}/\textbf{0.026})]{
		\includegraphics[width=0.48\linewidth]{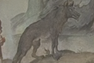}
	}
	\caption{Visual comparison on KODAK-24 with synthetic Brown Gaussian noise. The PSNR/PI/LPIPS results are provided in the brackets. Compared with N2C, N2N and BM3D, our method not only achieves better perceptual quality, but also achieves higher PSNR. The advantage over N2N and BM3D is especially conspicuous.}
\end{figure}

\section{Synthetic Noisy Depth Images }

\begin{figure}[H]
	\centering
	\subfigure[Noisy (20.34 dB)]{
		\includegraphics[width=0.42\linewidth]{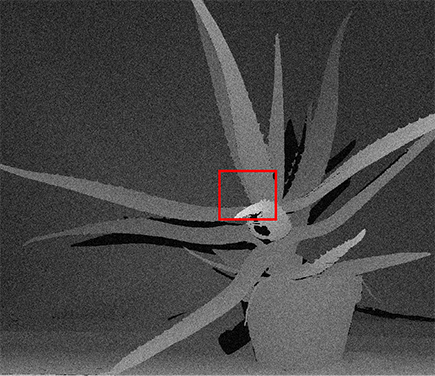}
	}
	\subfigure[Ground truth]{
		\includegraphics[width=0.42\linewidth]{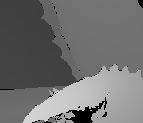}
	}
	\subfigure[N2C (\textbf{36.54 dB})]{
		\includegraphics[width=0.42\linewidth]{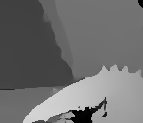}
	}
	\subfigure[N2N (36.43 dB)]{
		\includegraphics[width=0.42\linewidth]{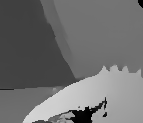}
	}
	\subfigure[BM3D (36.23 dB)]{
		\includegraphics[width=0.42\linewidth]{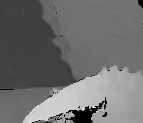}
	}
	\subfigure[Ours (36.27 dB)]{
		\includegraphics[width=0.42\linewidth]{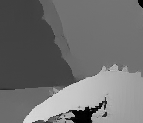}
	}
	\caption{Visual comparison on synthetic noisy depth image with Gaussian noise ($\sigma=25$) (Example 1: Aloe). Our method achieves distinctly better perceptual quality than N2C, N2N and
		BM3D, as it preserves more details while N2C, N2N and
		BM3D yield over-smoothed restoration.}
\end{figure}
\clearpage

\begin{figure}[H]
	\centering
	\subfigure[Noisy (20.32 dB)]{
		\includegraphics[width=0.42\linewidth]{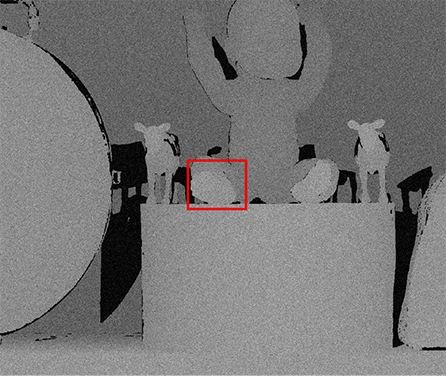}
	}
	\subfigure[Ground truth]{
		\includegraphics[width=0.42\linewidth]{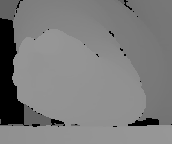}
	}
	\subfigure[N2C (\textbf{36.98 dB})]{
		\includegraphics[width=0.42\linewidth]{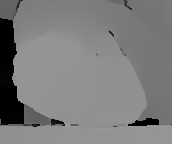}
	}
	\subfigure[N2N (36.78 dB)]{
		\includegraphics[width=0.42\linewidth]{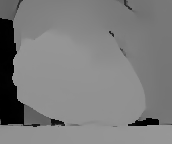}
	}
	\subfigure[BM3D (36.59 dB)]{
		\includegraphics[width=0.42\linewidth]{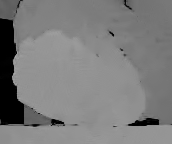}
	}
	\subfigure[Ours (36.01 dB)]{
		\includegraphics[width=0.42\linewidth]{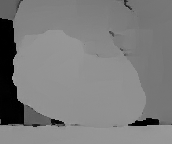}
	}
	\caption{Visual comparison on synthetic noisy depth image with  Gaussian noise ($\sigma=25$) (Example 2: Baby). Our method achieves distinctly better perceptual quality than N2C, N2N and
		BM3D, as it preserves more details while N2C, N2N and
		BM3D yield over-smoothed restoration.}
\end{figure}
\clearpage

\begin{figure}[H]
	\centering
	\subfigure[Noisy (20.21 dB)]{
		\includegraphics[width=0.42\linewidth]{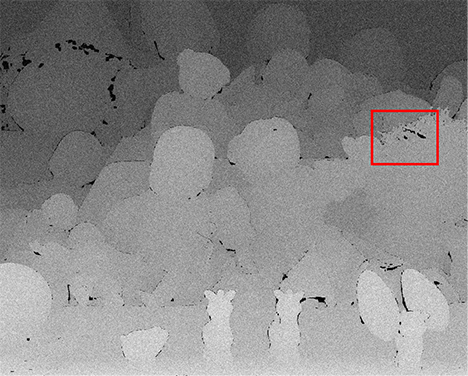}
	}
	\subfigure[Ground truth]{
		\includegraphics[width=0.42\linewidth]{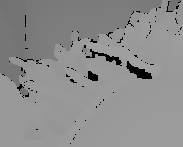}
	}
	\subfigure[N2C (\textbf{32.85 dB})]{
		\includegraphics[width=0.42\linewidth]{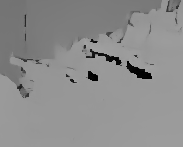}
	}
	\subfigure[N2N (32.75 dB)]{
		\includegraphics[width=0.42\linewidth]{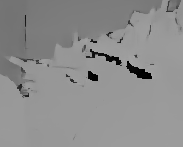}
	}
	\subfigure[BM3D (32.40 dB)]{
		\includegraphics[width=0.42\linewidth]{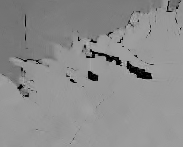}
	}
	\subfigure[Ours (31.98 dB)]{
		\includegraphics[width=0.42\linewidth]{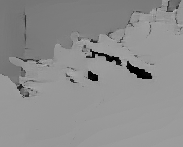}
	}
	\caption{Visual comparison on synthetic noisy depth image with Gaussian noise ($\sigma=25$) (Example 3: Dolls). Our method achieves distinctly better perceptual quality than N2C, N2N and BM3D, as it preserves more details while N2C, N2N and
		BM3D yield over-smoothed restoration.}
\end{figure}

\section{Real-world Microscopy Images}

\begin{figure}[H]
	\centering
	\subfigure[Noisy (26.55 dB)]{
		\includegraphics[width=0.38\linewidth]{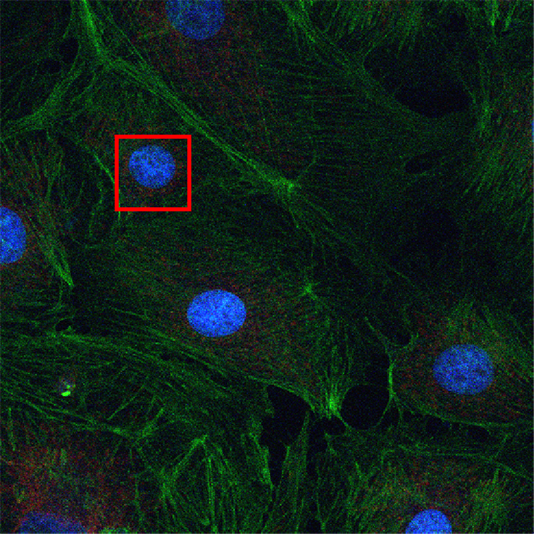}
	}
	\subfigure[Ground truth (Average)]{
		\includegraphics[width=0.38\linewidth]{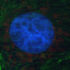}
	}
	\subfigure[N2C (34.50/5.40/0.072)]{
		\includegraphics[width=0.38\linewidth]{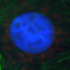}
	}
	\subfigure[N2N (34.36/5.27/0.070)]{
		\includegraphics[width=0.38\linewidth]{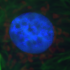}
	}
	\subfigure[VST+BM3D (33.87/7.37/0.099)]{
		\includegraphics[width=0.38\linewidth]{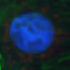}
	}
	\subfigure[Ours (\textbf{34.69}/\textbf{5.16}/\textbf{0.048})]{
		\includegraphics[width=0.38\linewidth]{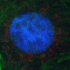}
	}
	\caption{Visual comparison on real-world microscopy images (Example 1). The PSNR/PI/LPIPS results are provided in the brackets. Our method
		not only achieves the highest PSNR, but also yields a restoration much clear than that of VST+BM3D, N2C and N2N.}
\end{figure}
\clearpage

\begin{figure}[H]
	\centering
	\subfigure[Noisy (26.27 dB)]{
		\includegraphics[width=0.38\linewidth]{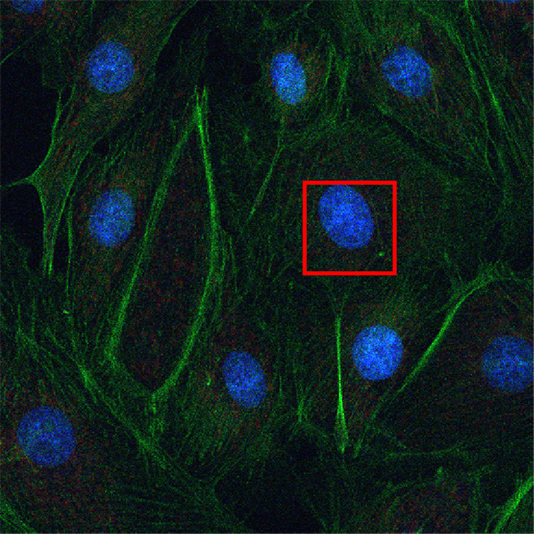}
	}
	\subfigure[Ground truth (Average)]{
		\includegraphics[width=0.38\linewidth]{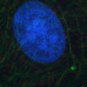}
	}
	\subfigure[N2C (\textbf{36.10}/5.41/0.074)]{
		\includegraphics[width=0.38\linewidth]{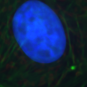}
	}
	\subfigure[N2N (35.64/5.18/0.067)]{
		\includegraphics[width=0.38\linewidth]{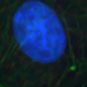}
	}
	\subfigure[VST+BM3D (35.27/8.18/0.502)]{
		\includegraphics[width=0.38\linewidth]{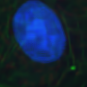}
	}
	\subfigure[Ours (35.58/\textbf{5.11}/\textbf{0.041})]{
		\includegraphics[width=0.38\linewidth]{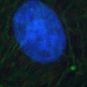}
	}
	\caption{Visual comparison on real-world microscopy images (Example 2). The PSNR/PI/LPIPS results are provided in the brackets. Our method yields a restoration much clear than that of VST+BM3D, N2C and N2N.}
\end{figure}
\clearpage

\begin{figure}[H]
	\centering
	\subfigure[Noisy (26.31 dB)]{
		\includegraphics[width=0.38\linewidth]{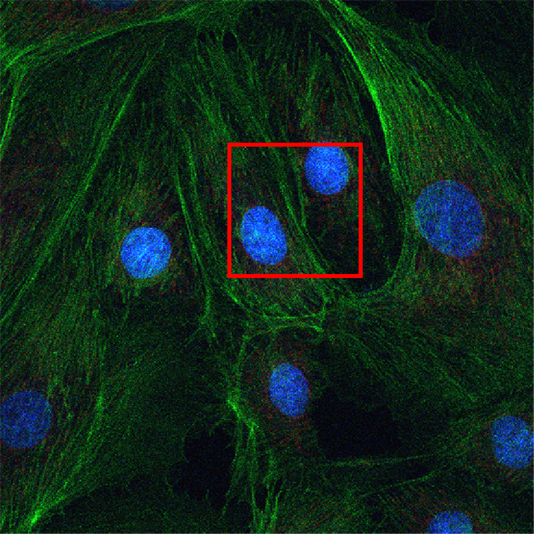}
	}
	\subfigure[Ground truth (Average)]{
		\includegraphics[width=0.38\linewidth]{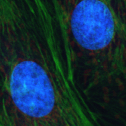}
	}
	\subfigure[N2C (34.58/4.87/0.065)]{
		\includegraphics[width=0.38\linewidth]{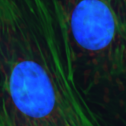}
	}
	\subfigure[N2N (34.28/4.97/0.063)]{
		\includegraphics[width=0.38\linewidth]{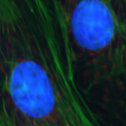}
	}
	\subfigure[VST+BM3D (33.75/7.57/0.471)]{
		\includegraphics[width=0.38\linewidth]{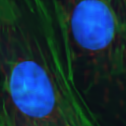}
	}
	\subfigure[Ours (\textbf{34.67}/\textbf{4.40}/\textbf{0.041})]{
		\includegraphics[width=0.38\linewidth]{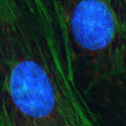}
	}
	\caption{Visual comparison on real-world microscopy images (Example 3). The PSNR/PI/LPIPS results are provided in the brackets. Our method
		not only achieves the highest PSNR, but also yields a restoration much clear than that of VST+BM3D, N2C and N2N.}
\end{figure}

\section{Real-World Photographic Images }

\begin{figure}[H]
	\centering
	\subfigure[Noisy (30.10 dB)]{
		\includegraphics[width=0.38\linewidth]{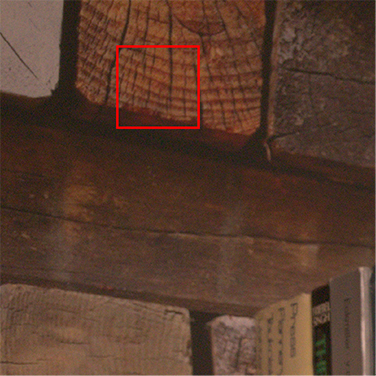}
	}
	\subfigure[Ground truth (Average)]{
		\includegraphics[width=0.38\linewidth]{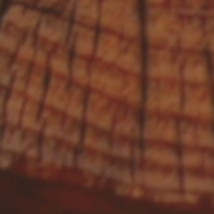}
	}
	\subfigure[N2C (\textbf{41.04}/7.90/0.013)]{
		\includegraphics[width=0.38\linewidth]{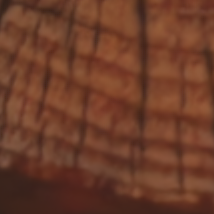}
	}
	\subfigure[N2N (40.91/8.03/0.014)]{
		\includegraphics[width=0.38\linewidth]{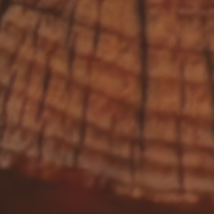}
	}
	\subfigure[BM3D (38.87/8.04/0.028)]{
		\includegraphics[width=0.38\linewidth]{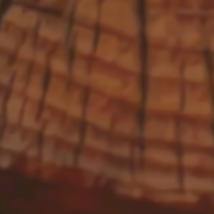}
	}
	\subfigure[Ours (40.48/\textbf{7.48}/\textbf{0.009})]{
		\includegraphics[width=0.38\linewidth]{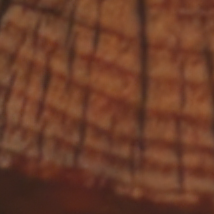}
	}
	\caption{Visual comparison on real-world photographic images (Example 1). The PSNR/PI/LPIPS results are provided in the brackets.}
\end{figure}
\clearpage

\begin{figure}[H]
	\centering
	\subfigure[Noisy (30.87 dB)]{
		\includegraphics[width=0.38\linewidth]{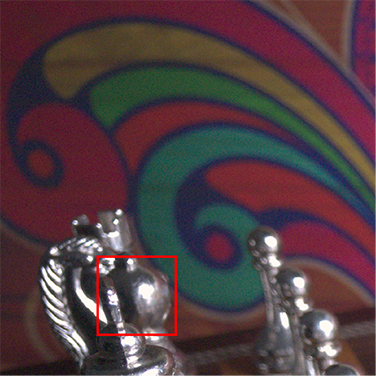}
	}
	\subfigure[Ground truth (Average)]{
		\includegraphics[width=0.38\linewidth]{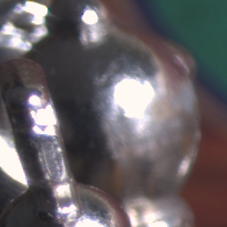}
	}
	\subfigure[N2C (\textbf{39.90}/7.79/0.031)]{
		\includegraphics[width=0.38\linewidth]{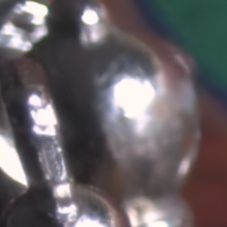}
	}
	\subfigure[N2N (39.79/7.82/0.030)]{
		\includegraphics[width=0.38\linewidth]{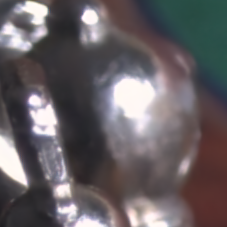}
	}
	\subfigure[BM3D (38.89/7.55/0.063)]{
		\includegraphics[width=0.38\linewidth]{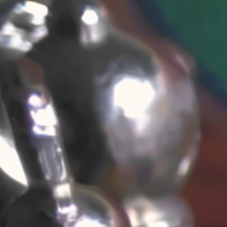}
	}
	\subfigure[Ours (38.85/\textbf{7.44}/\textbf{0.022})]{
		\includegraphics[width=0.38\linewidth]{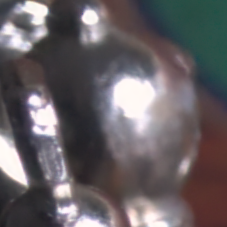}
	}
	\caption{Visual comparison on real-world photographic images (Example 2). The PSNR/PI/LPIPS results are provided in the brackets.}
\end{figure}
\clearpage

\begin{figure}[H]
	\centering
	\subfigure[Noisy (30.98 dB)]{
		\includegraphics[width=0.38\linewidth]{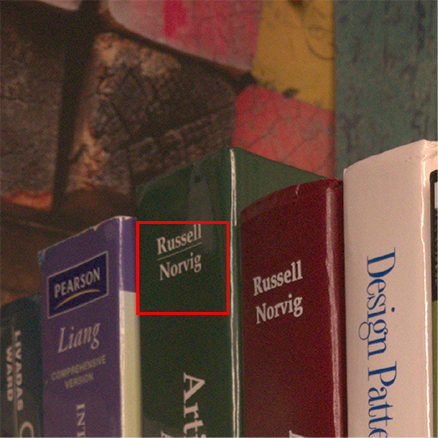}
	}
	\subfigure[Ground truth (Average)]{
		\includegraphics[width=0.38\linewidth]{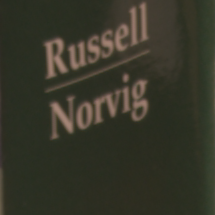}
	}
	\subfigure[N2C (\textbf{39.52}/7.50/0.027)]{
		\includegraphics[width=0.38\linewidth]{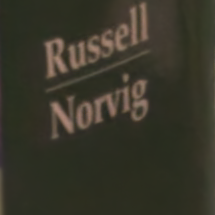}
	}
	\subfigure[N2N (38.98/7.51/0.028)]{
		\includegraphics[width=0.38\linewidth]{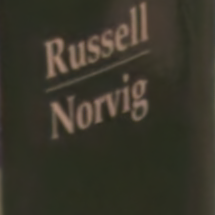}
	}
	\subfigure[BM3D (38.89/7.22/0.045)]{
		\includegraphics[width=0.38\linewidth]{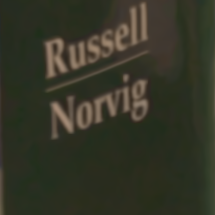}
	}
	\subfigure[Ours (39.49/\textbf{6.98}/\textbf{0.020})]{
		\includegraphics[width=0.38\linewidth]{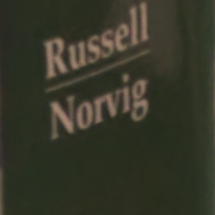}
	}
	\caption{Visual comparison on real-world photographic images (Example 3). The PSNR/PI/LPIPS results are provided in the brackets.}
\end{figure}

\section{Real-World Depth Images }

\begin{figure}[H]
	\centering
	\subfigure[RGB of the scene]{
		\includegraphics[width=0.48\linewidth]{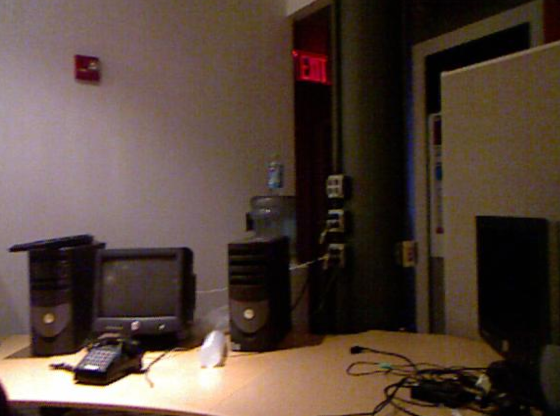}
	}
	\subfigure[Depth of Kinect]{
		\includegraphics[width=0.48\linewidth]{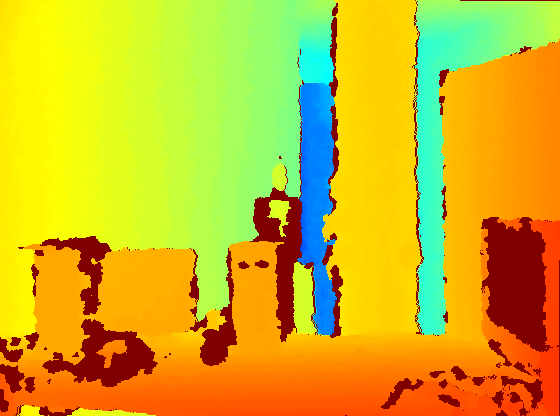}
	}
	\subfigure[Colorization]{
		\includegraphics[width=0.48\linewidth]{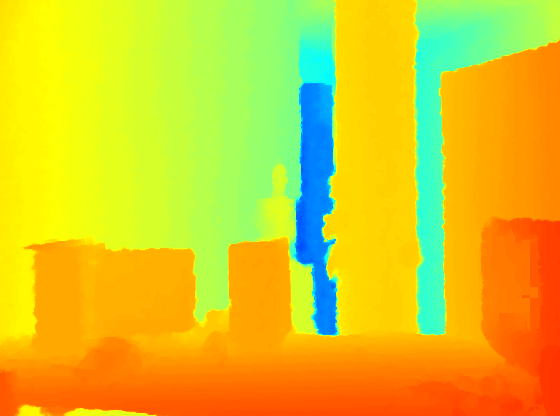}
	}
	\subfigure[Cross-bilateral filter]{
		\includegraphics[width=0.48\linewidth]{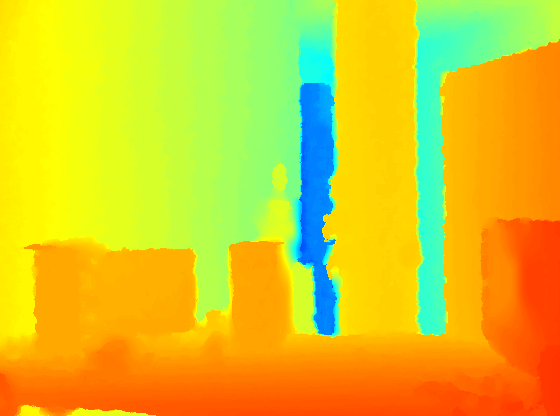}
	}
	\subfigure[Nearest-neighbor]{
		\includegraphics[width=0.48\linewidth]{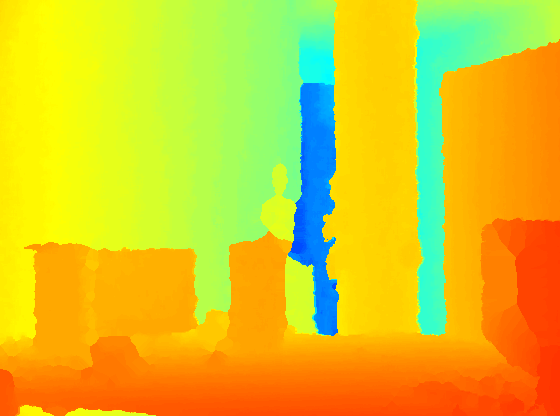}
	}
	\subfigure[Ours]{
		\includegraphics[width=0.48\linewidth]{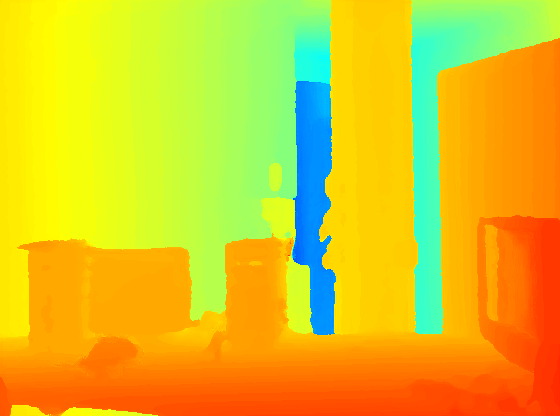}
	}
	\caption{Visual comparison on real-world depth images (Example 1). Our method significantly outperforms the compared methods, as its restoration has straighter and sharper edges, which is more consistent with the geometry of the scene (please see the RGB of the scene). The colorization and cross-bilateral filter methods are RGB-D fusion methods.}
\end{figure}
\clearpage

\begin{figure}[H]
	\centering
	\subfigure[RGB of the scene]{
		\includegraphics[width=0.48\linewidth]{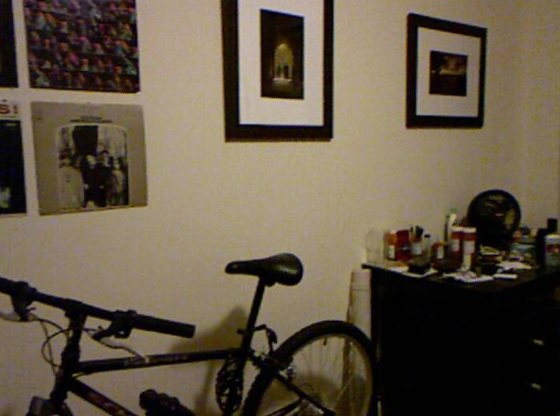}
	}
	\subfigure[Depth of Kinect]{
		\includegraphics[width=0.48\linewidth]{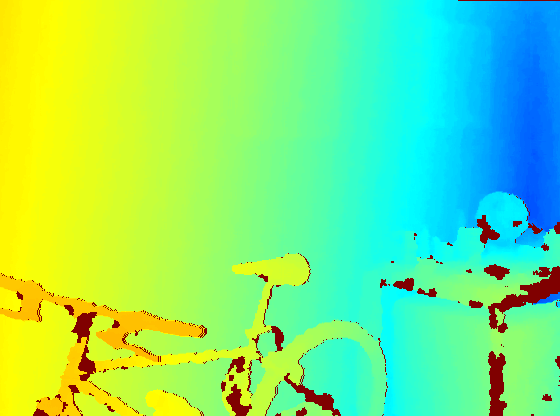}
	}
	\subfigure[Colorization]{
		\includegraphics[width=0.48\linewidth]{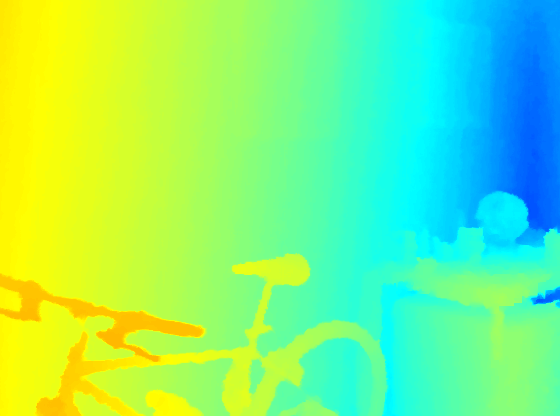}
	}
	\subfigure[Cross-bilateral filter]{
		\includegraphics[width=0.48\linewidth]{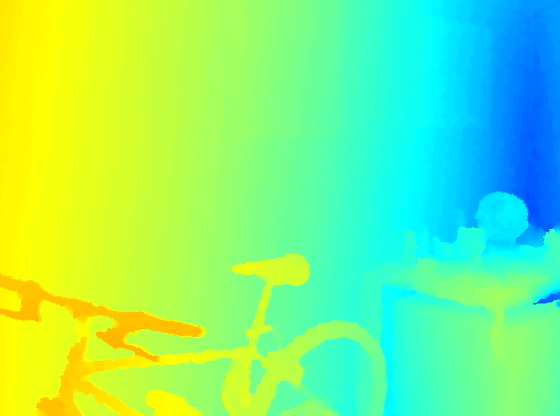}
	}
	\subfigure[Nearest-neighbor]{
		\includegraphics[width=0.48\linewidth]{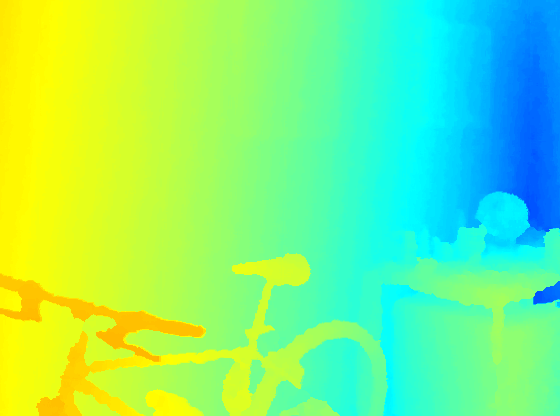}
	}
	\subfigure[Ours]{
		\includegraphics[width=0.48\linewidth]{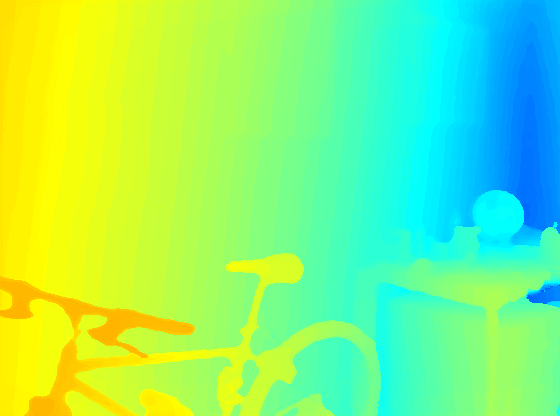}
	}
	\caption{Visual comparison on real-world depth images (Example 2). Our method significantly outperforms the compared methods, as its restoration has straighter and sharper edges, which is more consistent with the geometry of the scene (please see the RGB of the scene). The colorization and cross-bilateral filter methods are RGB-D fusion methods.}
\end{figure}
\clearpage

\begin{figure}[H]
	\centering
	\subfigure[RGB of the scene]{
		\includegraphics[width=0.48\linewidth]{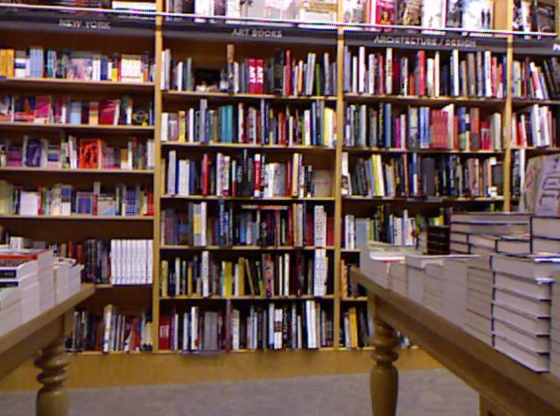}
	}
	\subfigure[Depth of Kinect]{
		\includegraphics[width=0.48\linewidth]{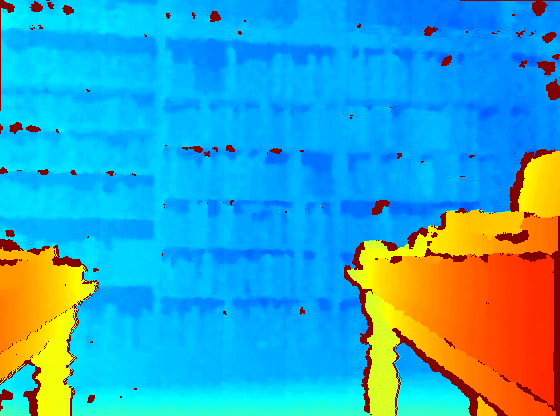}
	}
	\subfigure[Colorization]{
		\includegraphics[width=0.48\linewidth]{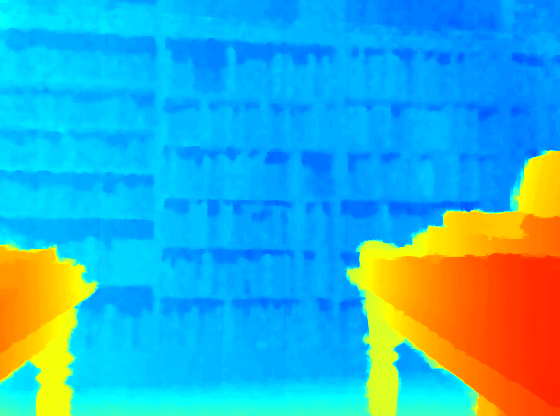}
	}
	\subfigure[Cross-bilateral filter]{
		\includegraphics[width=0.48\linewidth]{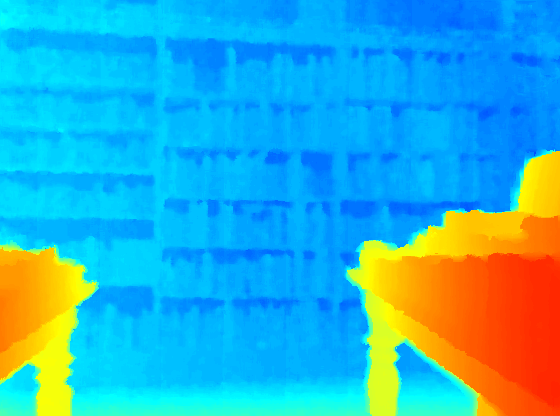}
	}
	\subfigure[Nearest-neighbor]{
		\includegraphics[width=0.48\linewidth]{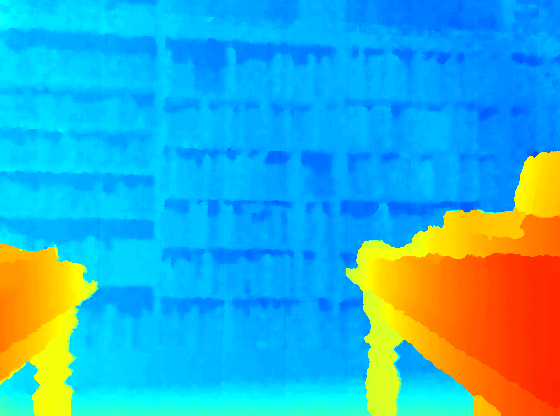}
	}
	\subfigure[Ours]{
		\includegraphics[width=0.48\linewidth]{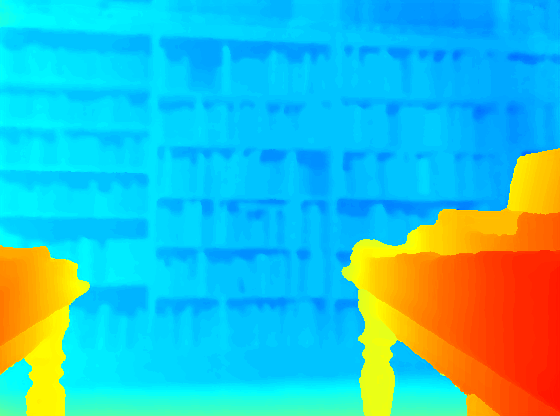}
	}
	\caption{Visual comparison on real-world depth images (Example 3). Our method significantly outperforms the compared methods, as its restoration has straighter and sharper edges, which is more consistent with the geometry of the scene (please see the RGB of the scene). The colorization and cross-bilateral filter methods are RGB-D fusion methods.}
\end{figure}

\section{Real-World Raw Depth Images }

\begin{figure}[H]
	\centering
	\subfigure[Raw depth]{
		\includegraphics[width=0.48\linewidth]{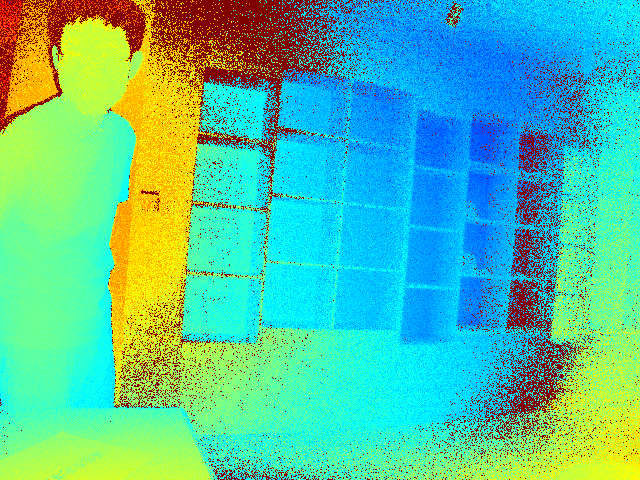}
	}
	\subfigure[Nearest-neighbor (NN)]{
		\includegraphics[width=0.48\linewidth]{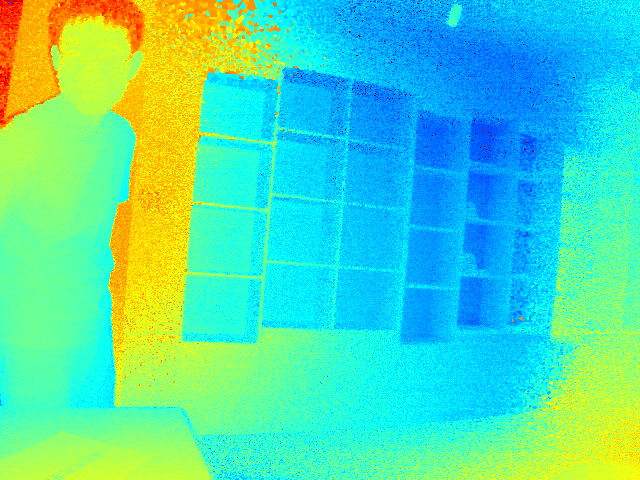}
	}
	\subfigure[NN+BM3D]{
		\includegraphics[width=0.48\linewidth]{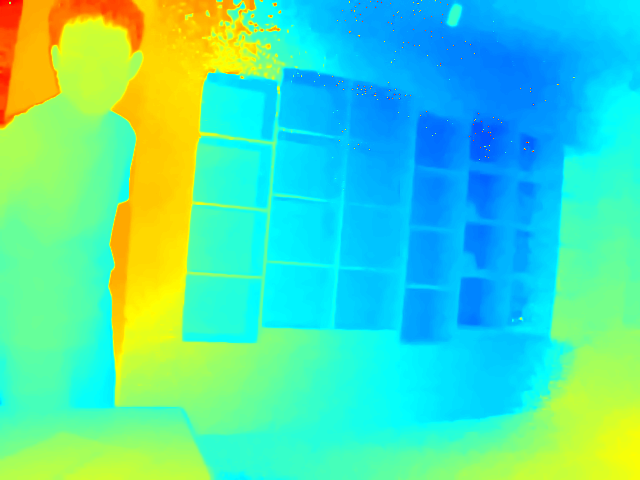}
	}
	\subfigure[Ours]{
		\includegraphics[width=0.48\linewidth]{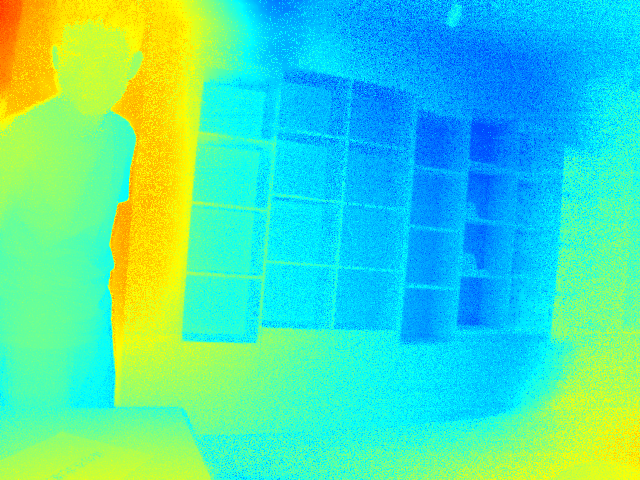}
	}
	\subfigure[Ours+BM3D]{
		\includegraphics[width=0.48\linewidth]{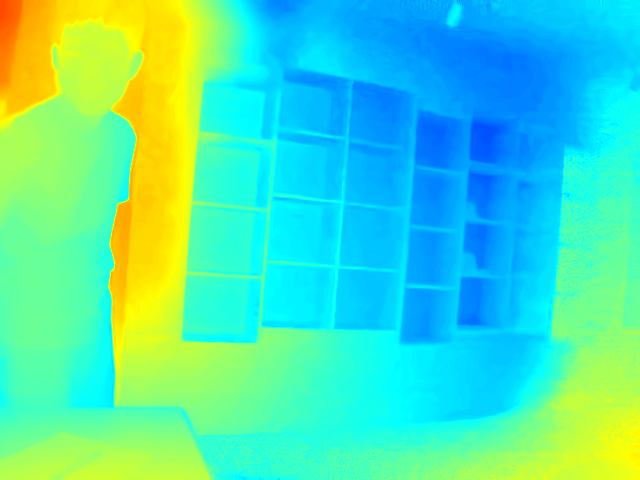}
	}
	\subfigure[Ours+OT denoising]{
		\includegraphics[width=0.48\linewidth]{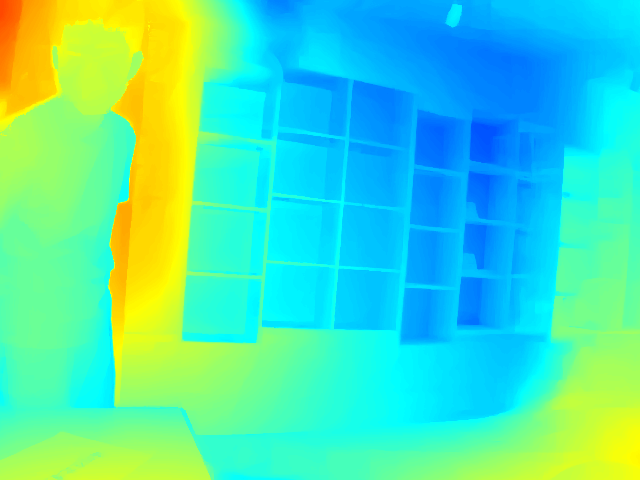}
	}
	\caption{Visual comparison on real-world raw depth images (Example 1). Our method achieves remarkable performance and significantly outperforms the NN and NN+BM3D methods, as it can fill in the holes more naturally while preserving more edge details of the images, especially when it is paired with the OT denoiser (our denoising model trained in Section 5.3 for denoising Gaussian noise).}
\end{figure}
\clearpage

\begin{figure}[H]
	\centering
	\subfigure[Raw depth]{
		\includegraphics[width=0.48\linewidth]{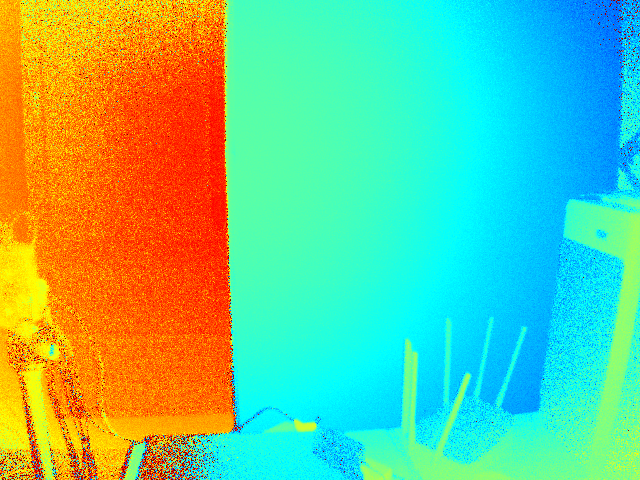}
	}
	\subfigure[Nearest-neighbor (NN)]{
		\includegraphics[width=0.48\linewidth]{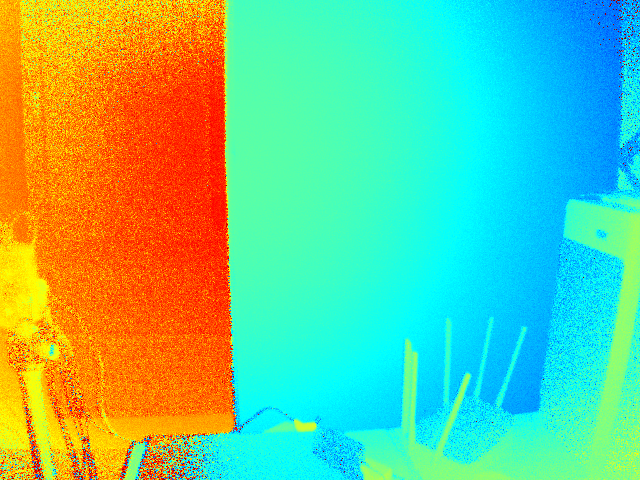}
	}
	\subfigure[NN+BM3D]{
		\includegraphics[width=0.48\linewidth]{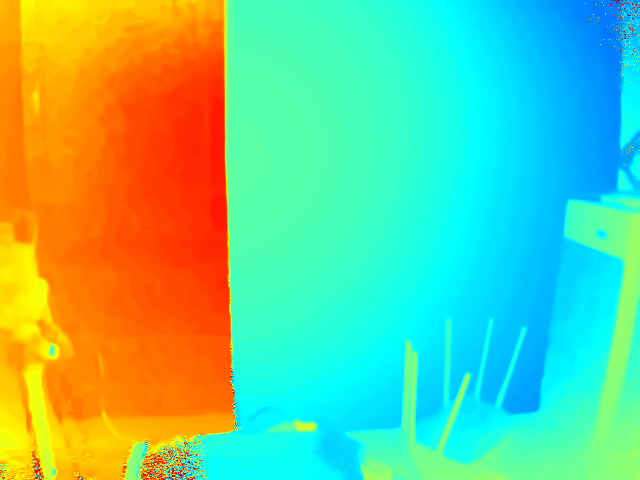}
	}
	\subfigure[Ours]{
		\includegraphics[width=0.48\linewidth]{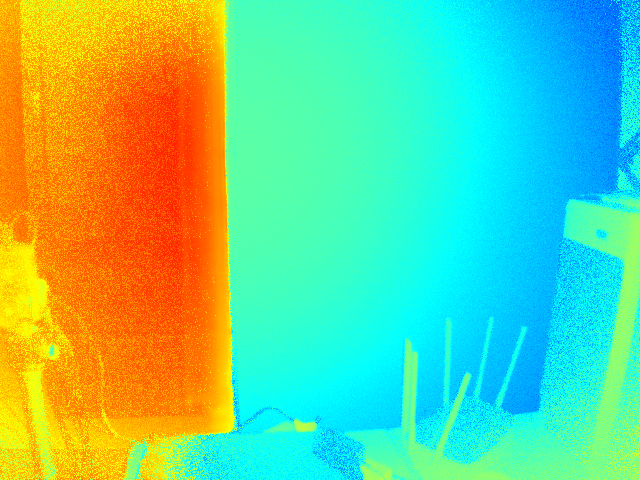}
	}
	\subfigure[Ours+BM3D]{
		\includegraphics[width=0.48\linewidth]{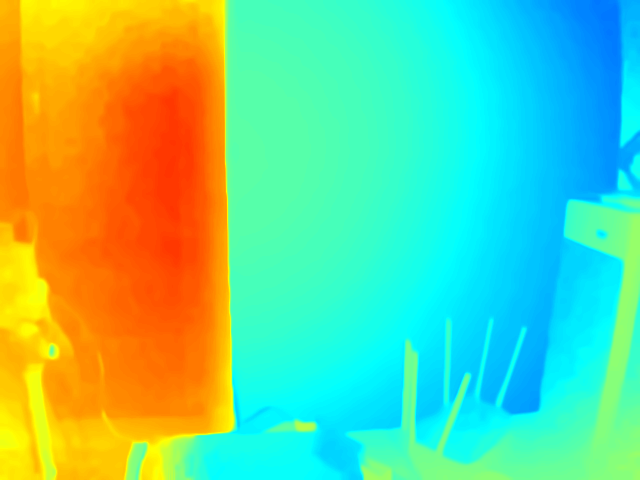}
	}
	\subfigure[Ours+OT denoising]{
		\includegraphics[width=0.48\linewidth]{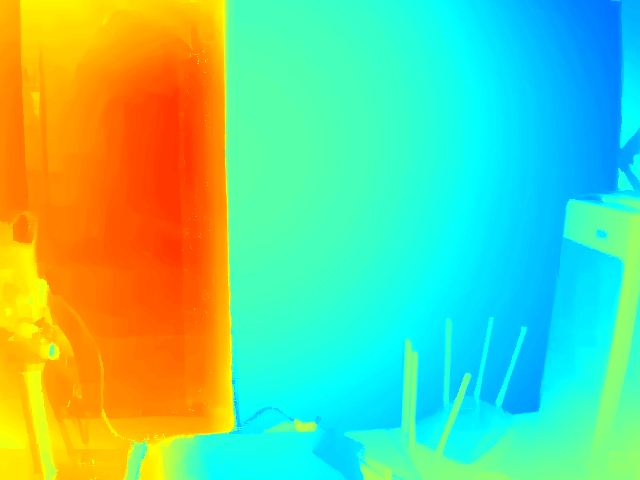}
	}
	\caption{Visual comparison on real-world raw depth images (Example 2). Our method achieves remarkable performance and significantly outperforms the NN and NN+BM3D methods, as it can fill in the holes more naturally while preserving more edge details of the images, especially when it is paired with the OT denoiser (our denoising model trained in Section 5.3 for denoising Gaussian noise).}
\end{figure}
\clearpage

\begin{figure}[H]
	\centering
	\subfigure[Raw depth]{
		\includegraphics[width=0.48\linewidth]{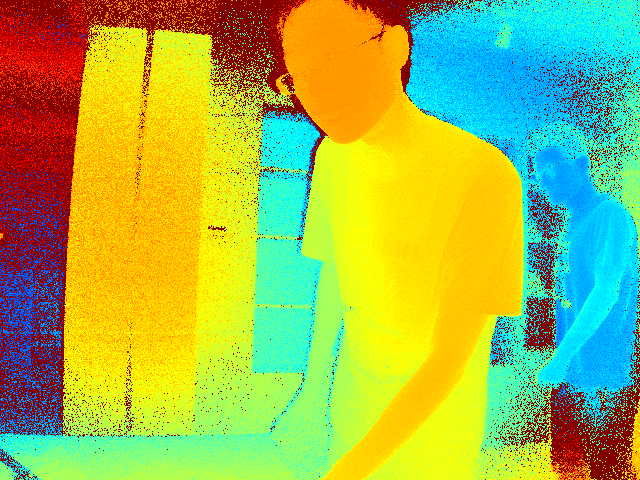}
	}
	\subfigure[Nearest-neighbor (NN)]{
		\includegraphics[width=0.48\linewidth]{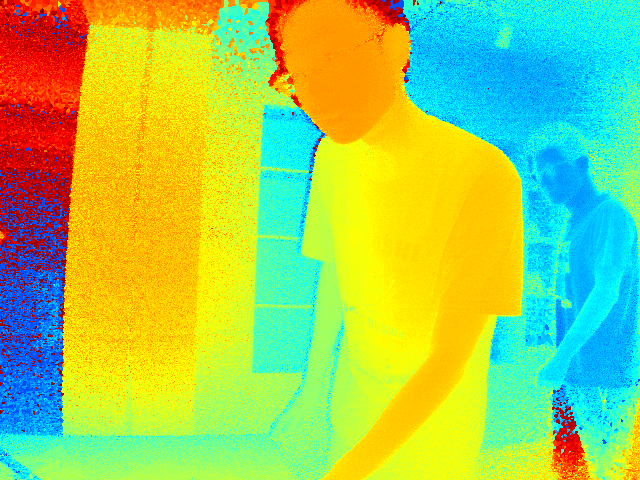}
	}
	\subfigure[NN+BM3D]{
		\includegraphics[width=0.48\linewidth]{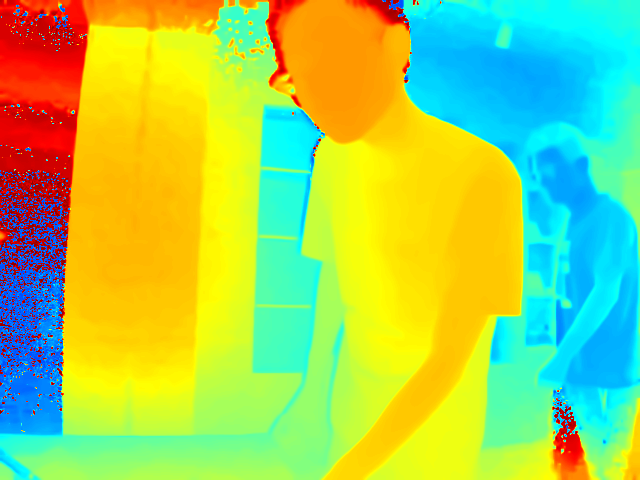}
	}
	\subfigure[Ours]{
		\includegraphics[width=0.48\linewidth]{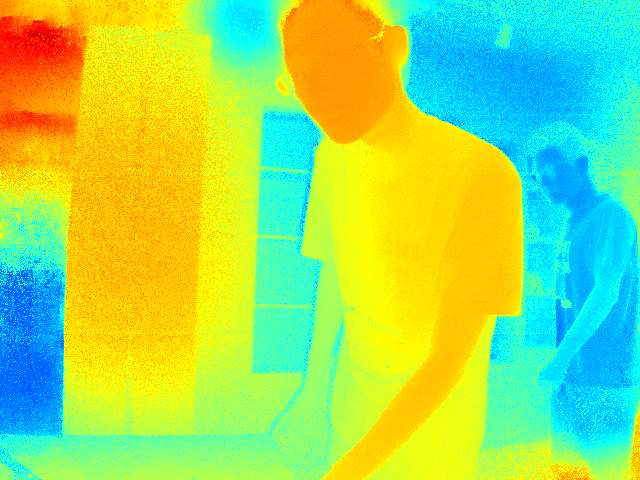}
	}
	\subfigure[Ours+BM3D]{
		\includegraphics[width=0.48\linewidth]{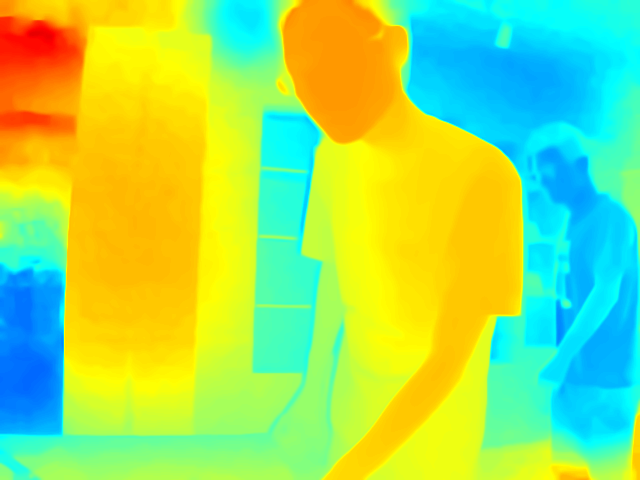}
	}
	\subfigure[Ours+OT denoising]{
		\includegraphics[width=0.48\linewidth]{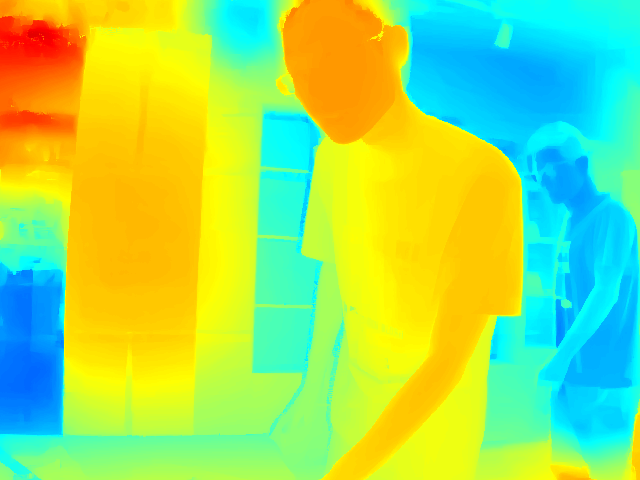}
	}
	\caption{Visual comparison on real-world raw depth images (Example 3). Our method achieves remarkable performance and significantly outperforms the NN and NN+BM3D methods, as it can fill in the holes more naturally while preserving more edge details of the images, especially when it is paired with the OT denoiser (our denoising model trained in Section 5.3 for denoising Gaussian noise).}
\end{figure}
\clearpage

\end{document}